\documentclass[]{aa}
\usepackage{graphicx}
\usepackage{media9}
\usepackage[varg]{txfonts}
\usepackage{url}
\newcommand{\kms}{km\,s$^{-1}$}
\usepackage{natbib}
\bibpunct{(}{)}{;}{a}{}{,}
\usepackage{multirow}
\usepackage{tabularx}
\usepackage{amsmath}
\usepackage{makecell}
\usepackage{multicol}
\usepackage{pdfpages}
\usepackage{tablefootnote}
\usepackage{filecontents,catchfile,booktabs}
\usepackage{placeins}
\usepackage{sidecap}
\begin{document} 

\title{A search for cool molecular gas\\ in GK Persei and other classical novae}
   \author{Tomek Kami\'nski\inst{1} 
          \and Helena J. Mazurek\inst{2}
          \and Karl M.\ Menten\inst{3}
          \and Romuald Tylenda\inst{1}
          }
\institute{\centering
Nicolaus Copernicus Astronomical Center, Polish Academy of Sciences, Rabia{\'n}ska 8, 87-100 Toru{\'n}, \email{tomkam@ncac.torun.pl}\label{inst1}
\and Institute of Astronomy, Nicolaus Copernicus University, Grudzi\k{a}dzka 5, 87-100 Toru{\'n}, Poland\label{inst2}
\and Max-Planck-Institut f\"ur Radioastronomie, Auf dem H\"ugel 69, 53-121 Bonn, Germany \label{inst3}
}
\authorrunning{Kami\'nski et al.}
\titlerunning{Cool molecular gas in novae}
\abstract{Detecting molecular line emission from classical nova remnants has the potential of revealing information on the composition of the ejecta, in particular, it can deliver accurate isotopic ratios in the matter processed by a thermonuclear runaway. We conducted searches toward more than 100 classical novae for emission in lines of the CO or HCN molecules using single-dish telescopes and interferometric arrays at  millimeter and submillimeter wavelengths. The survey demonstrates that classical novae, young or old, are not strong sources of molecular emission at submillimeter and millimeter wavelengths. Additionally, we mapped CO emission around Nova Persei 1901 (GK\,Per), earlier claimed to be circumstellar in origin. Our measurements indicate that the observed  emission is from the interstellar medium. Although no molecular emission at millimeter and submillimeter wavelengths has been found in classical novae, it is still likely that some will be detected with high-sensitivity interferometers such as ALMA.}

\maketitle

\section{Introduction}\label{intro}
Classical novae are close binary systems where a white dwarf accretes material from a companion, typically a late-type dwarf. Accretion via a disk accumulates hydrogen in a degenerate layer on the surface of the white dwarf \citep{CVbook,CNbook}. At some point, the pressure and temperature at the bottom of the accreted envelope become high enough for hydrogen burning to start. This initiates the thermonuclear runaway process
that is directly responsible for classical novae outbursts at luminosities of 10$^4$--10$^5$\,L$_{\sun}$. 
The energetic explosion on the surface of a white dwarf ejects the accumulated material, and, according to some models, may even remove some material forming the outer layers of the white dwarf. A typical mass lost during a single nova explosion is 10$^{-7}$--10$^{-3}$\,M$_{\sun}$. A few percent of this material was processed by the thermonuclear burning. 

Although studied for decades, classical novae are still not fully understood. In particular, the physics of thermonuclear runaway and dust formation associated with novae remain unclear. It is very difficult to model the thermonuclear runaway, mainly because (1) we do not know the mechanism that mixes the accreted envelope and the outermost shells of the white dwarf; (2) we do not fully understand the process of material ejection; and (3) nuclear reaction rates are uncertain \citep[][]{hix}. In consequence, it is not known how effective classical novae are in enhancing the ejected material with products of nuclear burning. Knowing the composition of classical novae ejecta is of great importance because the material eventually reaches the interstellar medium, enhancing it with the nuclear ashes. Classical novae are currently thought to be the major source of $^{13}$C, $^{15}$N, and $^{17}$O in the Galaxy \citep{isotopesISM1,isotopesISM2} and are likely to significantly contribute to the Galactic content of $^7$Li, $^{19}$F, and $^{26}$Al  \citep{Gehrz, izzo}. 

There is thus a great need for robust measurements of the composition of classical novae ejecta. The so-called presolar ``nova grains'', should they indeed originate in classical novae outburst, provide only a limited and biased information on the ejecta composition \citep{amari2001, iliadis}. To date, only a handful of observations has been made whose results permit direct tests of theoretical predictions for the isotopic yields of nova thermonuclear runaway. There is thus a great need for observational data from which elemental isotopic abundances could be inferred \citep[e.g.][]{black} and compared to models. Observational astronomy often relies on molecular spectra to constrain the isotopic composition of objects.

Molecules are expected to form in ejecta of classical novae \citep{PR2004,Derdzinski}. Observations collected in the last decades corroborate this notion. First, it is clear from observations that 30--40\% of classical novae produce dust 10--100 days after their outbursts \citep{CNbook}. Dust cannot be produced without the presence of parent molecules or nucleation seeds, whose formation requires simple diatomic species. Furthermore, dust provides protection for the molecules from the Galactic ultraviolet (UV) field and from the harsh radiation of the nova explosion. Second, molecules are observed in spectra of classical novae at optical and near-infrared (NIR) wavelengths, including bands of C$_2$, CN, and CO \citep[e.g.,][ and references therein]{detectionCN,detectionC2,banerjee2016,fujii}. At these wavelengths, the isotopic shift between different isotopologs is typically too small to be resolved, owing to generally modest spectral resolutions and high intrinsic line widths (500--5000\,\kms). The spectral confusion hampers  stringent observational tests for nova nucleosynthesis. A straightforward method of measuring isotopic ratios, successfully applied to the envelopes of evolved stars and the interstellar medium, are ground-based observations of molecular rotational transitions at millimeter (mm) and submillimeter (submm) wavelengths. 


There have been attempts to detect mm-wave molecular emission in classical novae, which almost all (see Sect.\,\ref{GKPer}) turned out to be unsuccessful  or resulted in detections in ambient material unrelated to the nova phenomenon \citep[for a review see Sect. 13 in][]{CNbook}. Past attempts resulted in upper limits on CO emission obtained with the Swedish-ESO Submillimeter Telescope (SEST) \citep{weight1993,NS2003}, the IRAM 30 m telescope \citep{SB1992}, and with the University of Texas Millimeter Wave Observatory \citep{AE1989}. The reports from before $\sim$1995 are however unreliable and often contradictory. The outflow velocities in novae are of 500--5000\,\kms\ and are thus expected to have very broad lines. which require receivers with a broad bandwidth, and with flat and stable spectroscopic baselines, which in the past were difficult to attain. Most of the published negative results were obtained with instruments which did not fulfill these requirements. Some claims of nonexistent emission have been recently debunked. For instance, a lack of detection of CO(1--0) was reported for Nova\,1670 (also known as CK\,Vul) by Weight\,et\,al. Observed in 2014--2015 with modern receiver systems, CK\,Vul showed clear emission lines of CO for $J_{\rm up}$\,$<$7 \citep{KamiNat, KamiCKSingleDish}. However, CK\,Vul turned out to not be a classical nova but a red nova, that is a stellar merger product, although with similarly broad lines as expected for classical novae, that is $\approx$300\,\kms. Another example is $\eta$\,Carinae, which is not a nova but a luminous blue variable with broad lines. $\eta$\,Car was long claimed to be free of molecular mm-wave emission \citep[e.g.][]{CB}. However, when observed with APEX/FLASH in 2011, it showed a plethora of molecular features \citep{Lonard}. Another example is the first observation of rotational lines of SiO and CO in the remnant of SN\,1987A \citep{sn1987a} using the ALMA interferometer. The lines of the supernova remnant are very broad, FWHM$\approx$2200\,\kms, but were recovered within the ALMA correlator bandwidth, even though the lines were broader than the covered spectral band. For classical novae, spectral lines can be as broad as a few GHz in the mm range, and one needs a receiver system with a large instantaneous bandwidth to be even able to recognize a line and to distinguish its presence from an elevated  continuum level. 

Encouraged by the discoveries in CK\,Vul, $\eta$\,Car, and SNR\,1987A, we decided to revise the view on the presence of molecular emission in classical novae by undertaking an observation campaign aimed to detect molecules at mm wavelengths using modern instruments, including single-dish telescopes and interferometers. The search for molecular emission in the common isotopologs was meant as the first step towards providing isotopic ratios in thermonuclear ashes of classical novae. In Sect.\,\ref{sec-apex} and \ref{sec-iram}, we present observations with APEX and IRAM 30\,m telescopes for a large sample of classical novae and related objects. In Sect.\,\ref{GKPer}, we revise the nature of molecular emission near nova remnant GK\,Per based on mapping observations from the IRAM telescope. In Sect.\,\ref{interer}, we present our attempts to detect molecular emission with the SMA and ALMA interferometers. Finally, in Sect.\,\ref{summary} we summarize the observational material and conclude our findings.

\section{Single-dish observations}\label{obs}
In 2014--2016, we conducted surveys of old classical nova with the APEX and IRAM telescopes, targeting emission lines of CO and HCN in novae on both hemispheres. The idea was to observe many sources at a sensitivity level comparable to that at which emission lines would be easily detectable in CK\,Vul. This typically required 5--20 min of integration on source. We searched for lowest rotational transitions of CO and HCN, because these are usually the strongest lines in cool circumstellar\footnote{By circumstellar we mean in this paper material that is directly originating from stellar outflows.} envelopes. Particularly, emission of HCN might  be expected to be strong if the gas is enhanced in CNO-processed material, as expected for classical novae. 

The sources were selected from the \textit{Central Bureau for Astronomical Telegrams (CBAT) List of Novae in the Milky Way}\footnote{\url{http://www.cbat.eps.harvard.edu/nova_list.html}}, which includes many historical novae. Individual targets were chosen for observations based on their availability during periods assigned to us at the observatories. The lists of sources observed with the APEX and the IRAM telescopes are given in Tables\,\ref{tab-apex-sources} and \ref{tab-iram-sources}, respectively. There, we classified the observed objects using the SIMBAD astronomical database\footnote{\url{http://simbad.u-strasbg.fr/simbad/}} of Centre de Donne\'es Astronomiques de Strasbourg (CDS). According to this classification, as of August 2020, eleven of the observed sources are not novae despite being listed as such in the CBAT catalog. Twelve other objects are dwarf novae. We decided to include them in our survey summary nevertheless. Technical details on the observations are given below.

\begin{table*}
\footnotesize\centering
\caption{Objects observed with APEX.}\label{tab-apex-sources}
    \begin{tabular}{@{}llllll@{}}
    \toprule
    Object & ID & CDS type & Observation date & RA (J2000) & Dec (J2000)\\
\hline
NOVA Cir 1914                        &  AICIR           &  Nova           &  13-Jul-2016  &  14:49:31.2  &  --68:51:35.9  \\
V* AP Cru 	&APCRU	&Nova		& 09-Jun-2015&   12:31:20.5 &  --64:26:25.2  \\
V* AT Sgr                            &  ATSGR           &  Nova           &  14-Jul-2016  &  18:03:30.8  &  --26:28:28.5  \\
V* BD Pav 	&BDPAV	 &CataclyV*	& 10-Jun-2015&   18:43:11.9 &  --57:30:44.9  \\
    V* CG CMa\tablefootmark{a}\tablefootmark{b}                          &  CG-CMA       &  DwarfNova  &  24-Aug-2014  &  07:04:05.0  &  --23:45:34.6  \\
    NOVA Vel 1905\tablefootmark{a}                        &  CN-VEL       &  Nova       &  24-Aug-2014, 09-Jun-2015  &  11:02:38.5  &  --54:23:09.5  \\
    V* CP Pup 	&CPPUP	&Nova		& 08-Jun-2015&   08:11:46.1 &  --35:21:04.9  \\
    NOVA Vel 1940\tablefootmark{a}                          &  CQ-VEL       &  Nova       &  26-Aug-2014  &  08:58:50.9  &  --53:20:17.8  \\
    V* DY Pup\tablefootmark{a}                              &  DY-PUP       &  Nova       &  25-Aug-2014, 08-Jun-2015  &  08:13:48.5  &  --26:33:56.5  \\
NOVA Sgr 1926                        &  FMSGR           &  Nova           &  13-Jul-2016  &  18:17:18.1  &  --23:38:27.0  \\
    NOVA Mus 1983\tablefootmark{a}                          &  GQ-MUS       &  Nova       &  28-Aug-2014  &  11:52:02.3  &  --67:12:20.2  \\
V* GR Sgr                            &  GRSGR           &  Nova           &  13-Jul-2016  &  18:22:58.5  &  --25:34:47.3  \\
    V* GU Mus\tablefootmark{a}\tablefootmark{b}                          &  GU-MUS       &  HMXB       &  24-Aug-2014  &  11:26:26.6  &  --68:40:32.3  \\
NOVA Sgr 1900                        &  HSSGR           &  Nova           &  13-Jul-2016  &  18:28:03.4  &  --21:34:24.7  \\
NOVA Nor 1893                        &  ILNOR           &  Nova           &  07-Jul-2016  &  15:29:23.1  &  --50:35:00.7  \\
NOVA Nor 1920                        &  IMNOR           &  Nova           &  07-Jul-2016  &  15:39:26.4  &  --52:19:17.9  \\
    V* LZ Mus\tablefootmark{a}                             &  LZ-MUS       &  Nova       &  24-Aug-2014  &  11:56:09.2  &  --65:34:20.1  \\
NOVA Cen 1931                        &  MTCEN           &  Nova           &  30-Jun-2016  &  11:44:00.8  &  --60:33:39.5  \\
    V* RR Pic\tablefootmark{a}                              &  RR-PIC       &  Nova       &  25-Aug-2014  &  06:35:36.0  &  --62:38:24.3  \\
    V* RR Tel 	&RRTEL	 &Symbiotic*& 10-Jun-2015&   20:04:18.5 &  --55:43:33.2  \\
NOVA Car 1895                        &  RSCAR           &  Nova           &  30-Jun-2016  &  11:08:06.6  &  --61:56:04.6  \\
    NOVA Pyx 1890\tablefootmark{a}                        &  T-PYX        &  Nova       &  24-Aug-2014  &  09:04:41.5  &  --32:22:47.5  \\
    V* TV Crv\tablefootmark{a}                              &  TV-CRV       &  DwarfNova  &  24-Aug-2014  &  12:20:24.1  &  --18:27:02.0  \\
    V* U Sco 	&USCO	&CataclyV*	& 10-Jun-2015&   16:22:30.8 &  --17:52:43.2  \\
NOVA Sgr 1905                        &  V1015SGR         &  Nova           &  06-Jul-2016  &  18:09:02.0  &  --32:28:32.0  \\
NOVA Sgr 1899                        &  V1016SGR         &  Nova           &  13-Jul-2016  &  18:19:57.6  &  --25:11:14.6  \\
V* V1017 Sgr                         &  V1017SGR         &  Nova           &  13-Jul-2016  &  18:32:04.4  &  --29:23:12.5  \\
    NOVA Cen 2007\tablefootmark{a}                          &  V1065CEN    &  Nova       &  24-Aug-2014  &  11:43:10.3  &  --58:04:04.3  \\
    V* V1148 Sgr                         &  V1148SGR    &  Nova       &  04-Apr-2015  &  18:09:05.8  &  --25:59:08.0  \\
NOVA Sgr 1945 a                      &  V1149SGR         &  Nova           &  13-Jul-2016  &  18:18:30.4  &  --28:17:17.0  \\
V* V1151 Sgr                         &  V1151SGR         &  Nova           &  14-Jul-2016  &  18:25:23.7  &  --20:11:59.3  \\
NOVA Sgr 1928                        &  V1583SGR         &  Nova           &  14-Jul-2016  &  18:15:26.3  &  --23:23:18.0  \\
    V* V351 Car\tablefootmark{a}                            &  V351-CAR     &  Mira       &  28-Aug-2014  &  10:45:19.1  &  --72:03:56.0  \\
    V* V359 Cen\tablefootmark{a}                            &  V359-CEN     &  DwarfNova  &  24-Aug-2014  &  11:58:15.3  &  --41:46:08.4  \\
NOVA Sgr 1927                        &  V363SGR          &  Nova           &  11-Jun-2015, 01-Jul-2016  &  19:11:16.3  &  --29:50:00.0  \\
V* V365 Car                          &  V365CAR          &  Nova           &  09-Jun-2015, 30-Jun-2016  &  11:03:16.7  &  --58:27:24.9  \\
NOVA Sco 1901\tablefootmark{a}                         &  V382SCO          &  Nova           &  13-Jul-2016  &  17:51:56.1  &  --35:25:05.4  \\
    NOVA Vel 1999                        &  V382-VEL     &  Nova       &  28-Aug-2014, 08-Sep-2014  &  10:44:48.3  &  --52:25:30.7  \\
NOVA Sgr 1893                        &  V384SCO          &  Nova           &  14-Jul-2016  &  18:01:43.1  &  --35:39:27.8  \\
V* V522 Sgr                          &  V522SGR          &  Nova           &  11-Jun-2015, 01-Jul-2016  &  18:48:00.4  &  --25:22:21.9  \\
NOVA Oph 1940                        &  V553OPH          &  Nova           &  01-Jul-2016  &  17:42:53.5  &  --24:51:26.2  \\
    NOVA Pup 2007 b\tablefootmark{a}                        &  V598-PUP     &  Nova       &  25-Aug-2014  &  07:05:42.5  &  --38:14:39.4  \\
NOVA Sco 1944                        &  V696SCO          &  Nova           &  13-Jul-2016  &  17:53:11.5  &  --35:50:14.4  \\
NOVA Sco 1941                        &  V697SCO          &  Nova           &  13-Jul-2016  &  17:51:21.8  &  --37:24:55.2  \\
NOVA Sco 1922                        &  V707SCO          &  Nova           &  06-Jul-2016  &  17:48:26.3  &  --36:37:54.9  \\
NOVA Sco 1906                        &  V711SCO          &  Nova           &  14-Jul-2016  &  17:54:06.1  &  --34:21:15.5  \\
V* V729 Sco                          &  V729SCO          &  V*             &  10-Jun-2015, 01-Jul-2016  &  17:22:02.6  &  --32:05:48.8  \\
NOVA Sgr 1936                        &  V732SGR          &  Nova           &  13-Jul-2016  &  17:56:07.5  &  --27:22:16.1  \\
V* V733 Sco                          &  V733SCO          &  $\rm Candidate {Mi*}$  &  06-Jul-2016  &  17:39:42.8  &  --35:52:38.4  \\
NOVA Sgr 1933                        &  V737SGR          &  Nova           &  14-Jul-2016  &  18:07:08.6  &  --28:44:52.3  \\
V* V745 Sco                          &  V745SCO          &  LPV*           &  13-Jul-2016  &  17:55:22.2  &  --33:14:58.5  \\
NOVA Sgr 1937                        &  V787SGR          &  Nova           &  14-Jul-2016  &  18:00:02.2  &  --30:30:31.0  \\
V* V794 Oph                          &  V794OPH          &  Nova           &  01-Jul-2016  &  17:38:49.2  &  --22:50:48.9  \\
V* V840 Oph 	&V840OPH&Nova	& 10-Jun-2015&   16:54:43.9 &  --29:37:26.8  \\
NOVA Sgr 1941                        &  V909SGR          &  Nova           &  13-Jul-2016  &  18:25:52.3  &  --35:01:27.0  \\
V* V941 Sgr                          &  V941SGR          &  Mira           &  13-Jul-2016  &  18:34:43.4  &  --29:34:49.1  \\
NOVA Sgr 1910                        &  V999SGR          &  Nova           &  13-Jul-2016  &  18:00:05.5  &  --27:33:14.0  \\
    V* VX For\tablefootmark{a}                              &  VX-FOR       &  DwarfNova  &  24-Aug-2014  &  03:26:45.7  &  --34:26:25.2  \\
    V* WX Cet\tablefootmark{a}                              &  WX-CET       &  DwarfNova  &  24-Aug-2014  &  01:17:04.1  &  --17:56:23.0  \\

    \bottomrule
    \end{tabular}
    \tablefoot{Objects were observed in the CO(3--2), CO(3--2) OSB, CO(4--3), CO(4--3) OSB setups. \tablefoottext{a}{Observation reported in \cite{KamiNat}.} \tablefoottext{b}{ Observed only in CO(3--2), CO(3--2) OSB setups.}}
\end{table*}

\begin{table*}
\footnotesize\centering
\caption{Objects observed with IRAM.}\label{tab-iram-sources}
    \begin{tabular}{@{}lllllll@{}}
    \toprule
    Object & ID & CDS type & Observation date & RA (J2000) & Dec (J2000) & Target lines\\
\hline
V* BC Cas                            &  BC-CAS       &  Nova        &  10-Aug-2015  &  23:51:17.4  &  60:18:10.0  &  $ \rm HCN(1-0), ^{12}$CO$(1-0)$    \\
NOVA Aql 1917                        &  CI-AQL       &  Nova        &  09-Aug-2015  &  18:52:03.5  &  01:28:39.4  &  $ \rm HCN(1-0)$      \\
V* CI Gem                            &  CI-GEM       &  DwarfNova   &  10-Aug-2015  &  06:30:05.8  &  22:18:50.7  &  $ \rm HCN(1-0)$      \\
V* CK Vul                            &  CK-VUL       &  CataclyV*   &  07-Aug-2015  &  19:47:37.9  &  27:18:48.0  &  $ \rm HCN(1-0)$      \\
V* CP Lac                            &  CP-LAC       &  Nova        &  09-Aug-2015  &  22:15:41.0  &  55:37:01.3  &  $ \rm HCN(1-0), ^{12}$CO$(1-0)$    \\
V* DI Lac                            &  DI-LAC       &  Nova        &  10-Aug-2015  &  22:35:48.4  &  52:42:59.6  &  $ \rm HCN(1-0), ^{12}$CO$(1-0)$    \\
NOVA Gem 1903                        &  DM-GEM       &  Nova        &  10-Aug-2015  &  06:44:12.0  &  29:56:41.8  &  $ \rm HCN(1-0)$      \\
V* DN Gem                            &  DN-GEM       &  Nova        &  10-Aug-2015  &  06:54:54.3  &  32:08:27.9  &  $ \rm HCN(1-0)$      \\
NOVA Ser 1960                        &  DZ-SER       &  Nova        &  07-Aug-2015  &  18:00:58.8  &  10:33:52.1  &  $ \rm HCN(1-0)$      \\
V* EL Aql                            &  EL-AQL       &  Nova        &  09-Aug-2015  &  18:56:02.0  &  03:19:20.5  &  $ \rm HCN(1-0)$      \\
V* GK Per                            &  GK-PER       &  Nova        &  07-Aug-2015  &  03:31:12.0  &  43:54:15.4  &  $ \rm HCN(1-0)$      \\
NOVA Ori 1916                        &  GR-ORI       &  Nova        &  11-Aug-2015  &  05:21:35.2  &  01:10:08.6  &  $ \rm HCN(1-0)$      \\
NOVA Sge 1977                        &  HS-SGE       &  Nova        &  10-Aug-2015  &  19:39:22.0  &  18:07:53.9  &  $ \rm HCN(1-0)$      \\
NOVA Mon 1942                        &  KT-MON       &  Nova        &  11-Aug-2015  &  06:25:18.4  &  05:26:31.7  &  $ \rm HCN(1-0)$      \\
NOVA And 1986                        &  OS-AND       &  Nova        &  11-Aug-2015  &  03:12:05.7  &  47:28:19.7  &  $ \rm HCN(1-0)$      \\
V* PQ And                            &  PQ-AND       &  DwarfNova   &  07-Aug-2015  &  02:29:29.6  &  40:02:40.9  &  $ \rm HCN(1-0)$      \\
V* Q Cyg                             &  Q-CYG     &  Nova        &  09-Aug-2015  &  21:41:43.9  &  42:50:29.0  &  $ \rm HCN(1-0)$      \\
V* RS Oph                            &  RS-OPH       &  Nova        &  07-Aug-2015  &  17:50:13.1  &  06:42:28.4  &  $ \rm HCN(1-0)$      \\
NOVA UMi 1956                        &  RW-UMI       &  Nova        &  10-Aug-2015  &  16:47:54.7  &  77:02:12.1  &  $ \rm HCN(1-0)$      \\
NOVA Sge 1916                        &  SS-SGE       &  Nova        &  10-Aug-2015  &  19:39:08.3  &  16:42:40.5  &  $ \rm HCN(1-0)$      \\
NOVA Gem 1857                        &  SY-GEM       &  Nova        &  10-Aug-2015  &  06:40:39.1  &  31:10:55.3  &  $ \rm HCN(1-0)$      \\
NOVA Per 1853                        &  SZ-PER       &  Nova        &  07-Aug-2015  &  03:47:06.3  &  34:19:17.9  &  $ \rm HCN(1-0)$      \\
V* UW Per                            &  UW-PER       &  DwarfNova   &  07-Aug-2015  &  02:12:29.5  &  57:05:19.7  &  $ \rm HCN(1-0)$      \\
V* UZ Tri                            &  UZ-TRI       &  Nova        &  10-Aug-2015  &  01:58:24.6  &  33:31:30.0  &  $ \rm HCN(1-0)$      \\
V* V1059 Sgr                         &  V1059-SGR    &  Nova        &  09-Aug-2015  &  19:01:50.5  &  13:09:41.9  &  $ \rm HCN(1-0)$      \\
V* V1229 Aql                         &  V1229-AQL    &  Nova        &  09-Aug-2015  &  19:24:44.5  &  04:14:48.6  &  $ \rm HCN(1-0)$      \\
V* V1378 Aql                         &  V1378-AQL    &  Nova        &  09-Aug-2015  &  19:16:35.4  &  03:43:26.3  &  $ \rm HCN(1-0)$      \\
V* V1449 Cyg                         &  V1449-CYG    &  DwarfNova   &  09-Aug-2015  &  19:49:16.5  &  34:10:49.1  &  $ \rm HCN(1-0)$      \\
V* V1500 Cyg                         &  V1500-CYG    &  Nova        &  09-Aug-2015  &  21:11:36.3  &  48:09:05.8  &  $ \rm HCN(1-0)$      \\
NOVA Cyg 1978                        &  V1668-CYG    &  Nova        &  09-Aug-2015  &  21:42:35.2  &  44:01:54.9  &  $ \rm HCN(1-0)$      \\
V* V1697 Cyg                         &  V1697-CYG    &  DwarfNova   &  09-Aug-2015  &  20:43:16.9  &  42:42:38.9  &  $ \rm HCN(1-0)$      \\
V* V1974 Cyg                         &  V1974-CYG    &  Nova        &  09-Aug-2015  &  20:30:31.6  &  52:37:51.3  &  $ \rm HCN(1-0)$      \\
NOVA Oph 1994                        &  V2313-OPH    &  Nova        &  08-Aug-2015  &  17:35:44.6  &  19:19:34.0  &  $ \rm HCN(1-0)$      \\
NOVA Cyg 2005                        &  V2361-CYG    &  Nova        &  09-Aug-2015  &  20:09:19.0  &  39:48:52.9  &  $ \rm HCN(1-0)$      \\
NOVA Cyg 2006                        &  V2362-CYG    &  Nova        &  09-Aug-2015  &  21:11:32.3  &  44:48:03.6  &  $ \rm HCN(1-0)$      \\
V* V2467 Cyg                         &  V2467-CYG    &  Nova        &  09-Aug-2015  &  20:28:12.4  &  41:48:36.5  &  $ \rm HCN(1-0)$      \\
PN K 3-25                            &  V352-AQL     &  Symbiotic*  &  09-Aug-2015  &  19:13:33.6  &  02:18:12.9  &  $ \rm HCN(1-0)$      \\
NOVA Her 1892                        &  V360-HER     &  Nova        &  08-Aug-2015  &  17:16:38.0  &  24:26:47.2  &  $ \rm HCN(1-0)$      \\
V* V368 Aql                          &  V368-AQL     &  Nova        &  10-Aug-2015  &  19:26:34.4  &  07:36:13.8  &  $ \rm HCN(1-0)$      \\
NOVA Per 1974                        &  V400-PER     &  Nova        &  07-Aug-2015  &  03:07:38.1  &  47:07:38.8  &  $ \rm HCN(1-0)$      \\
V* V446 Her                          &  V446-HER     &  Nova        &  08-Aug-2015  &  18:57:21.5  &  13:14:29.8  &  $ \rm HCN(1-0)$      \\
NOVA Cyg 1948                        &  V465-CYG     &  Nova        &  09-Aug-2015  &  19:52:37.6  &  36:33:52.6  &  $ \rm HCN(1-0)$      \\
V* V476 Cyg                          &  V476-CYG     &  Nova        &  09-Aug-2015  &  19:58:24.4  &  53:37:07.4  &  $ \rm HCN(1-0)$      \\
NOVA Aql 1943                        &  V500-AQL     &  Nova        &  10-Aug-2015  &  19:52:27.8  &  08:28:46.3  &  $ \rm HCN(1-0)$      \\
NOVA Ori 1667                        &  V529-ORI     &  Nova        &  10-Aug-2015  &  05:58:20.1  &  20:15:45.4  &  $ \rm HCN(1-0)$      \\
HD 176779                            &  V604-AQL     &  Nova        &  09-Aug-2015  &  19:02:06.3  &  04:26:43.2  &  $ \rm HCN(1-0)$      \\
HD 181419                            &  V606-AQL     &  Nova        &  09-Aug-2015  &  19:20:24.2  &  00:08:07.0  &  $ \rm HCN(1-0)$      \\
V* V630 Cas                          &  V630-CAS     &  DwarfNova   &  10-Aug-2015  &  23:48:51.9  &  51:27:39.1  &  $ \rm HCN(1-0), ^{12}$CO$(1-0)$    \\
V* V705 Cas                          &  V705-CAS     &  Nova        &  11-Aug-2015  &  23:41:47.1  &  57:30:59.5  &  $ \rm HCN(1-0)$      \\
V* V723 Cas                          &  V723-CAS     &  Nova        &  10-Aug-2015  &  01:05:05.3  &  54:00:40.2  &  $ \rm HCN(1-0), ^{12}$CO$(1-0)$    \\
NOVA Her 1991                        &  V838-HER     &  Nova        &  08-Aug-2015  &  18:46:31.5  &  12:14:00.6  &  $ \rm HCN(1-0)$      \\
NOVA Aql 1951                        &  V841-AQL     &  Nova        &  09-Aug-2015  &  19:07:39.7  &  10:29:43.7  &  $ \rm HCN(1-0)$      \\
NOVA Oph 1919                        &  V849-OPH     &  Nova        &  08-Aug-2015  &  18:14:07.1  &  11:36:42.7  &  $ \rm HCN(1-0)$      \\
NOVA Per 1887                        &  V-PER        &  Nova        &  10-Aug-2015  &  02:01:53.7  &  56:44:04.0  &  $ \rm HCN(1-0), ^{12}$CO$(1-0)$    \\
V* VY Aqr                            &  VY-AQR       &  DwarfNova   &  06-Aug-2015  &  21:12:09.2  &  08:49:36.7  &  $ \rm HCN(1-0)$      \\
NOVA Gem 1856                        &  VZ-GEM       &  Nova        &  11-Aug-2015  &  08:07:47.0  &  30:50:54.0  &  $ \rm HCN(1-0)$      \\
NOVA Ari 1855                        &  W-ARI        &  Nova        &  07-Aug-2015  &  03:20:45.1  &  28:57:13.9  &  $ \rm HCN(1-0)$      \\
NOVA Sge 1783                        &  WY-SGE       &  Nova        &  10-Aug-2015  &  19:32:43.8  &  17:44:55.8  &  $ \rm HCN(1-0)$      \\
    \bottomrule
    \end{tabular}
\end{table*}

\subsection{APEX observations and data reduction}\label{sec-apex}
The fifty-nine objects listed in Table\,\ref{tab-apex-sources} were observed with the Atacama Pathfinder Experiment (APEX) 12-m submillimeter telescope \citep{gusten}. Forty-five of these objects are classified as classical novae. The observations took place in 24--28 July 2014, 4 April 2015, 8--11 June 2015, 10 July 2015, 30 July 2016, and 1, 6, 13 and 14 July 2016. For frequencies between 278 and 492\,GHz, we used the FLASH$^+$ receiver \citep{flash}, which operated simultaneously in two atmospheric bands at about 345\,GHz and 460\,GHz. The version of FLASH$^+$ separates the two heterodyne sidebands in each of its two modules, which cover the 345 and 460\,GHz bands, giving four spectra, each 4\,GHz wide. The image-band rejection factors used are all above 10\,dB and at most frequencies the rejection is higher than 20\,dB. In all APEX observations, we used the eXtended Fast Fourier Transform Spectrometer \citep[XFFTS;][]{klein}, which typically provided us with resolutions better than 0.05\,\kms. To decrease the noise to levels appropriate for the expected broad lines, we rebinned the spectra to a much lower resolution, typically near 30\,\kms. In all observations, we used wobbler switching with the maximal azimuthal wobbler throw of 300\arcsec\ (typically throws of 60--100\arcsec\ were used), which resulted in flat baselines.

Overall, with APEX we covered four spectral ranges characterized in Table\,\ref{spectral-setups}. The table includes the nominal FWHM beam size and  the main-beam efficiency ($\eta_{mb}$). In the two out of four units of FLASH$^+$, we obtained spectra centered on the CO(3--2) and (4--3) lines with rest frequencies of 345.79598990 and 461.0407682 GHz, respectively.  Also covered were the H$^{13}$CO(4--3), SiS(19--18), and SiO(8--7), transition which are readily apparent in the spectrum of one of our the sources we observed to check the telescope's pointing performance, that is IRC+10216 (cf. Fig.\,\ref{apex-co32}). We also registered the upper side band spectra (which we refer to as `CO(3--2) OSB' and `CO(4--3) OSB'). We present all results in the antenna ($T_A^*$) temperature scale. Velocities are expressed in the local standard of rest frame (LSR).

The APEX spectra were reduced in the CLASS package within GILDAS\footnote{\url{https://www.iram.fr/IRAMFR/GILDAS}}. The data reduction involved 0$\rm ^{th}$ order baseline subtraction and spectra averaging. Any faulty channels at spectral edges or caused by instrumental effects were blanked or interpolated between two nearest non-blanked channels. Sample APEX spectra are displayed in Figs.\,\ref{apex-co32}--\ref{apex-co43-osb} where we marked the corresponding 3$\sigma$ noise levels.  

APEX data for seventeen sources have been already reported in \citet{KamiNat}. They are included here for completeness and consistency, and were re-processed using the procedures presented here.

\begin{table}
\footnotesize\centering
\caption{Spectral setups used in the survey.}\label{spectral-setups}
    \begin{tabular}{cc ccc}
    \toprule
Telescope & Setup name & Range (GHz)& Beam (\arcsec) & $\eta_{\rm mb}$ \\
\hline
APEX & CO(3--2) OSB &  331.8--335.8 & 18.7 & 0.70 \\
APEX & CO(3--2)     &  343.8--348.8 & 18.0 & 0.69 \\
APEX & CO(4--3)     &  459.0--463.0 & 13.5 & 0.64 \\
APEX & CO(4--3) OSB &  471.0--475.0 & 13.2 & 0.63 \\[5pt]
IRAM & HCN(1--0)    &   83.1--91.0  & 27.8 & 0.85\\
IRAM & HCN(1--0) USB&   98.8--106.6 & 24.3 & 0.84\\
IRAM & CO(1--0) LSB &   94.1--101.9 & 25.5 & 0.84\\
IRAM & CO(1--0) USB &  109.8--117.5 & 21.3 & 0.83\\
    \bottomrule
    \end{tabular}
\end{table}

\subsection{IRAM observations and data reduction}\label{sec-iram}

Observations with the IRAM 30\,m telescope and the EMIR dual-sideband receiver \citep{emir} were obtained in 6--11 August 2015. Fifty-eight objects observed with IRAM are listed in Table\,\ref{tab-iram-sources} and include 49 classical novae. Spectra for each linear polarization covered the same spectral range. The Fast Fourier Transform Spectrometer \citep[FTS,][]{klein} and the WILMA correlator were used in parallel as backends. The FTS was used as the primary spectrometer, while WILMA data were used for reference since it covered a smaller part of the band (7.5\,GHz) than FTS (16\,GHz). WILMA spectra, as free from a platforming effect, were analyzed carefully, too. All IRAM observations were obtained using wobbler switching with OFF positions located azimuthaly 110\arcsec\ away from the central ON position.

The spectral setups used are listed in Table\,\ref{spectral-setups}. All objects were observed with the setup whose lower side band was centered on the HCN(1--0) transition at 88.63160230\,GHz; it serendipitously covered also the 1--0 transitions of H$^{13}$CN and HNC (cf. Fig.\,\ref{hcn-fts-low}). For six objects, additional spectra were obtained in the $^{12}$CO(1--0) transition at 115.2712018\,GHz; the corresponding lower sideband spectra covered also the $^{13}$CO(1--0) transition at 110.20135430\,GHz.

All data reduction procedures were performed using CLASS and involved a low-order baseline subtraction and averaging. The FTS spectra required applying a deplatforming routine. While searching for broad nova lines, we used Hanning smoothing to reduce the resolution to about 30\,\kms.  Spectra of all objects observed in the $^{12}$CO(1--0) transition and a sample of spectra with the HCN(1--0) setup are displayed in Figs.\,\ref{hcn-fts-low}--\ref{co-fts-high} together with marked 3$\sigma$ noise levels calculated for the entire spectrum. Owing to the changing atmospheric transparency with wavelength, the noise levels change over the spectrum, especially in the CO(1--0) setup (cf. Fig.\,\ref{co-fts-low}).

\subsection{Single-dish survey results: no detection of nova lines} 
To search for broad lines expected from classical novae outflows, we analyzed the spectra at a reduced spectral resolution of about 30\,\kms. Except for CK\,Vul, known to have broad molecular emission from earlier studies, none of our survey targets shows broad features that can be associated with a nova system. 

The measured rms noise levels for APEX spectra are listed in Table\,\ref{rms_apex}. The rms noise levels for the CO(3--2) spectra are relatively low, of 1--6\,mK.  The observations of CO(4--3) are much less constraining due to a lower atmosphere transparency at the respective frequencies. The IRAM observations in the CO(1--0) line for a handful of objects are comparatively deep to the best APEX spectra, with global rms noise levels listed in Table\,\ref{rms_all_CO}. The noise levels are several times smaller near the rest frequency of the CO(1--0). The upper limits on HCN emission can be derived from the rms values listed in Table\,\ref{rms_all_HCN}.

The southern part of the survey constitutes the largest sample of novae and related objects that were checked for the presence of rotational lines of CO. The noise levels achieved in our CO survey are one order of magnitude lower than in earlier observations of classical novae \citep[e.g.][]{SB1992}. The IRAM survey of northern sources in the HCN(1--0) transition is the first one where lines other than those of CO were targeted for a large sample of classical novae. 

\begin{table} 
    \footnotesize\centering
    \caption{Noise levels in APEX observations.}
     \label{rms_apex}
    \begin{tabular}{@{}llllll@{}}
    \toprule
    Object & \multicolumn{4}{c}{rms (mK)}\\
       &\small{CO32} & \small{CO32\,OSB} & \small{CO43} & \small{CO43\,OSB}\\
\midrule
NOVA Cir 1914    &  6.49  &  7.52  &  23.19  &  120.60\\
V* AT Sgr  &  3.55  &  4.17  &  9.48  &  33.44\\
V* CG CMa    &  1.57  &  1.87  &   & \\
NOVA Vel 1905  &  1.28 & 2.17 & 2.45 & 9.73 \\ 
NOVA Vel 1940  &  1.67  &  1.83  &  4.34  &  15.45\\
V* DY Pup    & 1.90 & 1.36 & 2.58 & 9.1 \\
NOVA Sgr 1926    &  3.87  &  4.08  &  8.79  &  35.42\\
NOVA Mus 1983  &  2.40  &  3.04  &  6.46  &  19.67\\
V* GR Sgr  &  2.56  &  3.29  &  8.87  &  35.24\\
V* GU Mus    &  2.11  &  2.53  &    &  \\
NOVA Sgr 1900    &  3.69  &  4.17  &  10.23  &  47.94\\
NOVA Nor 1893    &  3.86  &  3.92  &  10.74  &  65.30\\
NOVA Nor 1920    &  4.10  &  4.61  &  9.95  &  48.29\\
V* LZ Mus    &  2.45  &  2.82  &  10.86  &  43.82\\
NOVA Cen 1931    &  2.77  &  2.63  &  8.05  &  39.86\\
V* RR Pic    &  2.07  &  1.80  &  4.78  &  15.63\\
NOVA Car 1895    &  2.92  &  3.14  &  7.73  &  35.71\\
NOVA Pyx 1890  &  1.77  &  1.75  &  4.21  &  17.27\\
V* TV Crv    &  1.70  &  1.83  &  5.19  &  15.04\\
NOVA Sgr 1905    &  2.79  &  3.13  &  9.86  &  48.87\\
NOVA Sgr 1899    &  4.11  &  3.92  &  8.41  &  42.64\\
V* V1017 Sgr  &  2.43  &  2.91  &  6.09  &  22.74\\
NOVA Cen 2007  &  3.08  &  3.57  &  17.36  &  67.42\\
V* V1148 Sgr  &  1.75  &  1.58  &  2.97  &  9.31\\
NOVA Sgr 1945 a  &  3.33  &  3.40  &  5.36  &  16.59\\
V* V1151 Sgr  &  4.46  &  4.57  &  13.99  &  71.71\\
NOVA Sgr 1928    &  4.21  &  4.29  &  11.36  &  65.54\\
V* V351 Car  &  2.60  &  2.36  &  8.26  &  36.29\\
V* V359 Cen  &  1.68  &  1.85  &  4.35  &  13.84\\
NOVA Sgr 1927    &  3.07 & 3.63 & 17.23 & 107.46 \\ 
V* V365 Car  &  2.88 & 2.01 & 3.6 & 12.62 \\ 
NOVA Sco 1901    &  2.86  &  3.05  &  5.73  &  17.21\\
NOVA Vel 1999  &  1.06  &  2.03  &  4.06  &  33.58\\
NOVA Sgr 1893    &  3.94  &  4.82  &  13.10  &  64.73\\
V* V522 Sgr  &  4.62 & 5.05 & 28.57 & 120.05 \\ 
NOVA Oph 1940    &  5.81  &  6.83  &  30.70  &  197.26\\
NOVA Pup 2007 b  &  1.59  &  1.85  &  3.73  &  10.60\\
NOVA Sco 1944    &  3.27  &  3.34  &  6.98  &  26.70\\
NOVA Sco 1941    &  3.33  &  3.09  &  5.57  &  23.98\\
NOVA Sco 1922    &  4.03  &  3.99  &  13.37  &  82.88\\
NOVA Sco 1906    &  4.12  &  4.32  &  11.00  &  58.49\\
V* V729 Sco  &  2.58 & 3.75 & 6 & 17.57 \\ 
NOVA Sgr 1936    &  3.13  &  3.42  &  17.30  &  32.48\\
V* V733 Sco  &  4.05  &  4.94  &  12.78  &  69.67\\
NOVA Sgr 1933    &  4.26  &  3.69  &  9.47  &  36.23\\
V* V745 Sco  &  2.71  &  3.08  &  6.76  &  15.60\\
NOVA Sgr 1937    &  3.96  &  3.61  &  10.33  &  43.59\\
V* V794 Oph  &  5.07  &  7.11  &  27.27  &  184.94\\
NOVA Sgr 1941    &  3.17  &  3.59  &  8.13  &  30.68\\
V* V941 Sgr  &  3.90  &  4.12  &  11.39  &  55.70\\
NOVA Sgr 1910    &  3.55  &  3.75  &  7.11  &  34.40\\
V* VX For    &  2.05  &  2.34  &  5.87  &  24.96\\
V* WX Cet    &  1.84  &  1.68  &  4.91  &  12.56\\
V* CP Pup    &1.98 & 2.02 & 3.33 & 12.91 \\
V* AP Cru    &2.90 & 2.80 & 6.18 & 19.73 \\
V* U Sco     &3.13 & 4.25 & 7.34 & 25.34 \\
V* V840 Oph  &3.65 & 3.72 & 7.39 & 30.64 \\
V* BD Pav    &4.19 & 5.25 & 14.33 & 83.76 \\
V* RR Tel    &4.52 & 4.57 & 10.02 & 62.0 \\

    \bottomrule
    \end{tabular}
    \tablefoot{Root-mean-square values are given per velocity bins of 30\,\kms\ for CO(3--2) and CO(4--3), 31\,\kms\ for CO(3--2)\,OSB, and 29\,\kms\ for CO(4--3)\,OSB.}
\end{table}

\begin{table} 
    \footnotesize\centering
    \caption{Noise levels in IRAM spectra near the HCN(1--0) line.}
    \label{rms_all_HCN}
    \begin{tabular}{@{}lllll@{}}
    \toprule
           & \multicolumn{4}{c}{rms (mK)}\\
       & \multicolumn{2}{c}{WILMA}&\multicolumn{2}{c}{FTS}\\
Object &  LSB &  USB &  LSB &  USB\\
\midrule
V* BC Cas  &  4.26  &  5.25  &  4.44  &  5.47\\
NOVA Aql 1917   &  3.83  &  4.51  &  3.98  &  4.80\\
V* CI Gem  &  4.76  &  5.74  &  4.87  &  6.04\\
V* CK Vul  &  7.10  &  8.33  &  7.51  &  9.43\\
V* CP Lac  &  3.29  &  4.03  &  3.37  &  4.14\\
V* DI Lac  &  4.08  &  4.64  &  3.24  &  3.73\\
NOVA Gem 1903   &  3.93  &  5.05  &  4.42  &  5.24\\
V* DN Gem  &  3.87  &  4.81  &  4.07  &  4.76\\
NOVA Ser 1960   &  7.29  &  8.44  &  7.65  &  9.48\\
V* EL Aql  &  3.90  &  4.11  &  4.03  &  4.72\\
V* GK Per  &  2.61  &  2.80  &  2.67  &  3.00\\
NOVA Ori 1916   &  6.24  &  7.75  &  7.02  &  7.69\\
NOVA Sge 1977   &  4.74  &  5.72  &  5.19  &  6.11\\
NOVA Mon 1942   &  5.54  &  6.63  &  6.04  &  7.01\\
NOVA And 1986   &  3.38  &  4.06  &  3.51  &  4.01\\
V* PQ And  &  2.40  &  2.85  &  2.64  &  2.99\\
V* Q Cyg  &  3.24  &  3.60  &  3.60  &  4.00\\
V* RS Oph  &  6.71  &  8.48  &  7.09  &  8.94\\
NOVA UMi 1956   &  5.21  &  6.11  &  5.41  &  6.45\\
NOVA Sge 1916   &  4.56  &  5.84  &  5.07  &  6.08\\
NOVA Gem 1857   &  4.15  &  5.30  &  4.44  &  5.35\\
NOVA Per 1853   &  2.58  &  2.79  &  2.61  &  3.08\\
V* UW Per  &  3.14  &  3.62  &  3.38  &  3.81\\
V* UZ Tri  &  5.59  &  6.57  &  6.02  &  6.97\\
V* V1059 Sgr    &  3.50  &  3.83  &  3.63  &  4.09\\
V* V1229 Aql    &  4.01  &  4.71  &  4.29  &  5.08\\
V* V1378 Aql    &  3.97  &  4.48  &  4.25  &  5.05\\
V* V1449 Cyg    &  3.30  &  4.17  &  3.69  &  4.19\\
V* V1500 Cyg    &  3.20  &  3.70  &  3.68  &  3.96\\
NOVA Cyg 1978   &  3.30  &  3.94  &  3.43  &  3.95\\
V* V1697 Cyg    &  3.39  &  3.77  &  3.71  &  4.31\\
V* V1974 Cyg    &  3.41  &  4.03  &  3.75  &  4.22\\
NOVA Oph 1994   &  7.53  &  9.25  &  7.50  &  9.56\\
NOVA Cyg 2005   &  3.36  &  3.94  &  3.51  &  4.18\\
NOVA Cyg 2006   &  3.11  &  4.23  &  3.51  &  4.13\\
V* V2467 Cyg    &  3.23  &  3.89  &  3.65  &  4.18\\
PN K 3-25  &  2.78  &  3.45  &  2.75  &  3.55\\
NOVA Her 1892   &  9.47  &  11.77  &  9.27  &  12.39\\
V* V368 Aql  &  4.57  &  5.67  &  4.71  &  6.03\\
NOVA Per 1974   &  3.20  &  3.73  &  3.40  &  3.91\\
V* V446 Her  &  3.46  &  4.11  &  3.61  &  4.16\\
NOVA Cyg 1948   &  2.36  &  2.87  &  2.60  &  3.02\\
V* V476 Cyg  &  3.43  &  4.10  &  3.63  &  4.64\\
NOVA Aql 1943   &  4.54  &  5.89  &  4.74  &  5.76\\
NOVA Ori 1667   &  4.17  &  5.47  &  4.83  &  5.96\\
HD 176779  &  3.69  &  4.39  &  3.96  &  4.83\\
HD 181419  &  4.03  &  4.41  &  4.14  &  4.90\\
V* V630 Cas  &  4.37  &  4.59  &  4.32  &  5.06\\
V* V705 Cas  &  3.31  &  4.10  &  3.49  &  4.11\\
V* V723 Cas  &  3.75  &  4.75  &  4.33  &  4.91\\
NOVA Her 1991   &  3.53  &  4.11  &  3.84  &  4.17\\
NOVA Aql 1951   &  3.62  &  3.79  &  3.85  &  4.31\\
NOVA Oph 1919   &  17.22  &  20.94  &  16.91  &  22.81\\
NOVA Per 1887   &  4.13  &  4.87  &  4.47  &  5.45\\
V* VY Aqr  &  2.65  &  3.18  &  2.98  &  3.46\\
NOVA Gem 1856   &  3.63  &  4.19  &  3.83  &  4.71\\
NOVA Ari 1855   &  2.69  &  2.92  &  2.77  &  3.28\\
NOVA Sge 1783   &  4.73  &  5.67  &  4.98  &  5.97\\

    \bottomrule
    \end{tabular}
    \tablefoot{The noise rms values per 30\,\kms\ are given for both backends used (they differ in covarage and baseline quality). The LSB values correspond to the spectra which covered the HCN(1--0) transition.}
\end{table}

\begin{table} 
    \footnotesize\centering
    \caption{Noise levels in IRAM spectra near the CO(1--0) line.}
    \label{rms_all_CO}
    \begin{tabular}{@{\extracolsep{6pt}}llllll@{}}
    \toprule
       & \multicolumn{5}{c}{rms (mK)}\\ \cline{2-6}
       & \multicolumn{2}{c}{WILMA}&\multicolumn{3}{c}{FTS}\\
       \cline{2-3}\cline{4-6}
Object &  LSB &  USB &  LSB &  USB & CO10\tablefootmark{a}\\
\midrule
V* BC Cas   &  2.21  &  3.14  &  2.38  &  16.06  & 6.22 \\
V* CP Lac   &  2.29  &  3.47  &  2.22  &  10.39  & 6.87 \\
V* DI Lac   &  2.13  &  3.12  &  2.27  &  11.10  & 7.56 \\
V* V630 Cas &  1.62  &  2.28  &  1.57  &  6.23   & 4.71 \\
V* V723 Cas &  2.13  &  2.88  &  2.16  &  8.81   & 6.68 \\
NOVA Per 1887&  2.11  &  2.96  &  2.52  & 10.15  & 7.09 \\
    \bottomrule
    \end{tabular}
    \tablefoot{The $T_A^*$ noise rms values per 30\,\kms\ are given for both backends used (they differ in covarage and baseline quality). The USB values correspond to the entire spectra which covered the CO(1--0) transition.\tablefoottext{a}{The rms value measured in 114.27--116.27\,GHz, near the rest frequency of the CO(1--0) transition and ignoring the ISM features.}}
\end{table}

Our upper limits on the molecular emission are not easily convertible to the limits on the total molecular mass because no direct information about the flow velocity, envelope size, and gas excitation temperature are known for individual objects. Here, we make an attempt to make such a conversion for a model nova representative of our survey. By referring to Gaia distances to nearly half of our sources in \citet{distances}, we find that our typical classical nova is at a distance of 2\,kpc. We further assume a typical flow velocity of $\Delta V$=1000\,\kms. Then, with an age of our typical nova remnant of about 50\,yr, the shell has a size of 0.1\,pc. In the optically thin limit, the mass of molecular material can be calculated as
$$M_{\rm mol}=\frac{ 8 \pi k \mu m_{\rm H} Z(T_{\rm ex}) \nu^2 \exp(E_u/T_{\rm ex}) \eta_{\rm mb}}{h c^3 g_u A_{ul}}Wa \left( \frac{\theta_{\rm beam}}{\theta_{\rm source}}\right)^2,$$
where 
$$W=\int T_A^* d\varv \approx 3\times{\rm rms}\times 2\times\Delta V$$
is the integrated intensity within the line profile, assuming a rectangular feature with a peak intensity equal to three times the measured noise level. $k$, $h$, and $c$ are the Boltzmann and Planck constants, and the light speed, respectively; $A_{ul}$, $g_u$, $E_u$, and $\nu$ are Einstein coefficient, upper level statistical weight, energy (in K) of the upper level, and frequency of the considered transition; $Z$ is the partition function at the excitation temperature $T_{\rm ex}$; the angular beam size is designated as $\theta_{\rm beam}$ while the source angular size is $\theta_{\rm source}$; $\mu$ is the molecular weight and $m_{\rm H}$ is the mass of hydrogen atom; $a$ is the source projected area; both $a$ and  $\theta_{\rm source}$ depend on the distance. We considered excitation temperatures of 10, 75, and 150\,K to calculate rough upper limits on the total masses of CO from the CO(1--0) and (3--2) transitions and of HCN based on the HCN(1--0) spectra. The results are presented in Table\,\ref{tab-masses}.

\begin{table} 
    \footnotesize\centering
    \caption{Upper limits on molecular mass in a model classical nova.}
    \label{tab-masses}
    \begin{tabular}{@{}cccc@{}}
    \toprule
Observed   & 3$\times$rms & $T_{\rm ex}$ & Mass limit\\
transition & ($T_A^*$, mK)  & (K)          &(M$_{\sun}$)\\
           \midrule
CO(3--2)   &    6          & 10  & $M_{\rm CO}<9.0\times 10^{-5}$\\
           &               & 75  & $M_{\rm CO}<3.7\times 10^{-5}$\\
           &               &100  & $M_{\rm CO}<5.8\times 10^{-5}$\\[5pt]
CO(1--0)   &18          & 10  & $M_{\rm CO}<3.0\times 10^{-4}$\\
           &               & 75  & $M_{\rm CO}<1.4\times 10^{-3}$\\
           &               &100  & $M_{\rm CO}<2.6\times 10^{-3}$\\[5pt]
HCN(1--0)  &    9          & 10  & $M_{\rm HCN}<1.0\times 10^{-6}$\\
           &               & 75  & $M_{\rm HCN}<5.3\times 10^{-6}$\\
           &               &100  & $M_{\rm HCN}<1.0\times 10^{-5}$\\
    \bottomrule
    \end{tabular}
    \tablefoot{The 3$\times$rms values are based on Tables\,\ref{rms_apex}--\ref{rms_all_CO}.}
\end{table}

The most constraining were the sensitive observations in CO(3--2), which indicate that the CO mass must be lower than 10$^{-5}$\,M$_{\sun}$. For a CO fractional abundance relative to H\,I+H$_2$ of 10$^{-4}$ \citep[cf.][]{PR2004}, one obtains upper mass limits on molecular matter on the order of 0.1\,M$_{\sun}$. This value is not that constraining, given that a typical classical nova is thought to eject a total mass of $\lesssim$10$^{-3}$\,M$_{\sun}$. However, the CO fractional abundance in a gas heavily enhanced in CNO burning product does not have to be close to that known from other circumstellar environments. The fractional abundance of HCN in classical novae is even more uncertain \citep[cf.][]{HCN}, but the upper limits we get are comparable or worse in constraining the molecular mass than the CO(3--2) data. Molecular gas in classical novae is not easily observable with current single-dish telescopes. 

Our survey can also be compared to observations of the ro-vibrational band of CO at NIR wavelengths. As an example, we choose the case of V5668\,Sgr whose prominent CO first-overtone emission was observed about a week after maximum light and was analyzed in \citet{banerjee2016}. The emission is thought to arise from a source of the size of 42\,mas, in an outflow with and excitation temperature of about 4000\,K and an expansion velocity $\Delta V$=530\,\kms. At the CO column density of 2$\times$10$^{19}$\,cm$^{-2}$ and assuming LTE conditions, the corresponding mm-wave emission in CO(1--0) observed by IRAM would have a peak antenna temperature of 0.5\,$\mu$K. Analogously, the CO(3--2) line emission observed with APEX would reach 43\,$\mu$K  ($T_A^*$). These extremely low values are a consequence of the high temperature and small source size, illustrating how ambitious it is to detect molecular emission from a nova shortly after the outburst. Let us however assume that the CO gas observed at NIR wavelengths survives for the next 50 years, decreases its gas temperature to 30\,K, and expands into a source of a size of 2\farcs75. The size was calculated for free gas expansion at the distance of 2\,kpc \citep[cf.][]{banerjee2016}. For such a source, the predicted peak antenna temperatures are of 0.1 and 0.3\,K, well within the reach of our survey. Admittedly, V5668\,Sgr is the most optimistic case with a particularly high molecular abundance and we have arbitrarily chosen the excitation temperature at which mm and submm lines are strong. Given the large number of observed novae, we would expect however to see a few objects with emission above our detection limits. That we did not detect any such source is possibly due to the decay of CO in later phases of ejecta expansion and much smaller filling factors (or a smaller source size) than we assumed. Indeed, chemical models of CO evolution of \citet{PR2004} predict high CO abundances only till $\approx$10 days after the eruption, after which they drop by at least two orders of magnitude. With some caution, results of our survey give qualitative support to such behavior.

\subsection{Red-nova connection}
The survey can be interpreted also in the context of red novae research. Nova\,1670 or CK\,Vul had long been considered a classical nova until in 2014 strong mm-wave molecular emission was discovered in the object \citep{KamiNat}. This and other observational characteristics of the object led to the interpretation that instead of being a classical nova, CK\,Vul is a remnant of a stellar merger \citep{Kami26Al,Kami2021} and its 1670--72 outburst can be classified as belonging to the group of (luminous) red novae \citep{kamiSubmm, pastorello}. Our survey can then be treated as a search for CK\,Vul analogs among historical novae, some of which could have been misidentified, especially if the nova event was not followed up spectroscopically sufficiently long. Since this experiment gave a negative result, objects similar to CK\,Vul must be very rare in the Galaxy. Indeed, luminous red novae outbursts are expected to occur only twice a decade \citep{rates}. More systematic searches of red novae among historical objects classified as classical novae can improve our statistics of stellar mergers \citep[cf.][]{kochanek,howitt}.

Among the targets observed with APEX is V1148\,Sgr, a nova from 1943 reported by \citet{mayall} and whose observations suggest that it could have been a red nova \citep{Kimeswenger}. The exact position of the nova is uncertain. Although all the disputed positions of the remnant were well within the APEX beam, we do not detect any strong circumstellar emission in the CO transitions. This strongly advocates against a red nova classification of V1148\,Sgr.

\subsection{Observations of interstellar clouds}
\begin{figure}
    \centering
    \includegraphics[scale = 0.39, trim={15 55 165 5.2cm},clip, page=1]{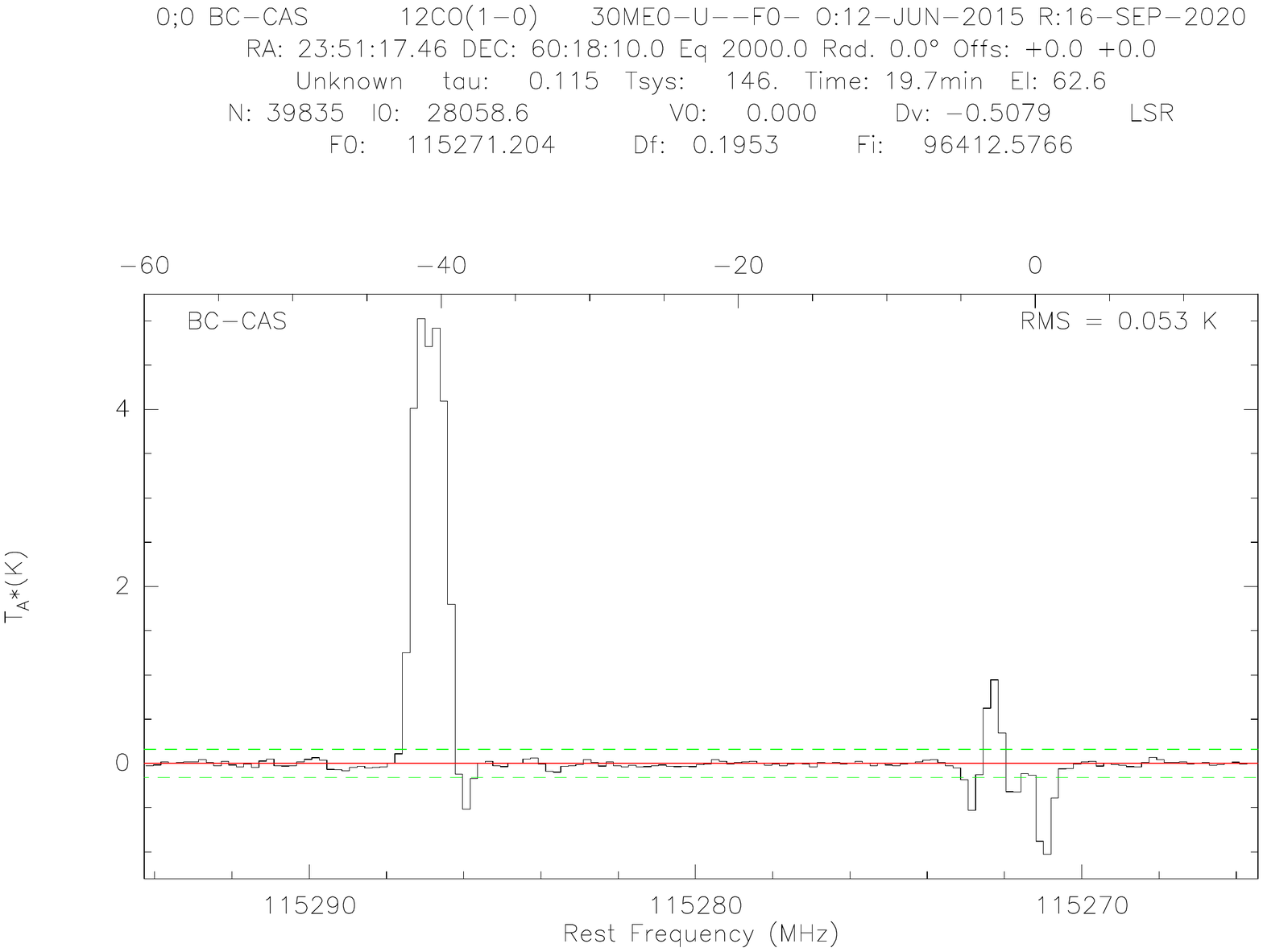}\\
\caption{The IRAM spectrum of BC\,Cas with ISM features of CO(1--0). The top axis gives LSR velocity with respect to the rest frequency of the CO transition. The dashed lines indicate the local 3$\sigma$ noise level.}
    \label{ism-co10}
\end{figure}

\begin{figure*}
    \includegraphics[scale = 0.39, trim={15 55 165 5.2cm},clip]{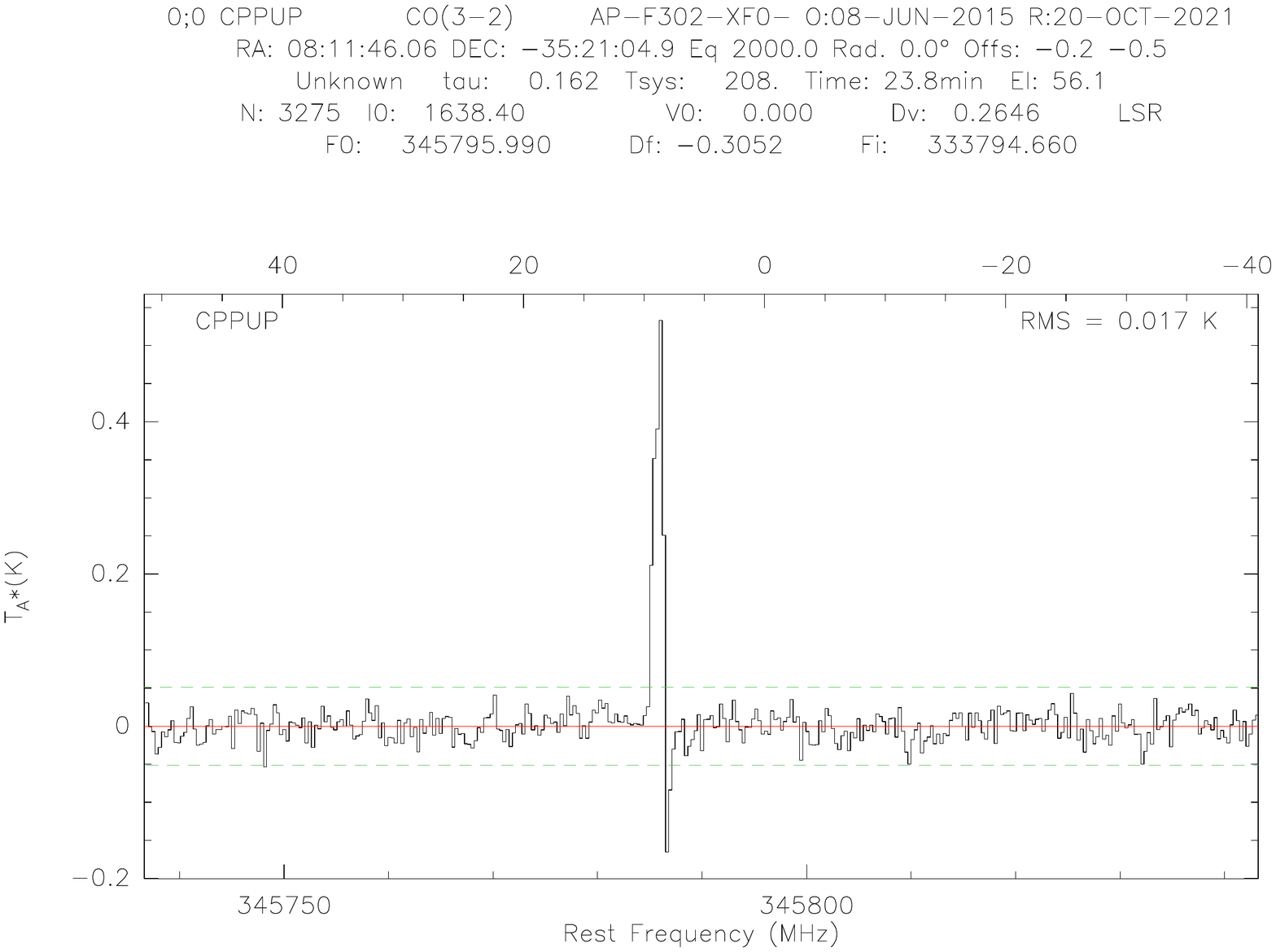}\\
    \includegraphics[scale = 0.39, trim={15 55 165 5.2cm},clip, page=2]{ism_candidates.pdf}
    \includegraphics[scale = 0.39, trim={15 55 165 5.2cm},clip, page=3]{ism_candidates.pdf}\\
    \includegraphics[scale = 0.39, trim={15 55 165 5.2cm},clip, page=4]{ism_candidates.pdf}
    \includegraphics[scale = 0.39, trim={15 55 165 5.2cm},clip]{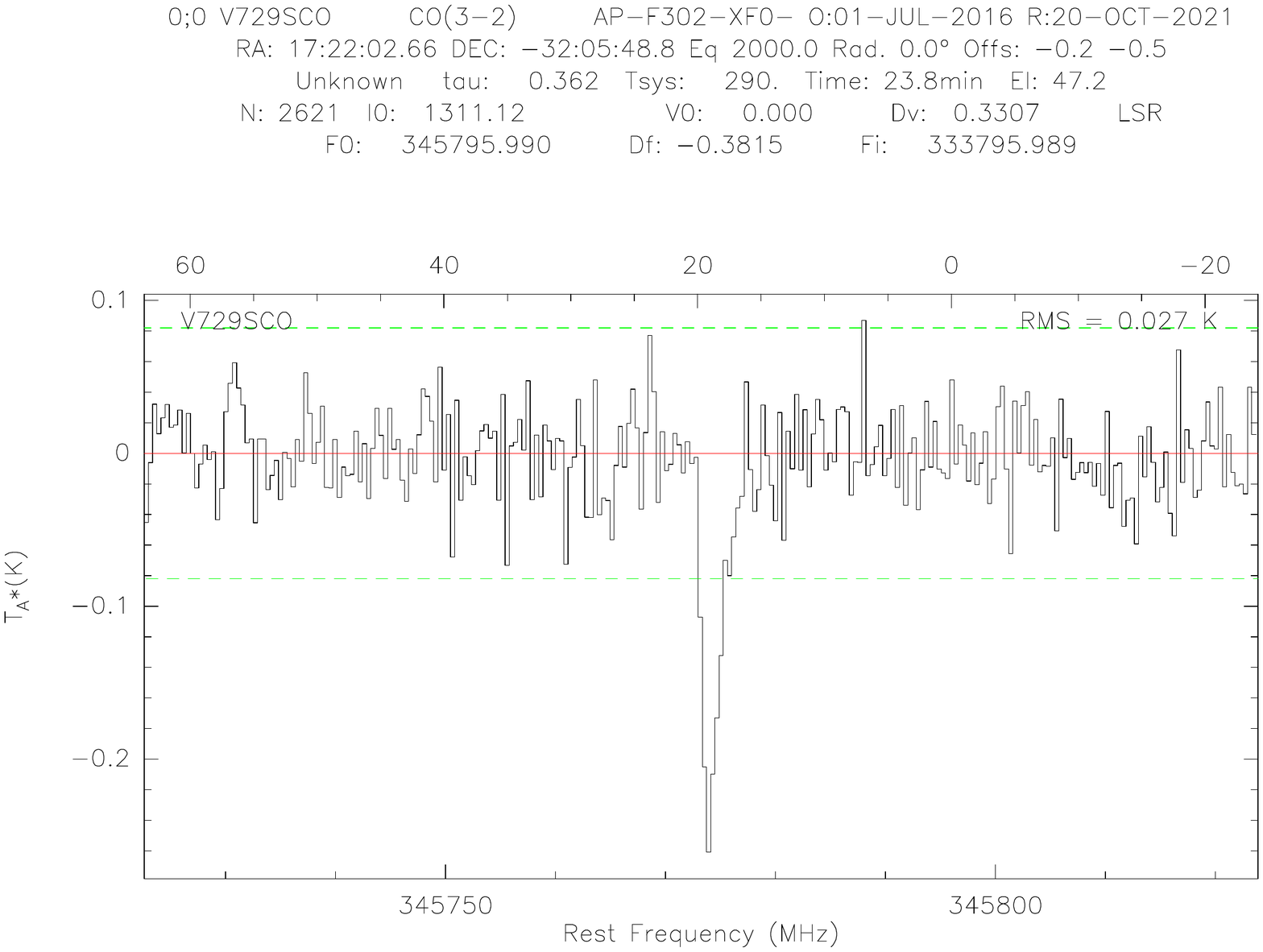}\\
    \includegraphics[scale = 0.39, trim={15 55 165 5.2cm},clip]{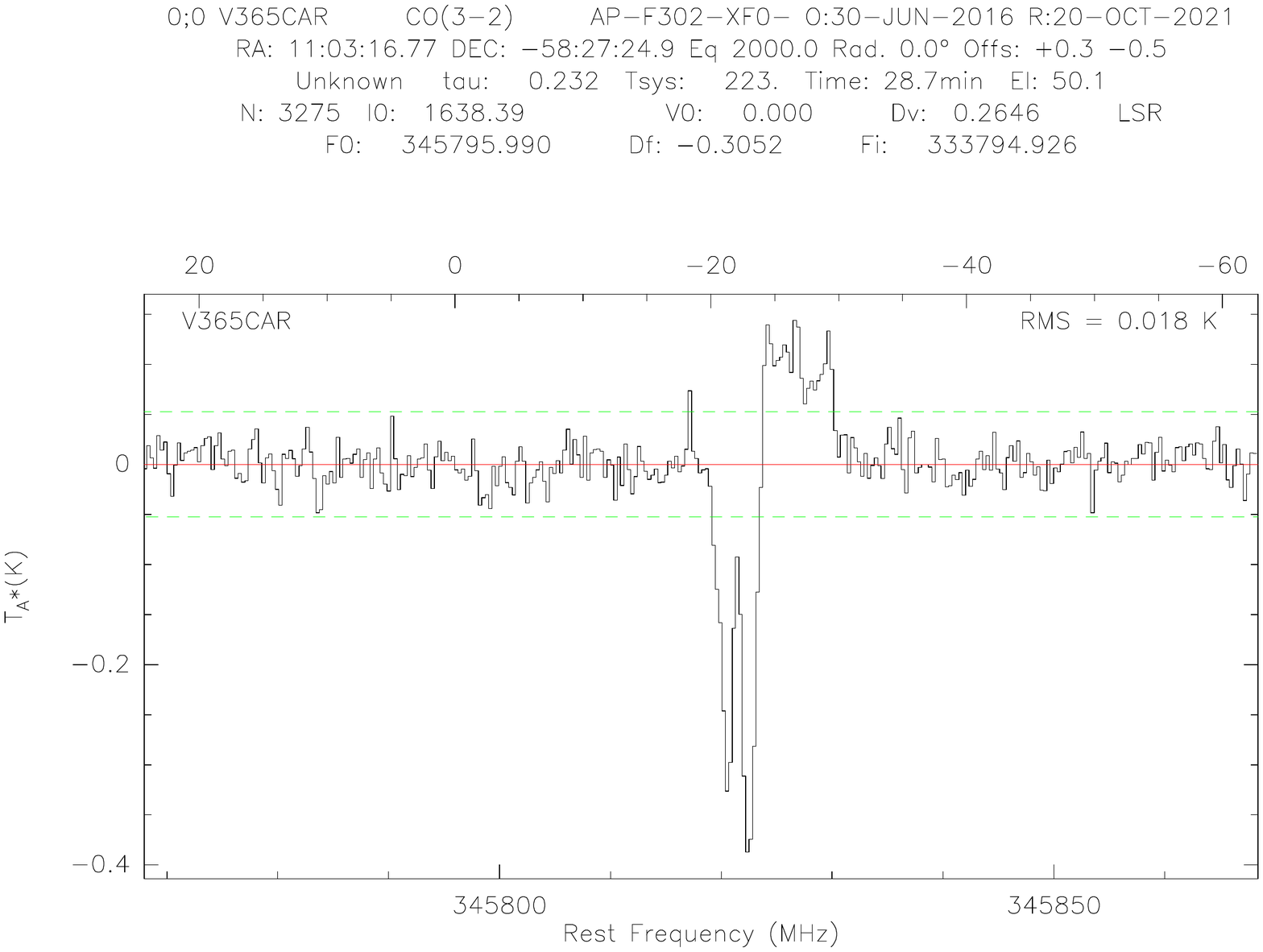}
    \includegraphics[scale = 0.39, trim={15 55 165 5.2cm},clip, page=5]{ism_candidates.pdf}
    \caption{APEX spectra with ISM features of CO(3--2). The top axes give LSR velocity with respect to the rest frequency of the corresponding CO transition. Dashed lines indicate local 3$\sigma$ noise levels. The rms values are provided for the baseline outside the ISM features and for the current binning.}
    \label{ism-co32}
\end{figure*}

\begin{figure*}
    \centering
    \includegraphics[scale = 0.39, trim={15 55 165 5.2cm},clip]{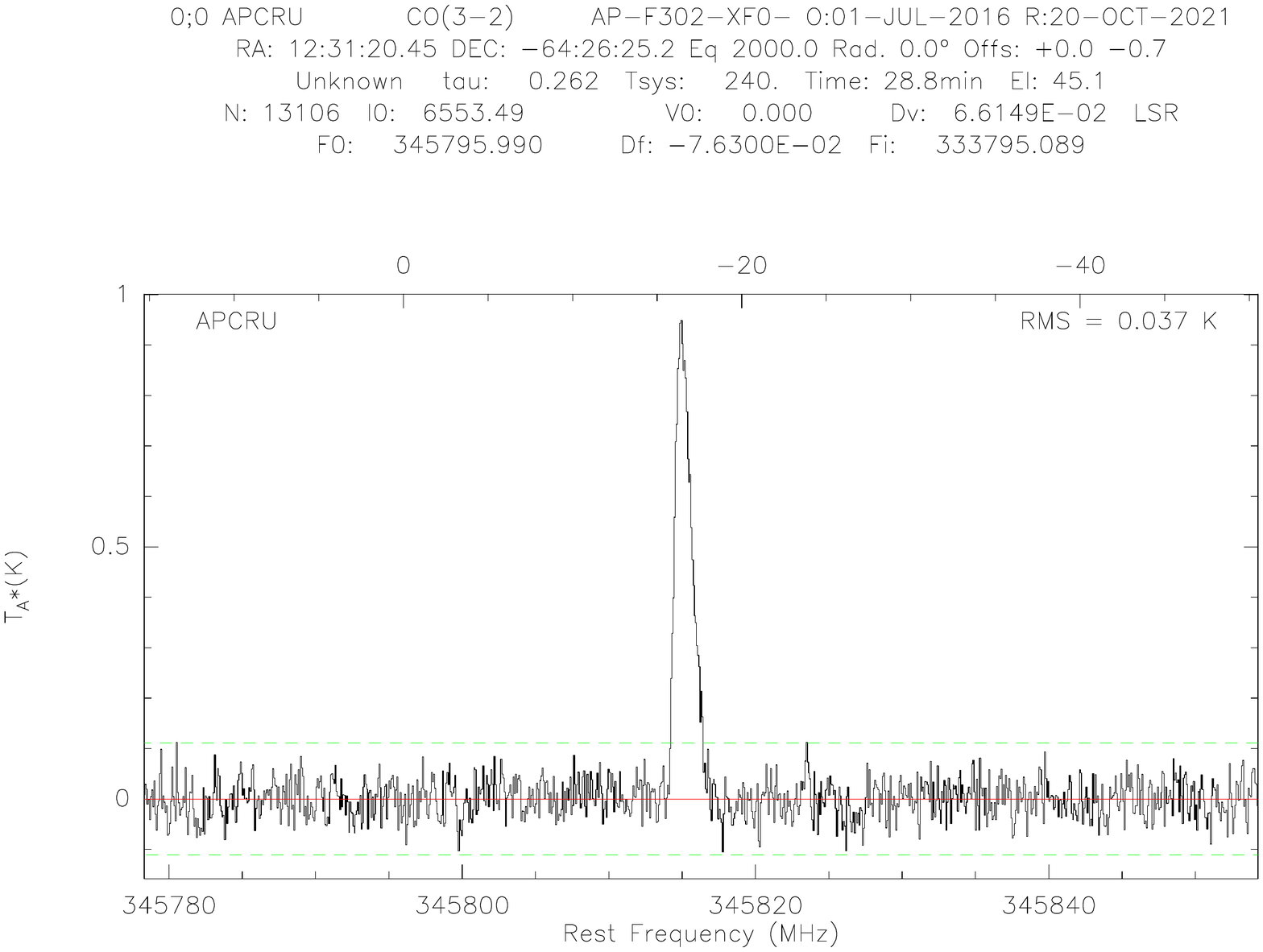}
    \includegraphics[scale = 0.39, trim={15 55 165 5.2cm},clip]{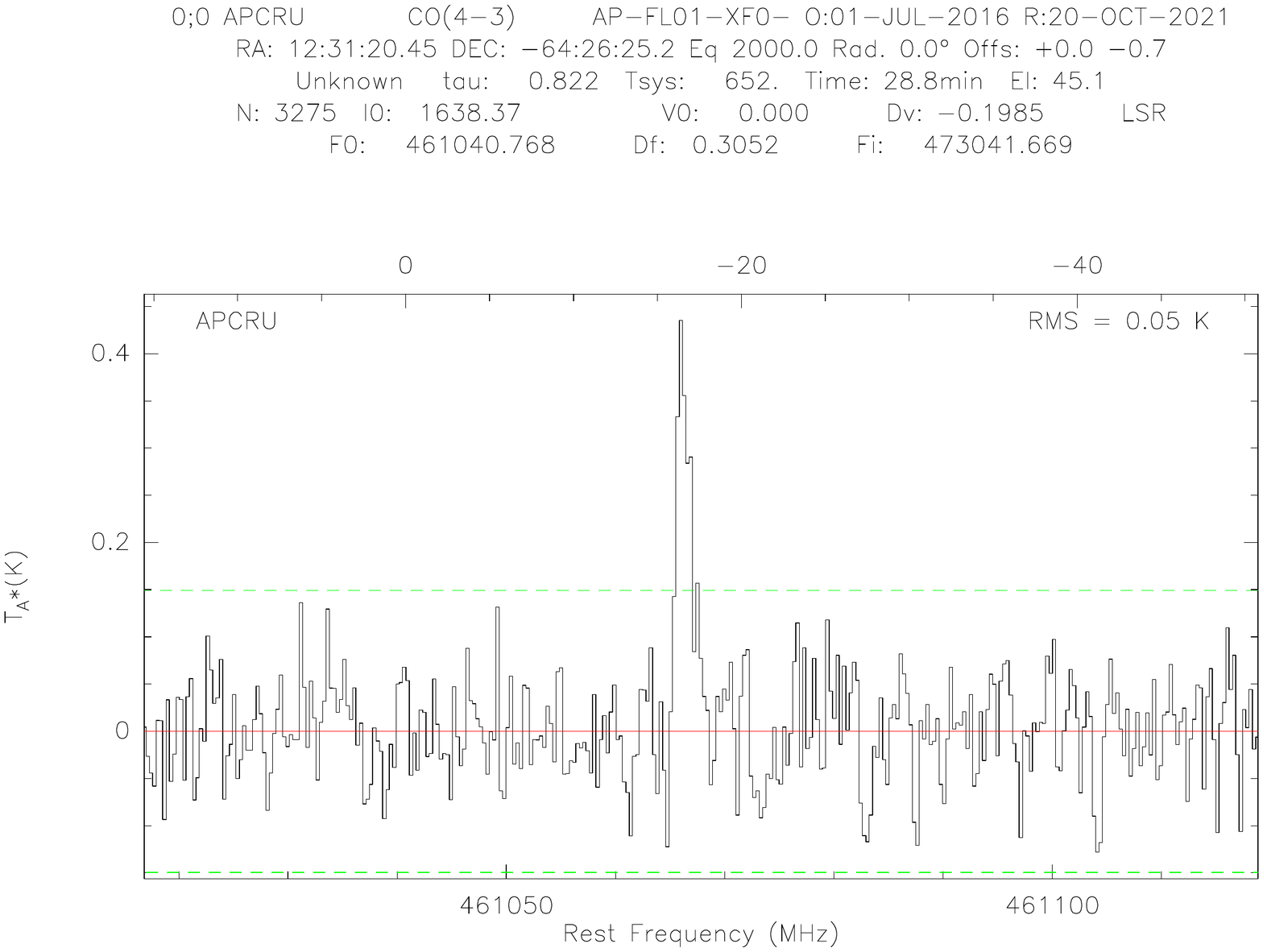}\\
    \includegraphics[scale = 0.39, trim={15 55 165 5.2cm},clip]{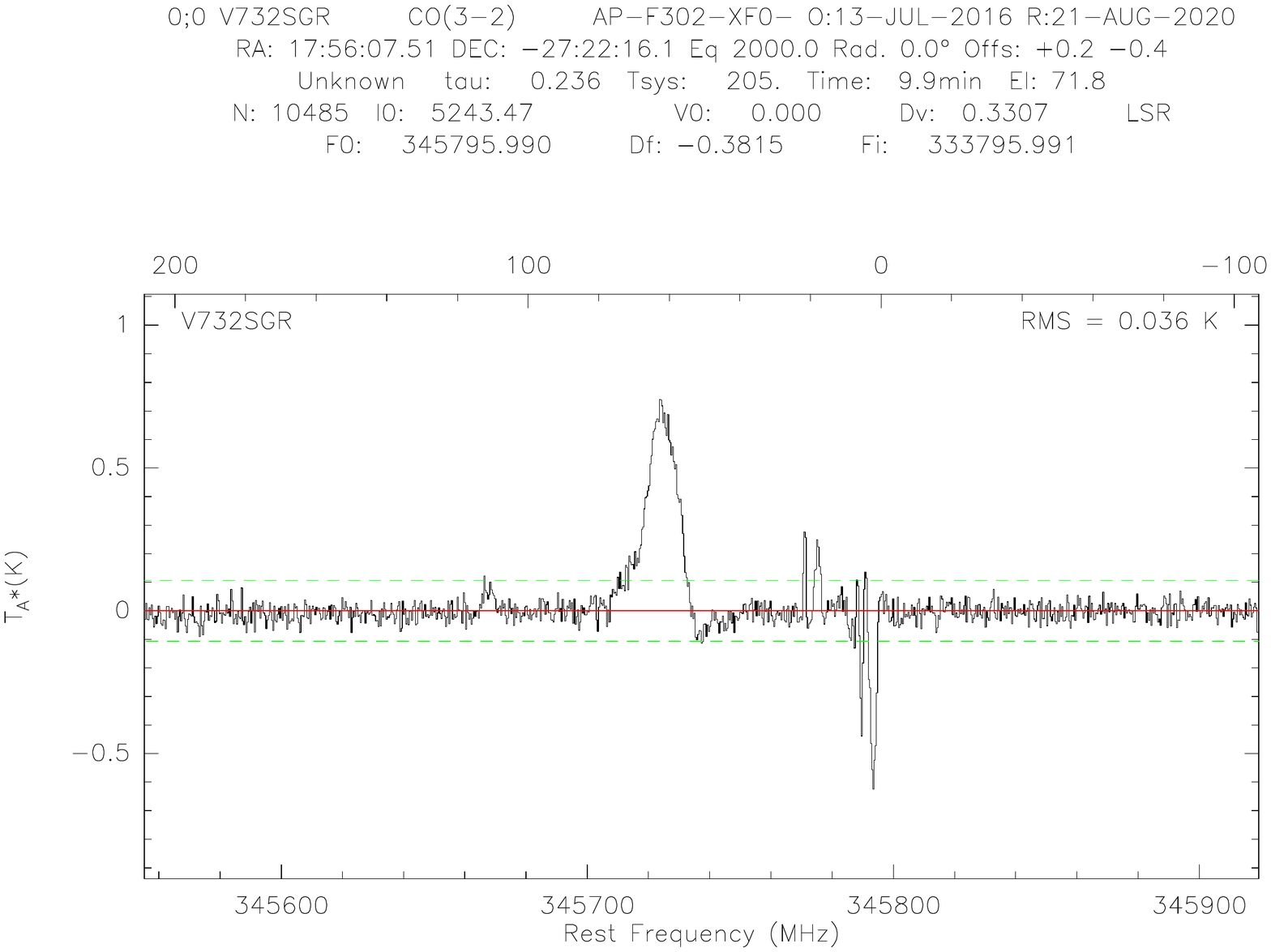}
    \includegraphics[scale = 0.39, trim={15 55 165 5.2cm},clip]{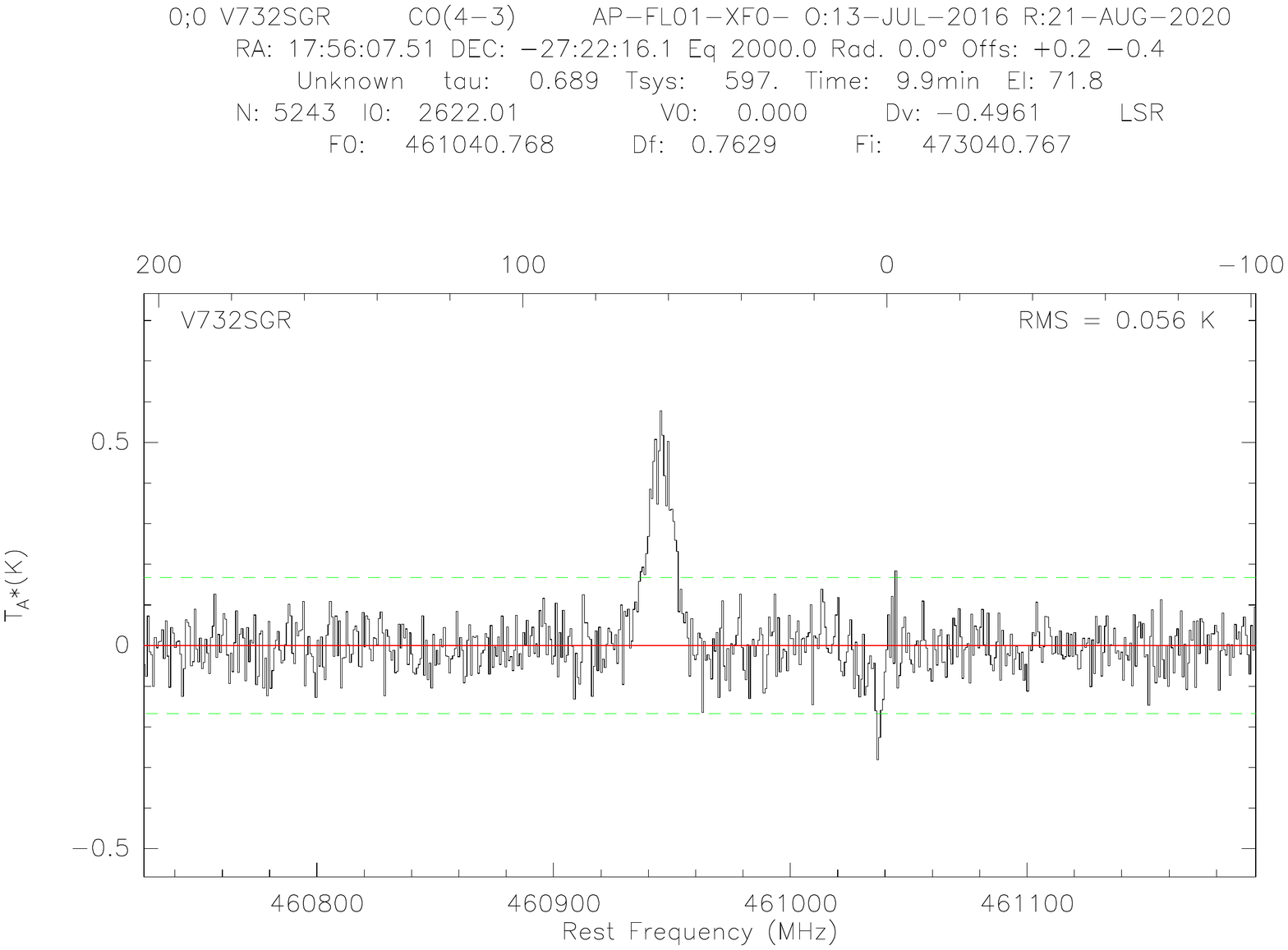}\\
    \caption{Spectra with ISM features detected on CO(3--2) (left) and CO(4--3) (right) in AP\,Cru (top) and V732\,Sgr (bottom). The top axes give LSR velocity with respect to the rest frequency of the corresponding CO transition. Dashed lines indicate local 3$\sigma$ noise levels.}
    \label{ism-co43}
\end{figure*}

Data gathered for our entire sample of targets resulted in no detection of broad molecular emission that would be expected from classical novae shells. However, in the spectra of ten targets we observed narrow features which most likely arise in the foreground or background interstellar medium (ISM). This is not surprising as these objects are located close to the Galactic plane. The spectra are shown in Figs.\,\ref{ism-co10}--\ref{ism-co43}. Some narrow spectral features appear in ``absorption'' when observed in one of the off-source wobbler positions. Below, we summarize the detected ISM features.
\begin{itemize}
    \item In the direction of BC\,Cas, we detected $^{12}$CO(1--0) features near --3 and --42\,\kms\ (Fig.\,\ref{ism-co10}), the stronger of which is also detected in the isotopolog line of $^{13}$CO(1--0) at $\approx$--42\,\kms (Fig.\,\ref{co-fts-high}). The main line has the highest peak intensity among all observed ISM features in our survey and is optically thick based on the intensity ratio of the two observed CO isotopologs. This is in accord with BC\,Cas being a highly-reddened nova ($A_V$=3.7\,mag) seen towards a star-forming region \citep{BCCas}.
    \item The spectra of CP\,Pup, RS\,Car, and AT\,Sgr show narrow emission and absorption features of CO(3--2) which combine into a P-Cyg type profiles (Fig.\,\ref{ism-co32}). This profile shape is a consequence of a small velocity difference between the ISM cloud velocity at the ON and OFF wobbler positions. The lines have a FWHM of 0.7--1.1\,\kms, typical for the translucent ISM \cite[cf][]{CloudReview}. Emission is observed at 8.8\,\kms\ in CP\,Pup, at --11.8\,\kms\ in RS\,Car, and at 12.0\,\kms\ in AT\,Sgr. The lack of the corresponding CO(4--3) lines at our sensitivities (implying a line ratio of $\gtrsim$2) is consistent with origin in a diffuse or translucent cloud.
    \item An absorption line centered at $\approx$--38\,\kms\ and with a FWHM of 2\,\kms\ is visible in the CO(3--2) spectrum of the recurrent nova IM\,Nor. A similar CO(3--2) feature is seen in the spectrum of V729\,Sco at 19\,\kms. The responsible ISM clouds are located in one of the wobbler OFF positions and do not necessarily cross the direction towards the novae.  
    \item In the spectrum of a reddened nova V365\,Car \citep{V365}, we found emission and absorption combining into an inverse P-Cyg profile. This again is a consequence of the ISM cloud extending between the ON and OFF wobbler positions with a small velocity gradient. Both emission and absorption components appear double, but the absorption component (OFF) is stronger. The emission component is centered near --25\,\kms. No CO(4--3) emission is seen.
    \item The spectrum of V1148\,Sgr shows two emission components, one centered at 21\,\kms\ and with a FWHM of 3\,\kms, and another narrow feature combining absorption and emission in  6--9\,\kms. The emission lines are the weakest features detected in our survey.
    \item Two targets, AP\,Cru and V732\,Sgr, show ISM features in both the CO(3--2) and CO(4--3) transitions (Fig.\,\ref{ism-co43}). While the line intensity ratio in AP\,Cru is not unusual for cool low-density ISM, the emission towards V732\,Sgr is almost equal in both transitions, indicating a warmer interstellar environment. The case of V732\,Sgr is discussed in more detail below. 
\end{itemize}

\subsubsection{Emission towards V732\,Sgr}\label{sec-V732}
Spectra of V732\,Sgr centered at the CO(3--2) and (4--3) transitions are shown in the two bottom panels of Fig.\,\ref{ism-co43}. Both spectra display a broad (FWHM$\approx$\,12\,\kms) line centered near 62\,\kms\ and several very narrow lines that appear as emission or absorption between 0 and 22\,\kms. Additionally, the spectrum centered on the CO(3--2) transition shows a marginally significant feature near 111\,\kms. Given the relatively broad profile of the CO lines near 62\,\kms\ and a relatively high intensity of the CO(4--3) transition, this component is reminiscent of lines observed from envelopes of asymptotic-giant-branch (AGB) stars. However, CO emission towards V732\,Sgr is known since the early 1990s, when it was mapped with SEST in the CO(1--0) transition, to be of interstellar origin \citep{weight1993,Albinson}. The maps show that the molecular cloud is extended (10\arcmin$\times$3.5\arcmin) and massive ($10^4$\,M$_{\sun}$), and thus it cannot be directly associated with the nova. Interestingly, however, $\approx$100 days after V732\,Sgr erupted in 1936, a weak light echo was visible for about 30 days \citep{swope}. The echoing medium was most likely associated with the giant interstellar cloud observed with SEST with APEX. The presence of light echo make V732\,Sgr very similar to the nearby Nova Persei 1901 (GK\,Per), which we discuss in more detail in Sect.\,\ref{GKPer}. All the novae for which ISM emission has been detected \citep[including NQ\,Vul;][]{NS2003} are candidates to have been associated with a light echo (some could have been easily overlooked). Except GK\,Per and V732\,Sgr, none of our sources with ISM detection was targeted for an echo search \citep{bergh,schaefer}. The light echo can arise only in clouds relatively close, up to tens of parsecs \citep{schaefer}.

\FloatBarrier
\section{CO maps around GK\,Per}\label{GKPer}
\subsection{Motivation}
GK\,Per is a unique cataclysmic variable whose distance of 442\,pc \citep{distances} makes it the second-closest nova. The central binary consists of a magnetic white dwarf of 0.9--1.0\,M$_{\sun}$ \citep{wada,Alvarez} and a K1\,IV subgiant \citep{Alvarez}. It erupted as a classical nova in 1901 and, as mentioned, is one of two known classical novae to had been associated with light echos. The echoes reached a size of tens of arcseconds \citep{Ritchey}. Since the middle of the last century, GK\,Per has experienced dwarf-nova outbursts every 1--3 yr \citep{simon}. Nova ejecta was first observed in 1916 and remains visible to this day \citep{Liimets,Shara}. The Firework Nebula, as it is often referred to, has a size of nearly 100\arcsec\ and is considered the longest-lived remnant of a classical nova event. The highest gas velocities found in the remnant are of 2800\,\kms\ but typical values are of about 1100\,\kms.

GK\,Per is the only classical nova that has been claimed to have molecular gas in its immediate proximity long after its outburst\footnote{\citet{DQHer} claimed the presence of H$_2$ emission in DQ\,Her some 56 years after its eruption and based on imaging observations. The emission identification has never been confirmed spectroscopically, though.}. The nature of its observed CO emission has been somewhat controversial. Observations of the CO $J=2-1$ rotational line with the James Clerk Maxwell Telescope (JCMT) acquired in 1993 by \citet{Scott} revealed elongated symmetrical structures with two main ``concentrations'' connected by a bridge of emission $\approx$1\arcmin\ north of the object. The concentrations are located $3'$ east and west off GK\,Per. However, the emission region was not fully mapped; in particular, the concentrations are seen near the edges of the undersampled map, which puts into question the extent of the emission. Observed lines are narrow, with a FWHM of a few \kms, atypical for circumstellar envelopes. No large velocity gradients within the cloud were found. The molecular gas was proposed to be associated with a filamentary and dusty bipolar cloud of a size of 17\arcmin\ from which dust emission was first observed with the Infrared Astronomical Satellite (IRAS). This filamentary cloud was interpreted by \citet{BodeNat}, \citet{Seaquist}, and \citet{Dougherty} as a bipolar outflow or a planetary nebula (PN) from the nova system. Their argument was based mainly on the symmetry of the cloud's morphology  relative to GK\,Per. An atomic component of the cloud was also mapped in the 21\,cm line of neutral hydrogen (\ion{H}{I}) by \citet{Seaquist}. In order to explain the very large extent and the very high mass of 3.8\,M$_{\odot}$ derived for the cloud, both highly atypical for classical novae, Dougherty et al. invoked a rather elaborate scenario where mass loss involving the secondary had occurred prior to the Nova\,1901 event. They considered the binary to be then in the second Roche-lobe overflow phase, in which the mass transfer from the secondary, then of a mass of 1.3--1.5\,M$_{\sun}$, was sufficient to convert the white dwarf into a ``born again'' AGB star. Knowing the composition of the nebula would be of great value in order to verify this hypothesis. If the CO emission observed with the JCMT indeed represents such circumstellar gas, it could provide means to measure relevant isotopic ratios.

On the other hand, the CO emission near GK\,Per may be purely interstellar. In 1985 \citet{Hessman} mapped in the CO(1--0) line the large-scale structure that was identified in IRAS maps. With the 4.9-m telescope of the Millimeter Wave Observatory, he mapped a  much larger area than was covered with JCMT, roughly of 40\arcmin$\times$50\arcmin\ and at a resolution of 2\farcm6. He only found a weak correspondence between the CO emission and the IRAS emission, which would be atypical for any circumstellar material. Hessman concluded that most of the CO material belonged to an interstellar cirrus cloud that is unrelated to the nova. This interpretation is not favored in the recent literature on GK\,Per.

\begin{figure}
    \centering
    \includegraphics[width=0.9\columnwidth]{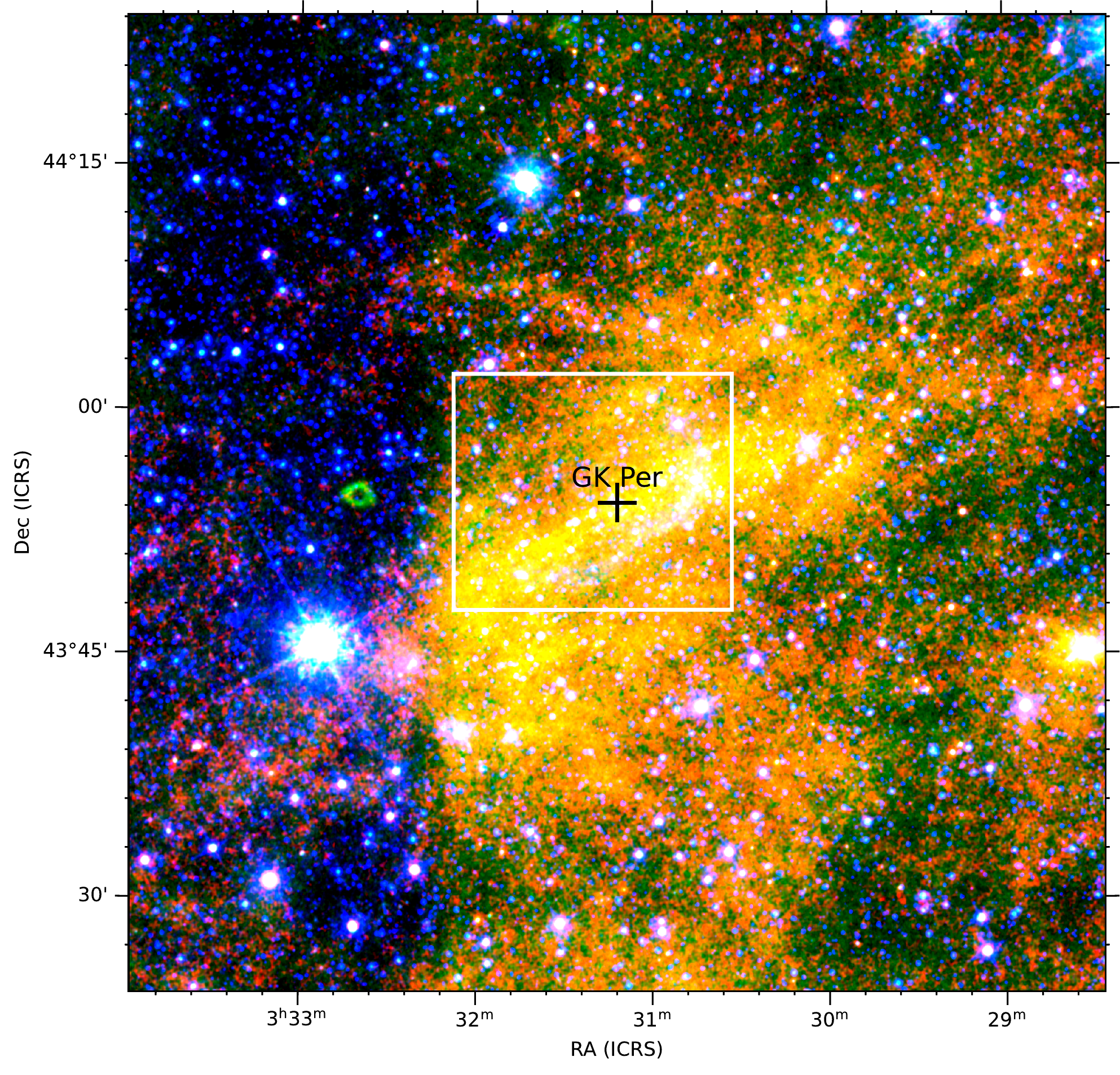}
    \includegraphics[width=0.9\columnwidth]{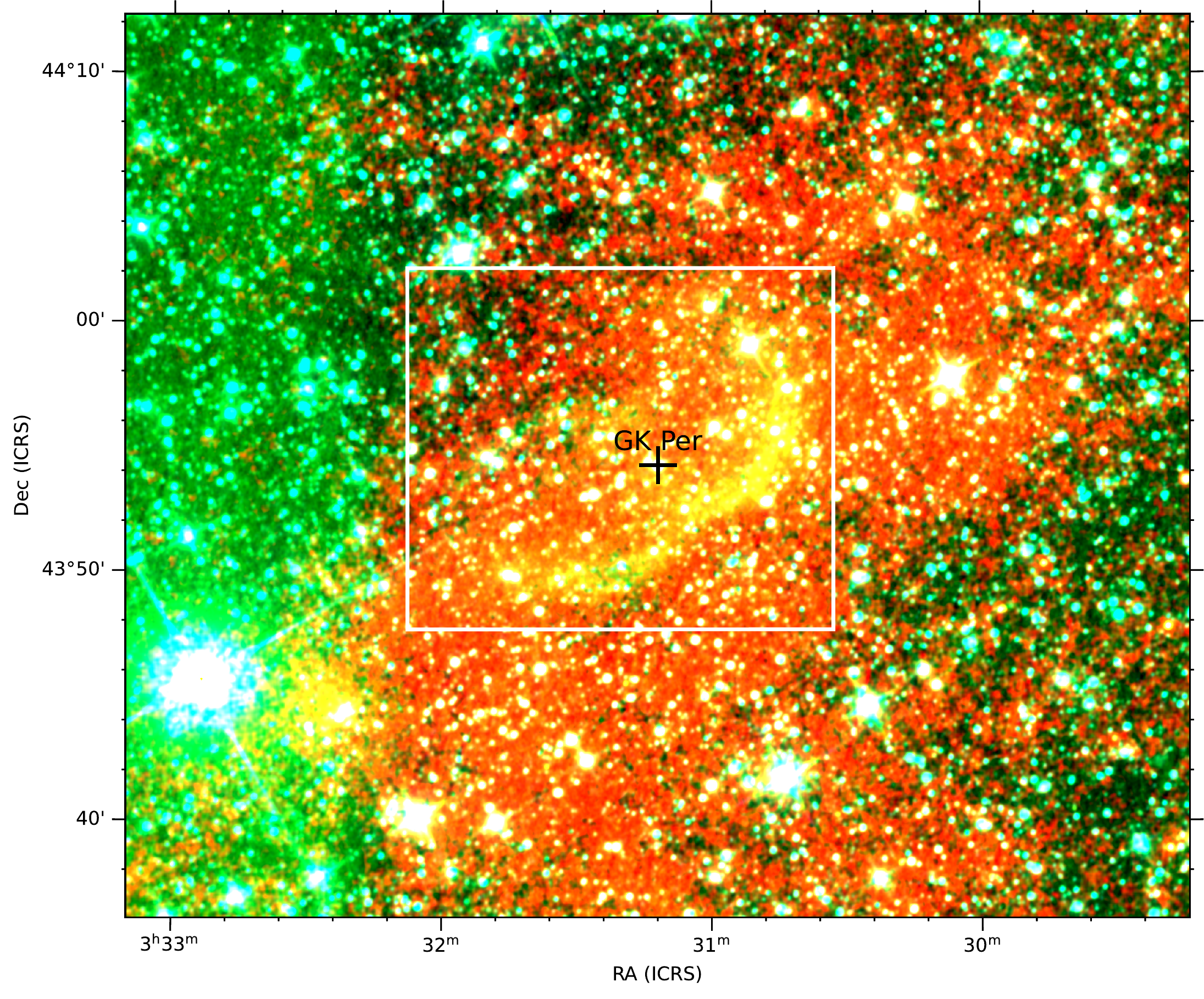}
    \includegraphics[width=0.9\columnwidth]{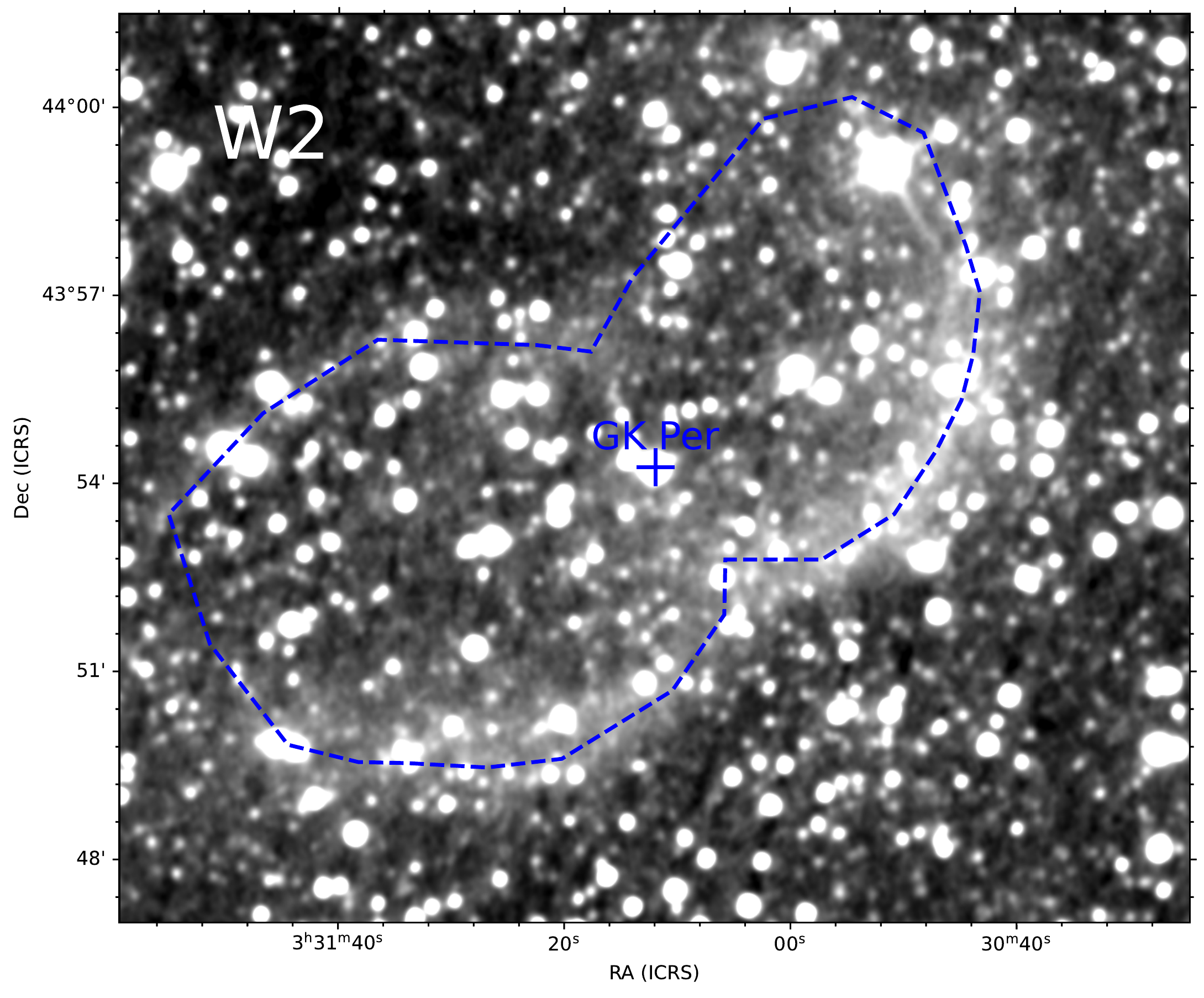}
    \caption{WISE images of the field around GK\,Per. Top: A wide-angle view (1\degr$\times$1\degr) with blue,  green, and red colors representing signal in $W2$, $W3$, and $W4$ filters, respectively. It shows the large cirrus IS cloud reported earlier in IRAS observations. Middle: Closeup view with blue, green, and red showing $W1$, $W2$, and $W4$ images, respectively. In yellow, a bipolar structure is seen near the position of GK\,Per. Bottom: $W2$ image of the bipolar structure. Its shape was arbitrarily outlined with a dashed line. The northern edge is not well-defined. The field shown (17\farcm0$\times$14\farcm5) corresponds to the white boxes drawn in two upper panels. All panels show intensity in linear scale.}
    \label{fig-wise}
\end{figure}

Most recently, the vicinity of GK\,Per was mapped with the  Wide-field Infrared Survey Explorer \citep[WISE][]{wiseSurv} at a higher sensitivity and with better angular resolutions than with IRAS. WISE maps in $W3$ and $W4$ bands, at 12 and 22\,$\mu$m, respectively, readily show the same extended dusty cloud that was identified in IRAS data  (Fig.~\ref{fig-wise}). It has the appearance of an interstellar cirrus. However, at shorter wavelengths, that is at 3.4\,$\mu$m ($W1$) and 4.6\,$\mu$m ($W2$), WISE maps display a smaller bubble-like, bipolar structure centered on GK\,Per. Although weak in both bands, it is most readily present in the $W2$ image. The longer axis of the nebula is of about 15\arcmin\ and is oriented at a position angle of $\approx$--30\degr. The structure appears irregular, with the southeastern lobe being bigger and of a width of about 7\arcmin. The same bipolar structure was observed earlier in the H$\alpha$ filter (covering the H$\alpha$ line and the [\ion{N}{II}] doublet) and in the [\ion{O}{III}] $\lambda$5007 line by \citet{Tweedy}\footnote{Their Fig.~1 has an incorrectly assigned orientation. The maps have east up and north right.} and \citet{Bode2004}. The observations imply that the optical emission is dominated by line emission of recombining low-density gas. Since the $W1$ and $W2$ bands also cover several emission lines, we cannot dismiss the possibility that the bipolar WISE structure is seen through IR emission lines. However, the southern part of the bipolar structure was seen in the light echoes in 1901--1902 and thus must contain dust grains. We assume that the mid-IR emission of this structure arises as continuum emission of relatively warm dust, warmer than the larger ISM cirrus seen in $W3$ and $W4$. Tweedy excluded the possibility that the observed feature represents a genuine PN, but later authors \citep[e.g.,][]{Bode2004, Harvey, Dougherty, Alvarez} challenged this conclusion. This nebula is smaller and much less massive than the dusty cloud seen in the IRAS images, but both structures have often been confused in the literature and named a planetary nebula interchangeably. Hereafter, we refer to the smaller structure as a planetary or bipolar nebula, and we call the larger-scale emission a cirrus. 


Based solely on the morphology of the smaller nebula, there is little doubt that it is a signature of mass-loss activity in GK\,Per before the Nova 1901 event. The longer axis of the structure aligns well with the larger interstellar cirrus, perhaps causing the confusion in the literature. The molecular gas mapped earlier with the JCMT and reported in \citet{Scott}, would be filling the bipolar lobes of the planetary nebula, with the concentration located closely to the main axis of the bipolar structure. This strongly suggests circumstellar origin of the CO gas, against the interpretation of \citet{Hessman}. With the presence of a PN now firmly established with the WISE maps, we decided to revisit the claim of the presence of circumstellar molecular gas around GK\,Per by obtaining new deeper maps of CO emission near GK\,Per.
 
\subsection{Mapping observations}
Mapping observations around GK\,Per were executed in 9, 10, 12 April 2021 and on 28 May 2021 with the IRAM 30\,m telescope. The data were acquired within the IRAM program 155-20 (PI: H. Mazurek). We used the EMIR E90 and E230 receivers simultaneously and centered them near the CO $J=1-0$ and $2-1$, transitions, which have rest frequencies of 115.2712 and 230.5380\,GHz, respectively. The FTS and WILMA backends were used as spectrometers with the FTS having a native resolution of 0.5 and 0.25\,\kms\ for the CO(1--0) and (2--1) spectra, respectively. Although effectively the spectral ranges of 113.522--117.570\,GHz and 228.240--236.016\,GHz were covered, no lines other than the targeted $^{12}$CO transitions were detected. Regular pointing observations were obtained at the beginning and during mapping runs. Focus calibration on a planet was secured before each observing session. 

The maps were obtained in the on-the-fly (OTF) mode, with a reference OFF position at (RA, Dec) offsets of (700\arcsec, 350\arcsec) from the position of GK\,Per. The reference position was checked earlier to be free of any molecular emission.  An area within a field of view of 1128\farcs8$\times$987\farcs0 was mapped in two smaller parts, each with multiple coverages obtained in orthogonal directions of the beam scanning. Relative to the position of GK\,Per, the map is composed of two rectangular regions at (RA, Dec) offsets of (610\arcsec\ to 10\arcsec, --410\arcsec\ to 210\arcsec) and (80\arcsec\ to --450\arcsec, --150\arcsec\ to 500\arcsec) and  with some overlap between the two areas near the position of the nova. The map sampling was set to 8\farcs5, which is optimal for the beam size of the CO(1--0) observations (21\farcs3). Owing to several coverages of each part of the map at a slightly different offset (of a few arcsec), the grid of observed positions is effectively denser, allowing also to obtain a map of CO(2--1) that, with the beam size of 10\farcs7, is only slightly undersampled. The data were calibrated every 8\,min and registered in $T_A^*$ units. The conversion of the CO(2--1) data to the main beam temperature units requires $\eta_{\rm mb}$=0.64 (cf. Table~\ref{spectral-setups}). All data were reduced in CLASS, including baseline subtraction, spectra averaging and rebinning, and map gridding. The final rms noise in the CO(1--0) cube is of 0.22\,K ($T_A^*$) at 0.45\kms\ binning; for the CO(2--1) cube, it is 0.41\,K per 0.45\,\kms. 

\subsection{Results of CO mapping}\label{sec-maps}
At the position of GK\,Per, no broad emission lines are found that could originate in the nova system. A narrow and weak CO(1--0) line is seen at this position and originates in the interstellar cloud.

\begin{figure*}
    \centering
    \includegraphics[width=0.5\textwidth]{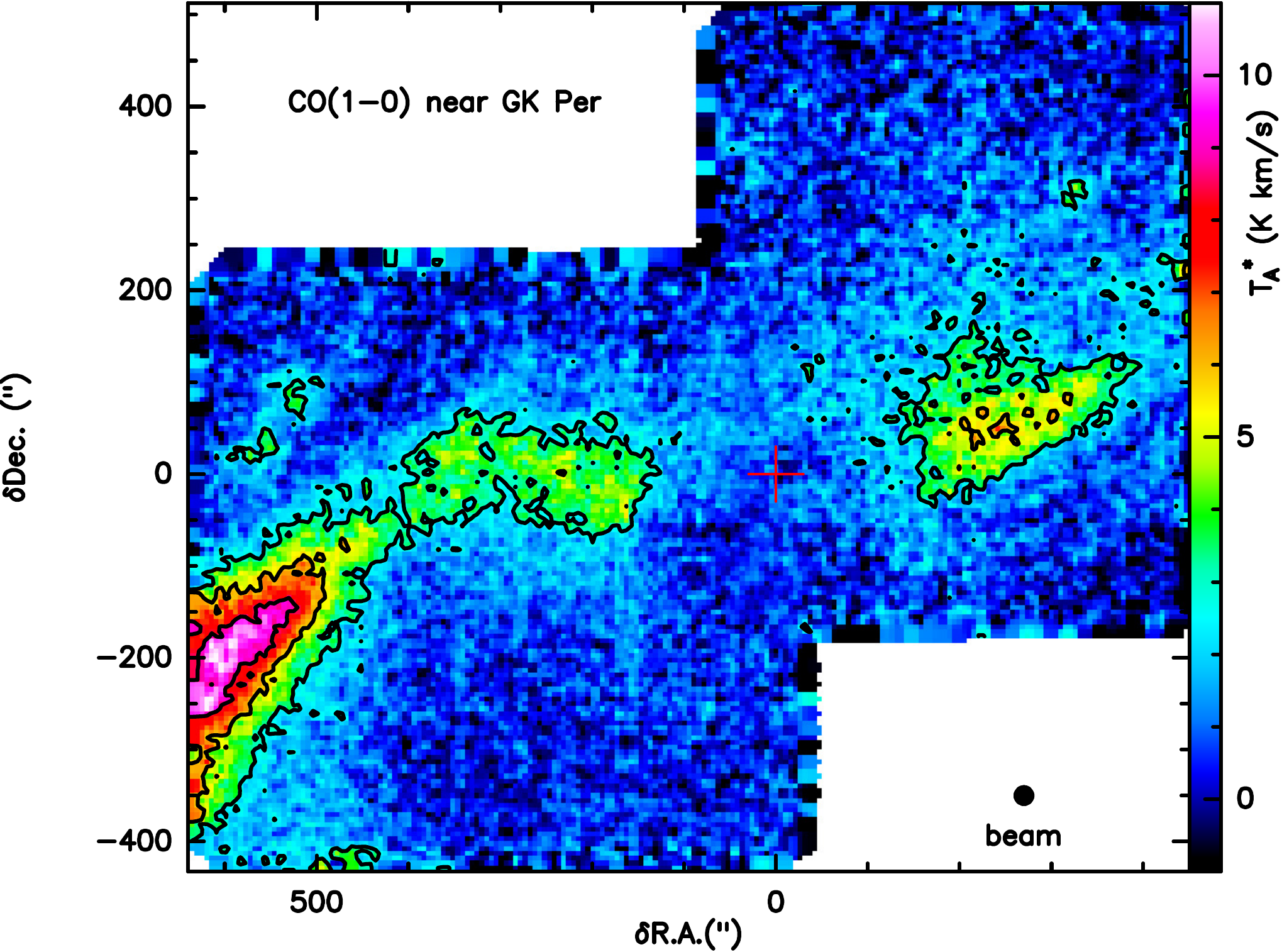}
    \includegraphics[width=0.44\textwidth]{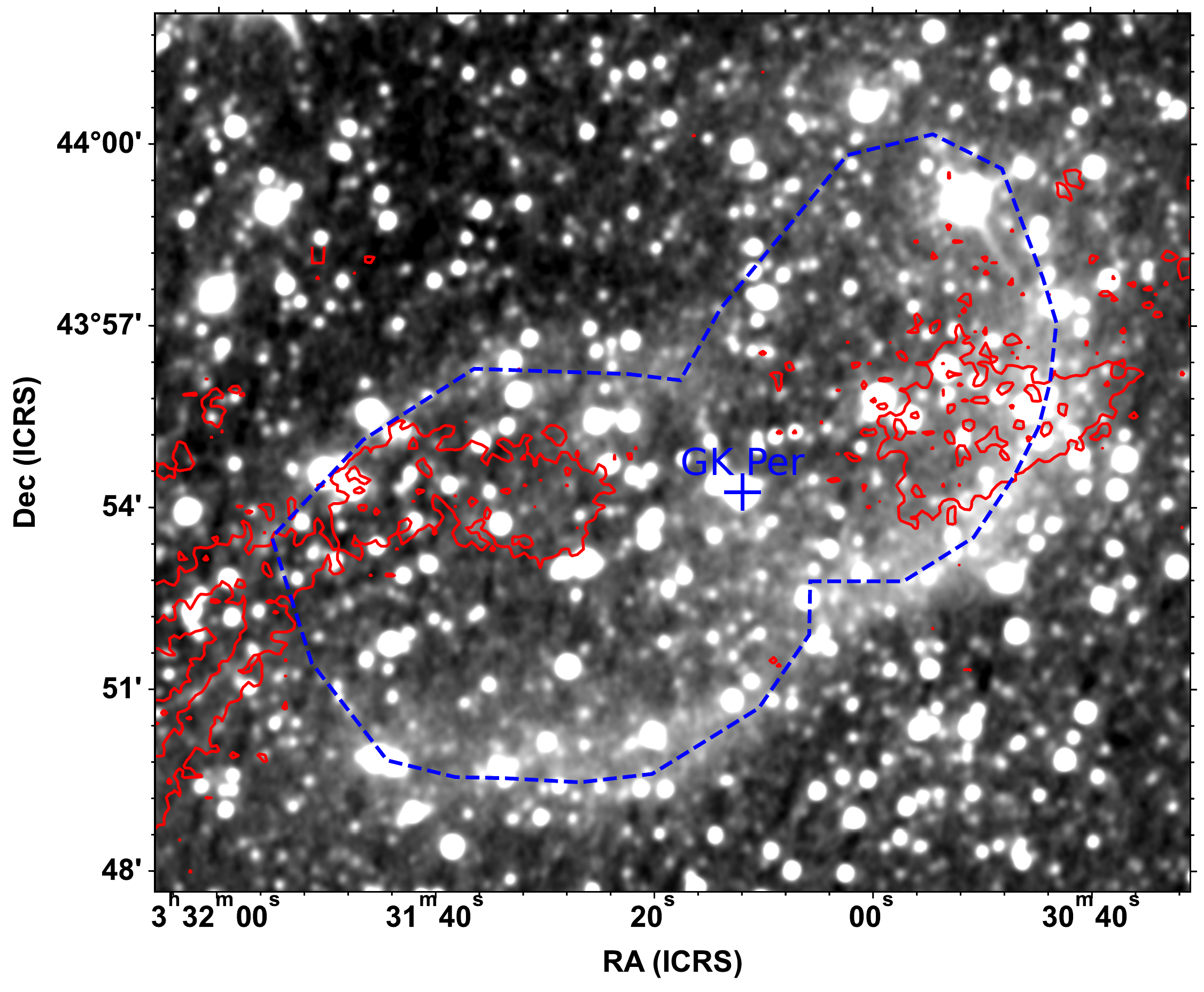}
    \caption{Total-intensity emission of CO(1--0) near GK\,Per. The cross marks the position of the nova. Contours are at 3, 6, 9 times the rms noise level. Left: Color map of the CO emission. Right: The same CO contours are plotted as red solid lines over the $W2$ image of the field (cf. Fig.~\ref{fig-wise}). The CO emission does not follow the dust emission seen in the $W2$ image.}
    \label{fig-ghperCO}
\end{figure*}

\begin{figure*}
    \includegraphics[width=12cm]{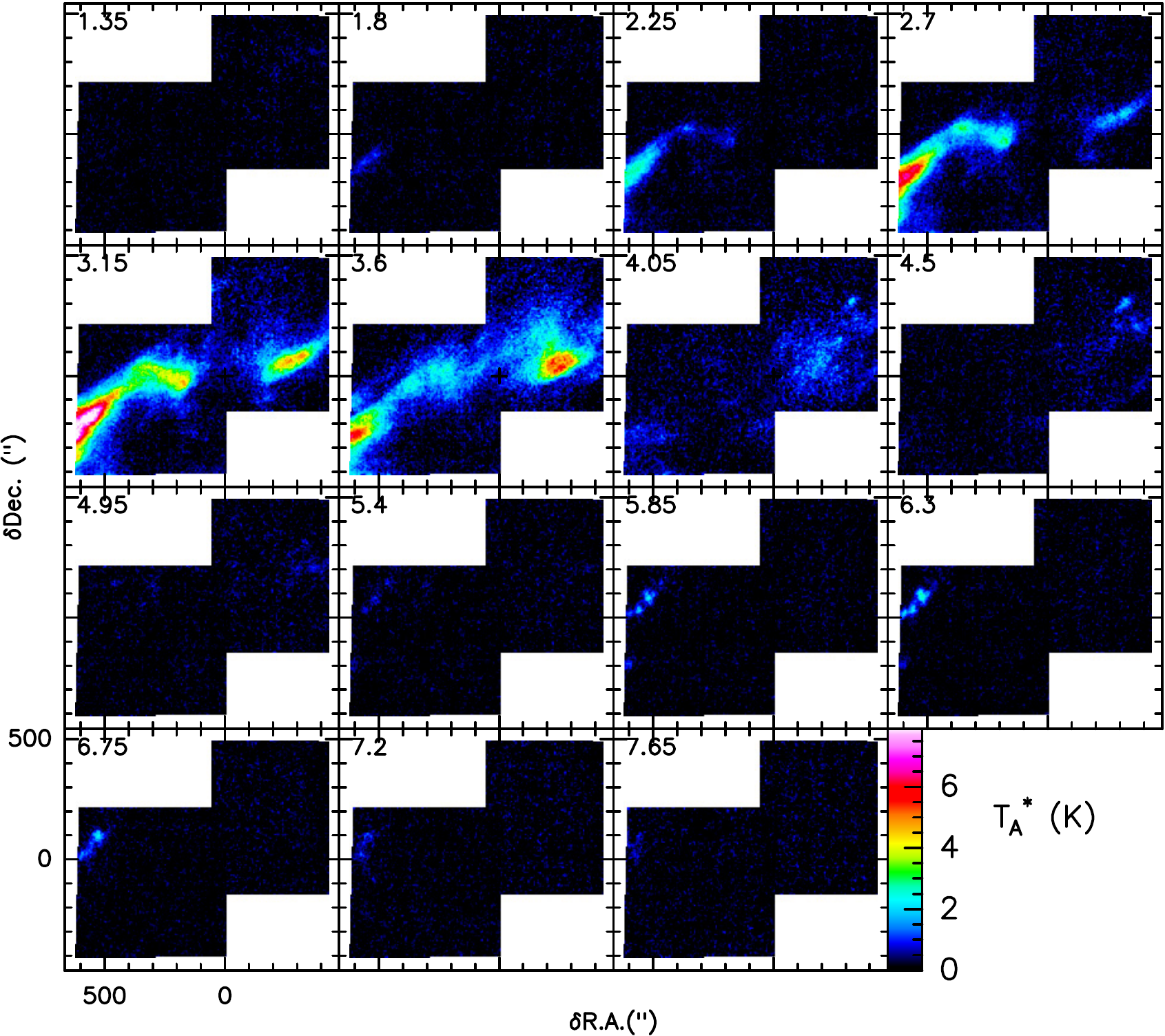}
    \includegraphics[width=6.5cm]{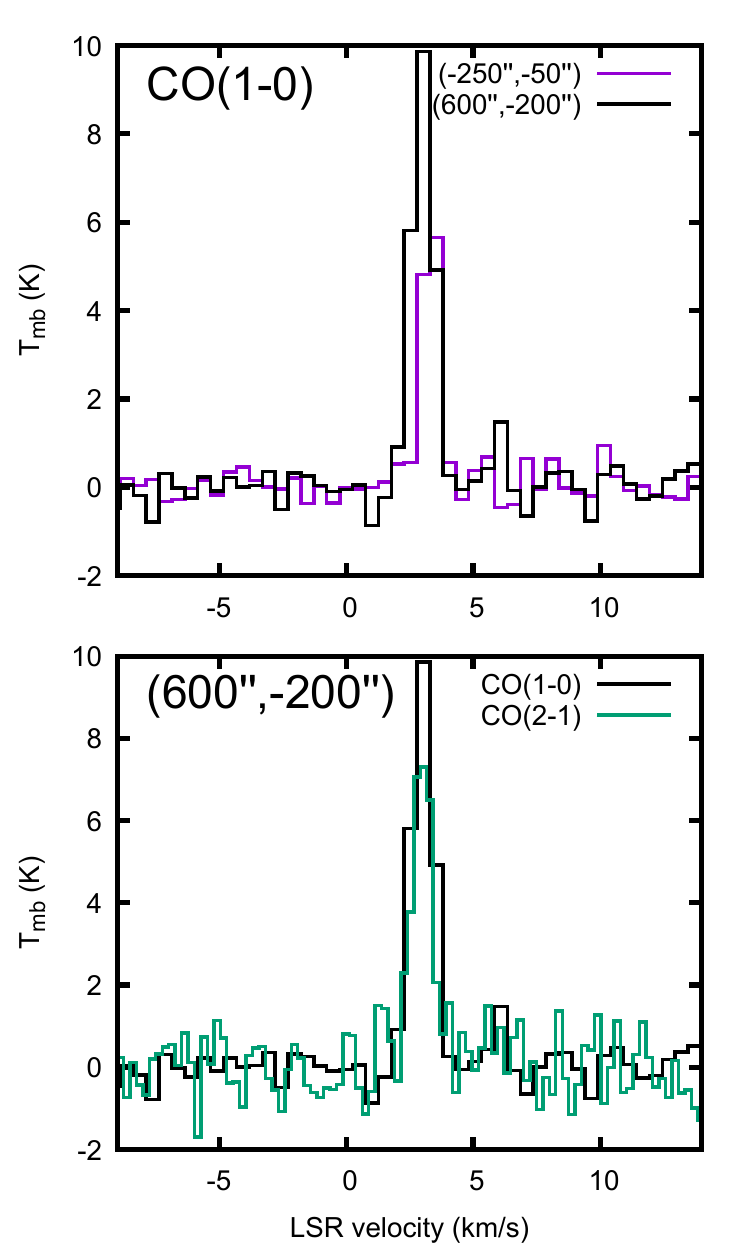}
    \caption{Left: Channel maps of CO(1--0). The map LSR velocity is given in the upper left corner of each panel. GK\,Per is located at the offset of (0, 0). The maps show two clouds at 1.8--4.95\,\kms\ and 5.4--7.65\,\kms. Right: Representative spectra. Top panel compares profiles of CO(1--0) at two positions east and west of GK\,Per as indicated in the legend. Hardly any shift is seen, unlike ones expected for any circumstellar material. Bottom panel: Line profiles of the two lowest CO transitions are compared for the eastern position where emission is the strongest.}
    \label{fig-channmaps}
\end{figure*}

Both maps show interstellar clouds emission over most of the mapped area and in the 1.5--8.0\,\kms\ range. Here we focus on the better sampled CO(1--0) map, as the CO(2--1) map looks qualitatively the same. The total intensity CO(1--0) map is shown in Fig.~\ref{fig-ghperCO}. It displays a ridge of molecular emission spreading over most of the mapped area. The ridge appears to be broken near the position of the nova. The orientation of the emission at RA offsets from 500\arcsec\ to the west-most edge of the map is close to horizontal. At larger offsets towards the east, the ridge turns southeast. This southeastern part is also the brightest fragment of the mapped filament. It very likely continues beyond the mapped region. Most of the emission belongs to the same kinematic structure within 1.3--4.5\,\kms. However, the channel maps in Fig.~\ref{fig-channmaps} show weak emission near offsets (500\arcsec, 50\arcsec) at higher velocities, 5.0--7.7\,\kms, thus revealing another overlapping cloud. The main ridge has a small velocity gradient, with velocity increasing from east to west. Only near 3.6\,\kms, very weak emission is seen close to the position of the nova. 

Our CO maps cover an area that is larger than the map of \citet{Scott} but smaller than that of \citet{Hessman}. The sampling, resolution, and sensitivities of our maps are far superior to the earlier studies and show the region in a greater detail. It is clear that the incomplete coverage of the field in the maps of \citet{Scott} lead them to believe that they saw a point-symmetric structure originating from GK\,Per. Our maps reveal that the emission they found is only a fragment of a longer ridge that most likely is much larger than the field covered by our map and is part of the large-scale complex mapped by Hessman.

The ratio of the main-beam intensities of the CO(1--0) to (2--1) emission integrated over the entire line profile is between around 1.2 and 1.5 over most of the ridge surface, with the lower values representing the brightest southeastern part. The varying ratio reflects temperature changes coupled with variations in the line optical depth. Without measurements in one extra transition, neither can be unambiguously determined. Under the assumption of the local-thermodynamic equilibrium, however, we calculate that the typical excitation temperature is of about 12\,K. Near the region with peak emission, southeast of GK\,Per, the CO column densities are  about 1.0$\times$10$^{16}$\,cm$^{-2}$ and optical depths are about 3 or lower. In the brightest patch west of the star, the column densities are about three times lower. If the emission arises in a diffuse medium of low density, CO may be subthermally excited and a non-LTE approach would necessary for column density determinations. More transitions of CO or extra measurements in $^{13}$CO would constrain the physical conditions better. We did not manage to measure $^{13}$CO emission, but Hessman found a $^{12}$CO to $^{13}$CO emission ratio of 15 south of the area mapped by us, where the emission is stronger and more optically thick. This ratio is thus consistent with the interstellar nature of the emitting medium. 

\subsection{Interpretation}
The mapped CO cloud has every characteristic of an interstellar translucent cloud \citep[cf.][]{CloudReview}. The mass of the molecular material seen in our maps can be estimated using the X-factor method. We use the $X_{\rm CO}$ value of 2.0$\times$10$^{20}$ derived for translucent clouds \citep{xfactor}. Within 5$\sigma$ contours of the total intensity maps of CO(1--0), which includes the western and eastern parts of the ridge, we measured an average intensity of 4.9\,K\,\kms\ (in $T_{\rm mb}$) and an area of 4.1$\cdot$10$^4$ arcsec$^2$ or 1.8$\cdot$10$^{36}$ cm$^2$ (at 442\,pc). This yields 1.8$\cdot$10$^{57}$ particles of hydrogen or a mass of 3\,M$_{\sun}$. As we conservatively included only emission above 5$\sigma$, this is only a lower limit on the mass. The entire complex mapped by Hessman must be several times more massive still. Such a massive complex could not originate from the low mass cataclysmic binary GK\,Per.

The narrow lines that we observe in the field of GK\,Per are very typical for the low-density interstellar medium and would be highly unusual for any circumstellar envelope, practically ruling out the latter interpretation. As shown in Fig.\,\ref{fig-channmaps}, the peak velocities almost do not change over the different parts of the filament. For a circumstellar outflow, this would only be possible if the filamentary cloud were flowing out of the system in exactly the sky plane, which is highly unlikely (as even then a non-zero opening angle would lead to some spread in velocity). The kinematics of the molecular region is fully consistent with a quiescent interstellar cloud. 

The distribution of CO emission does not correlate strongly with that of the dust emission seen in $W3$ and $W4$ maps, or with the emission of \ion{H}{I} \citep{Seaquist}. A strong correlation between the three phases is not expected for the low-density ISM \citep[e.g.,][]{PlanckHI, ISMphases}. For instance, for the translucent cloud responsible for the light echoes of V838\,Mon no such strong correlation is found either  \citep[see discussion in][]{KamiEcho}.

What is then the relative location of the interstellar molecular cloud relative to GK\,Per? Already \citet{Hessman} noticed that the harsh radiation of the nova remnant, including the dwarf novae outbursts, may photodissociate the local interstellar clouds out to a radius of 0.5\,pc or 3\arcmin. Were the gap in the filament that appears to be centered on GK\,Per not coincidental but caused by the nova radiation, the interstellar filament must be physically close to the nova system. This would be also consistent with the 1901 light echo arising in the dusty counterpart of the CO cloud. Since forward scattering is dominating the echo generation, the scattering cloud is likely located slightly closer to us than GK\,Per. As noted by earlier authors, observing an evolved object such as GK\,Per close to an interstellar cloud is rather unexpected and should be rare. On the other hand, a few  such cases are known, including the cepheid RS\,Pup \citep{RSpup} or V732\,Sgr (Sect.\,\ref{sec-V732}). 

The systemic heliocentric velocity of GK\,Per is 32 (28--41)\,\kms\ \citep[][and references therein]{Alvarez} or 24\,\kms\ in the LSR frame. This velocity is quite different  from that of the observed CO emission in the field (1.5--8\,\kms). If these velocities are used as distance indicators, the two objects could be considered as located far apart along the line of sight. On the other hand, GK\,Per may be moving through the local interstellar medium with a velocity that does not follow closely the overall Galactic rotation pattern. If physically related with the cloud, GK\,Per would have a relative radial velocity of $\approx$15\,\kms\ (plus unknown relative tangential motion) which is supersonic and sufficient to produce a shock. Based on the proper motion of the nova, \citet{Bode2004} found a space velocity (proper and radial combined) of 45\,\kms.  As was noticed by \citet{Bode2004}, the bipolar nebula bears some signatures of interactions with the interstellar medium, but we find no other evidence of strong interaction with the ISM. 

The bipolar planetary nebula seen in WISE maps and in optical emission lines is most likely circumstellar in origin. As we have shown, it does not possess a molecular component, which is common among genuine PNe of various types and ages \citep[e.g.][]{HCNinPNe,liz}. The material filling the bipolar shell may be interacting with the inner nova ejecta, as suggest in earlier studies \citep[e.g.,][]{Harvey}. In particular, the jet identified within the ejecta is directed near the plane of the equatorial waist of the WISE structure. 

In conclusion, near GK\,Per we find only CO emission associated with the surrounding ISM. Elucidating the nature of the bipolar (planetary) nebula associated with  GK\,Per is very important for understanding the full evolution of this rather bizarre cataclysmic system. It is thus very unfortunate that it cannot be studied through mm-wave molecular lines.  

\section{Interferometric search of molecules in novae}\label{interer}
It appears that all single-dish observations of classical novae have resulted in non-detections of molecular gas. This may be a consequence of insufficient sensitivities or problematic baselines. Millimeter-wave interferometric arrays are usually more sensitive and offer flat spectroscopic baselines, and thus provide the most optimal way to test classical novae for the presence of molecules (Sect.\,\ref{intro}). Here, we present a few experiments targeting classical novae  that were conducted with modern interferometric millimeter arrays. 

A dedicated search for CO and HCN emission in four classical novae, NQ\,Vul, LW\,Ser, V2676\,Oph, and V3662\,Oph, was performed with the Submillimeter Array (SMA). We used the SWARM correlator \citep{swarm} which effectively covered four spectral ranges 224.5--232.5, 240.5--248.5, 336.0--344.0, and 352.0--360.0 GHz. The spectra thus cover the CO(2--1) transition at 230.538\,GHz and the HCN(4--3) transition at 354.505\,GHz. Each source was observed interchangeably with two nearby quasars, whose observations were later used for phase calibration. In each observing run, a bright quasar or a planet were observed for bandpass and absolute-flux calibration. The SMA data were calibrated in MIR\footnote{\url{https://lweb.cfa.harvard.edu/rtdc/SMAdata/process/mir/}} and further processed (including data flagging and imaging) in CASA. We typically observed with 7--8 SMA antennas and at baselines from 9 to 77\,m. The details and results of these observations are listed in Table\,\ref{tab-interfer}. 

None of the novae observed with the SMA was detected in molecular lines. One object, V3662\,Oph or Nova Oph 2017, was observed 2 months after its discovery in outburst and about a month after detection of its near-IR first-overtone band of CO \citep{AtelCO}. In SMA data, only free-free continuum was detected in the 224--249\,GHz window for this nova \citep{ATel}. We achieved the best sensitivity, of 9.6\,mJy, in observations of V2676 Oph in the 224--249\,GHz spectra. This observation is several times deeper than our best APEX observations in CO(3--2) at an rms noise of 2\,mK or about 82\,mJy. 

\begin{table*}[]
    \centering
    \caption{Novae observed with the SMA.}
    \begin{tabular}{cc cc c cc cc}\hline
         Object & Alter. name & RA (J2000)& Dec (J2000)& Obs. dates & Beam$_{\rm CO}$ & Beam$_{\rm HCN}$ & rms$_{\rm CO}$ & rms$_{\rm HCN}$ \\
                &           &    &     & 2017 & (\arcsec) & (\arcsec) & \multicolumn{2}{c}{(mJy/beam)} \\    \hline  
         NQ Vul & Nova Vul 1976 & 19:29:14.75 & +20:27:59.63 & 19,20 Jun; 5,9 Jul & 3.7$\times$3.0 & 3.4$\times$2.3 & 13.9 & 15.2\\
         LW Ser & Nova Ser 1978 & 17:51:50.89 &--14:43:50.60 & 5 Jul              & 4.0$\times$3.0 & 2.6$\times$2.1 & 14.7 & 36.7\\
      V2676 Oph & Nova Oph 2012 & 17:26:07.08 &--25:51:45.40 & 19 Jun             & 4.5$\times$3.1 & 2.9$\times$2.0 & ~9.6 & 27.1\\
      V3662 Oph & Nova Oph 2017 & 17:39:46.08 &--24:57:55.50 & 20 Jun             & 4.4$\times$3.1 & 2.8$\times$2.0 & 13.2 & 31.9\\
    \hline \end{tabular}
    \label{tab-interfer}
    \tablefoot{Beam sizes are given as FWHM at natural weighting of visibilities. The rms values, rms$_{\rm CO}$ and rms$_{\rm HCN}$, are specified for a spectral bin of 30\,\kms\ near the rest frequencies of CO(2--1) and HCN(4--3), respectively.}
\end{table*}

ALMA offers the greatest sensitivity among all currently available mm-wave instruments, but, to our knowledge, no program dedicated to molecule detection in classical novae has ever been conducted with the array. In the ALMA archive, we find only two classical novae observed so far. Observations towards T\,Pyx (L. Schmidtobreick) were obtained with the ALMA Compact Array (ACA), that is, with a smaller subset of the ALMA array composed of 7\,m antennas. The observations were made in bands 4, 6, and 7, which covered several transitions commonly observed in circumstellar media, but our inspection of the available data show that no line was detected. For instance, near the SiO(3--2) line at 130.3\,GHz, we get an rms of 4.0\,mJy per beam of 12\farcs3$\times$7\farcs1 and per 30\,\kms. This is comparable to the sensitivity of our experiments at the SMA (given the smaller SMA beams). 

The main array of ALMA observed V5668\,Sgr (Nova Sgr 2015 b) in 2017 (PI: M. Diaz) in band 6. Emission in continuum and in H30\,$\alpha$ near 231.9\,GHz was reported by \citet{Diaz}. Although their data cover CO(2--1), no molecular features were found. In a spectrum extracted for the entire continuum source of a diameter of about 0\farcs5, the 3$\sigma$ upper limit on the CO emission is 0.54\,mJy. This is possibly the most stringent upper limit on CO emission in a classical nova to date.

The experiments to detect molecules in classical novae with mm-wave interferometers conducted for young and old novae have given negative results so far, but the number of observed sources is still small. Future similar experiments are encouraged.

\section{Summary and conclusions}\label{summary}
Our single-dish survey of the 94 old classical novae and several dwarf novae demonstrates that circumstellar envelopes of post-novae systems are poor in molecular gas. Radiation of the classical novae outburst and the interstellar radiation field must effectively destroy most molecules. Dust shielding, even if provided shortly after the eruption, must be insufficient in the later expansion phases of the nova ejecta. 

With new maps of CO emission near GK\,Per, we debunked the long-standing conviction that the remnant of Nova 1901 is associated with a  circumstellar molecular cloud. We find only interstellar emission in the direction of the nova. It is possible that the dwarf nova flashes of GK\,Per destroy molecules in the local ISM, reinforcing our conclusion of a deadly effect of novae on molecular material. 

The lack of molecular matter in classical and dwarf novae remnants show also the remarkable difference with the newly recognized group of eruptive variables of red novae, which are rich in molecules decades and centuries after their eruptions. While red nova outbursts are often practically indistinguishable from classical novae, the amount of molecular material produced by these two classes of objects is drastically different and may be used for new identifications of red nova in the future. 

The upper limits we derive on molecular emission in classical novae are relatively shallow and do not indicate that nova shells are completely devoid of molecular matter. There is a chance that sensitive interferometric observations will be able to detect this emission, especially given the superior sensitivity of ALMA. A few experiments with the SMA and ALMA arrays have resulted in negative results so far, but the sample of novae at different times since the last eruption should be enlarged, especially for novae actively forming dust. There is still hope for measuring accurate isotopic ratios in classical novae using mm-wave data. 

\begin{acknowledgements}
We thank the referee, Nye Evans, for constructive comments on the manuscript. 
T.K. and H.M. acknowledge funding from grant no 2018/30/E/ST9/00398 from the Polish National Science Center. R.T. acknowledge a support from grant 2017/27/B/ST9/01128 financed by the Polish National Science Center. 

This research has made use of the SIMBAD database, operated at CDS, Strasbourg, France. This research made use of hips2fits, \url{https://alasky.u-strasbg.fr/hips-image-services/hips2fits} a service provided by CDS.

This paper makes use of ALMA data ADS/JAO.ALMA\#2019.2.00172.S and ADS/JAO.ALMA\#2016.1.00682.S. ALMA is a partnership of ESO (representing its member states), NSF (USA) and NINS (Japan), together with NRC (Canada), MOST and ASIAA (Taiwan), and KASI (Republic of Korea), in cooperation with the Republic of Chile. The Joint ALMA Observatory is operated by ESO, AUI/NRAO and NAOJ. The National Radio Astronomy Observatory is a facility of the National Science Foundation operated under cooperative agreement by Associated Universities, Inc.


Based on observations with the APEX telescope under programme ID 094.F-9503(A). APEX is a collaboration between the Max-Planck-Institut f\"ur Radioastronomie, the European Southern Observatory, and the Onsala Observatory.

Part of this work is based on observations carried out with the IRAM 30m telescope. IRAM is supported by INSU/CNRS (France), MPG (Germany) and IGN (Spain).

\end{acknowledgements}

\bibliographystyle{aa}
\bibliography{papers.bib}

\begin{thebibliography}{74}
\expandafter\ifx\csname natexlab\endcsname\relax\def\natexlab#1{#1}\fi

\bibitem[{{Albinson} \& {Evans}(1989)}]{AE1989}
{Albinson}, J.~S. \& {Evans}, A. 1989, \mnras, 240, 47P

\bibitem[{{Albinson} {et~al.}(1994){Albinson}, {Evans}, {Krautter}, \&
  {Weight}}]{Albinson}
{Albinson}, J.~S., {Evans}, A., {Krautter}, J., \& {Weight}, A. 1994, \aap,
  284, 971

\bibitem[{{{\'A}lvarez-Hern{\'a}ndez}
  {et~al.}(2021){{\'A}lvarez-Hern{\'a}ndez}, {Torres}, {Rodr{\'\i}guez-Gil},
  {Shahbaz}, {Anupama}, {Gazeas}, {Pavana}, {Raj}, {Hakala}, {Stone}, {Gomez},
  {Jonker}, {Ren}, {Cannizzaro}, {Pastor-Marazuela}, {Goff}, {Corral-Santana},
  \& {Sabo}}]{Alvarez}
{{\'A}lvarez-Hern{\'a}ndez}, A., {Torres}, M.~A.~P., {Rodr{\'\i}guez-Gil}, P.,
  {et~al.} 2021, \mnras, 507, 5805

\bibitem[{{Amari} {et~al.}(2001){Amari}, {Gao}, {Nittler}, {Zinner},
  {Jos{\'e}}, {Hernanz}, \& {Lewis}}]{amari2001}
{Amari}, S., {Gao}, X., {Nittler}, L.~R., {et~al.} 2001, \apj, 551, 1065

\bibitem[{{Banerjee} {et~al.}(2016){Banerjee}, {Srivastava}, {Ashok}, \&
  {Venkataraman}}]{banerjee2016}
{Banerjee}, D.~P.~K., {Srivastava}, M.~K., {Ashok}, N.~M., \& {Venkataraman},
  V. 2016, \mnras, 455, L109

\bibitem[{{Black}(2005)}]{black}
{Black}, J. 2005, in High Resolution Infrared Spectroscopy in Astronomy, 3--14

\bibitem[{{Bode} \& {Evans}(2008)}]{CNbook}
{Bode}, M.~F. \& {Evans}, A. 2008, {Classical Novae}, Vol.~43

\bibitem[{{Bode} {et~al.}(2004){Bode}, {O'Brien}, \& {Simpson}}]{Bode2004}
{Bode}, M.~F., {O'Brien}, T.~J., \& {Simpson}, M. 2004, \apjl, 600, L63

\bibitem[{{Bode} {et~al.}(1987){Bode}, {Seaquist}, {Frail}, {Roberts},
  {Whittet}, {Evans}, \& {Albinson}}]{BodeNat}
{Bode}, M.~F., {Seaquist}, E.~R., {Frail}, D.~A., {et~al.} 1987, \nat, 329, 519

\bibitem[{{Bublitz} {et~al.}(2019){Bublitz}, {Kastner},
  {Santander-Garc{\'\i}a}, {Bujarrabal}, {Alcolea}, \& {Montez}}]{HCNinPNe}
{Bublitz}, J., {Kastner}, J.~H., {Santander-Garc{\'\i}a}, M., {et~al.} 2019,
  \aap, 625, A101

\bibitem[{{Carter} {et~al.}(2012){Carter}, {Lazareff}, {Maier}, {Chenu},
  {Fontana}, {Bortolotti}, {Boucher}, {Navarrini}, {Blanchet}, {Greve}, {John},
  {Kramer}, {Morel}, {Navarro}, {Pe{\~n}alver}, {Schuster}, \& {Thum}}]{emir}
{Carter}, M., {Lazareff}, B., {Maier}, D., {et~al.} 2012, \aap, 538, A89

\bibitem[{{Cox} \& {Bronfman}(1995)}]{CB}
{Cox}, P. \& {Bronfman}, L. 1995, \aap, 299, 583

\bibitem[{{Derdzinski} {et~al.}(2017){Derdzinski}, {Metzger}, \&
  {Lazzati}}]{Derdzinski}
{Derdzinski}, A.~M., {Metzger}, B.~D., \& {Lazzati}, D. 2017, \mnras, 469, 1314

\bibitem[{{Diaz} {et~al.}(2018){Diaz}, {Abraham}, {Ribeiro}, {Beaklini}, \&
  {Takeda}}]{Diaz}
{Diaz}, M.~P., {Abraham}, Z., {Ribeiro}, V. A.~R.~M., {Beaklini}, P. P.~B., \&
  {Takeda}, L. 2018, \mnras, 480, L54

\bibitem[{{Dougherty} {et~al.}(1996){Dougherty}, {Waters}, {Bode}, {Lloyd},
  {Kester}, \& {Bontekoe}}]{Dougherty}
{Dougherty}, S.~M., {Waters}, L.~B.~F.~M., {Bode}, M.~F., {et~al.} 1996, \aap,
  306, 547

\bibitem[{{Evans}(1997)}]{DQHer}
{Evans}, A. 1997, \apss, 251, 293

\bibitem[{{Fujii} {et~al.}(2021){Fujii}, {Arai}, \& {Kawakita}}]{fujii}
{Fujii}, M., {Arai}, A., \& {Kawakita}, H. 2021, \apj, 907, 70

\bibitem[{Gehrz {et~al.}(1998)Gehrz, Truran, Williams, \& Starrfield}]{Gehrz}
Gehrz, R.~D., Truran, J.~W., Williams, R.~E., \& Starrfield, S. 1998,
  Publications of the Astronomical Society of the Pacific, 110, 3

\bibitem[{{G{\"u}sten} {et~al.}(2006){G{\"u}sten}, {Nyman}, {Schilke},
  {Menten}, {Cesarsky}, \& {Booth}}]{gusten}
{G{\"u}sten}, R., {Nyman}, L.~{\r{A}}., {Schilke}, P., {et~al.} 2006, \aap,
  454, L13

\bibitem[{{Guzman-Ramirez} {et~al.}(2018){Guzman-Ramirez},
  {G{\'o}mez-Ru{\'\i}z}, {Boffin}, {Jones}, {Wesson}, {Zijlstra}, {Smith}, \&
  {Nyman}}]{liz}
{Guzman-Ramirez}, L., {G{\'o}mez-Ru{\'\i}z}, A.~I., {Boffin}, H.~M.~J.,
  {et~al.} 2018, \aap, 618, A91

\bibitem[{{Harvey} {et~al.}(2016){Harvey}, {Redman}, {Boumis}, \&
  {Akras}}]{Harvey}
{Harvey}, E., {Redman}, M.~P., {Boumis}, P., \& {Akras}, S. 2016, \aap, 595,
  A64

\bibitem[{{Hessman}(1989)}]{Hessman}
{Hessman}, F.~V. 1989, \mnras, 239, 759

\bibitem[{{Hix}(2001)}]{hix}
{Hix}, W.~R. 2001, {Nuclear Reaction Rate Uncertainties and Their Effects on
  Nova Nucleosynthesis Modeling}, NASA STI/Recon Technical Report N

\bibitem[{{Howitt} {et~al.}(2020{\natexlab{a}}){Howitt}, {Stevenson},
  {Vigna-G{\'o}mez}, {Justham}, {Ivanova}, {Woods}, {Neijssel}, \&
  {Mandel}}]{rates}
{Howitt}, G., {Stevenson}, S., {Vigna-G{\'o}mez}, A., {et~al.}
  2020{\natexlab{a}}, \mnras, 492, 3229

\bibitem[{{Howitt} {et~al.}(2020{\natexlab{b}}){Howitt}, {Stevenson},
  {Vigna-G{\'o}mez}, {Justham}, {Ivanova}, {Woods}, {Neijssel}, \&
  {Mandel}}]{howitt}
{Howitt}, G., {Stevenson}, S., {Vigna-G{\'o}mez}, A., {et~al.}
  2020{\natexlab{b}}, \mnras, 492, 3229

\bibitem[{{Iliadis} {et~al.}(2018){Iliadis}, {Downen}, {Jos{\'e}}, {Nittler},
  \& {Starrfield}}]{iliadis}
{Iliadis}, C., {Downen}, L.~N., {Jos{\'e}}, J., {Nittler}, L.~R., \&
  {Starrfield}, S. 2018, \apj, 855, 76

\bibitem[{{Izzo} {et~al.}(2015){Izzo}, {Della Valle}, {Mason}, {Matteucci},
  {Romano}, {Pasquini}, {Vanzi}, {Jordan}, {Fernandez}, {Bluhm}, {Brahm},
  {Espinoza}, \& {Williams}}]{izzo}
{Izzo}, L., {Della Valle}, M., {Mason}, E., {et~al.} 2015, \apjl, 808, L14

\bibitem[{{Joshi} \& {Banerjee}(2017)}]{AtelCO}
{Joshi}, V. \& {Banerjee}, D. P.~K. 2017, The Astronomer's Telegram, 10369, 1

\bibitem[{{Kamenetzky} {et~al.}(2013){Kamenetzky}, {McCray}, {Indebetouw},
  {Barlow}, {Matsuura}, {Baes}, {Blommaert}, {Bolatto}, {Decin}, {Dunne},
  {Fransson}, {Glenn}, {Gomez}, {Groenewegen}, {Hopwood}, {Kirshner},
  {Lakicevic}, {Marcaide}, {Marti-Vidal}, {Meixner}, {Royer}, {Soderberg},
  {Sonneborn}, {Staveley-Smith}, {Swinyard}, {Van de Steene}, {van Hoof}, {van
  Loon}, {Yates}, \& {Zanardo}}]{sn1987a}
{Kamenetzky}, J., {McCray}, R., {Indebetouw}, R., {et~al.} 2013, \apjl, 773,
  L34

\bibitem[{{Kami\'nski} \& {Gehrz}(2017)}]{ATel}
{Kami\'nski}, T. \& {Gehrz}, R. 2017, The Astronomer's Telegram, 10536, 1

\bibitem[{{Kami{\'n}ski} {et~al.}(2015){Kami{\'n}ski}, {Menten}, {Tylenda},
  {Hajduk}, {Patel}, \& {Kraus}}]{KamiNat}
{Kami{\'n}ski}, T., {Menten}, K.~M., {Tylenda}, R., {et~al.} 2015, \nat, 520,
  322

\bibitem[{{Kami{\'n}ski} {et~al.}(2017){Kami{\'n}ski}, {Menten}, {Tylenda},
  {Karakas}, {Belloche}, \& {Patel}}]{KamiCKSingleDish}
{Kami{\'n}ski}, T., {Menten}, K.~M., {Tylenda}, R., {et~al.} 2017, \aap, 607,
  A78

\bibitem[{{Kami{\'n}ski} {et~al.}(2021){Kami{\'n}ski}, {Steffen}, {Bujarrabal},
  {Tylenda}, {Menten}, \& {Hajduk}}]{Kami2021}
{Kami{\'n}ski}, T., {Steffen}, W., {Bujarrabal}, V., {et~al.} 2021, \aap, 646,
  A1

\bibitem[{{Kami{\'n}ski} {et~al.}(2018{\natexlab{a}}){Kami{\'n}ski}, {Steffen},
  {Tylenda}, {Young}, {Patel}, \& {Menten}}]{kamiSubmm}
{Kami{\'n}ski}, T., {Steffen}, W., {Tylenda}, R., {et~al.} 2018{\natexlab{a}},
  \aap, 617, A129

\bibitem[{{Kami{\'n}ski} {et~al.}(2011){Kami{\'n}ski}, {Tylenda}, \&
  {Deguchi}}]{KamiEcho}
{Kami{\'n}ski}, T., {Tylenda}, R., \& {Deguchi}, S. 2011, \aap, 529, A48

\bibitem[{{Kami{\'n}ski} {et~al.}(2018{\natexlab{b}}){Kami{\'n}ski}, {Tylenda},
  {Menten}, {Karakas}, {Winters}, {Breier}, {Wong}, {Giesen}, \&
  {Patel}}]{Kami26Al}
{Kami{\'n}ski}, T., {Tylenda}, R., {Menten}, K.~M., {et~al.}
  2018{\natexlab{b}}, Nature Astronomy, 2, 778

\bibitem[{{Kato} \& {Kojiguchi}(2020)}]{BCCas}
{Kato}, T. \& {Kojiguchi}, N. 2020, \pasj, 72, 98

\bibitem[{{Kervella} {et~al.}(2012){Kervella}, {M{\'e}rand}, {Szabados},
  {Sparks}, {Havlen}, {Bond}, {Pompei}, {Fouqu{\'e}}, {Bersier}, \&
  {Cracraft}}]{RSpup}
{Kervella}, P., {M{\'e}rand}, A., {Szabados}, L., {et~al.} 2012, \aap, 541, A18

\bibitem[{{Kimeswenger}(2007)}]{Kimeswenger}
{Kimeswenger}, S. 2007, in Astronomical Society of the Pacific Conference
  Series, Vol. 363, The Nature of V838 Mon and its Light Echo, ed. R.~L.~M.
  {Corradi} \& U.~{Munari}, 197

\bibitem[{{Klein} {et~al.}(2012){Klein}, {Hochg{\"u}rtel}, {Kr{\"a}mer},
  {Bell}, {Meyer}, \& {G{\"u}sten}}]{klein}
{Klein}, B., {Hochg{\"u}rtel}, S., {Kr{\"a}mer}, I., {et~al.} 2012, \aap, 542,
  L3

\bibitem[{{Klein} {et~al.}(2014){Klein}, {Ciechanowicz}, {Leinz}, {Heyminck},
  {Gusten}, {Kasemann}, {Wunsch}, {Maier}, \& {Sekimoto}}]{flash}
{Klein}, T., {Ciechanowicz}, M., {Leinz}, C., {et~al.} 2014, IEEE Transactions
  on Terahertz Science and Technology, 4, 588

\bibitem[{{Kochanek} {et~al.}(2014){Kochanek}, {Adams}, \&
  {Belczynski}}]{kochanek}
{Kochanek}, C.~S., {Adams}, S.~M., \& {Belczynski}, K. 2014, \mnras, 443, 1319

\bibitem[{Li {et~al.}(2016)Li, Zhu, Lü, \& Wang}]{isotopesISM1}
Li, F., Zhu, C., Lü, G., \& Wang, Z. 2016, Publications of the Astronomical
  Society of Japan, 68
  [\eprint{https://academic.oup.com/pasj/article-pdf/68/3/39/6847894/psw030.pdf}],
  39

\bibitem[{{Liimets} {et~al.}(2012){Liimets}, {Corradi},
  {Santander-Garc{\'\i}a}, {Villaver}, {Rodr{\'\i}guez-Gil}, {Verro}, \&
  {Kolka}}]{Liimets}
{Liimets}, T., {Corradi}, R.~L.~M., {Santander-Garc{\'\i}a}, M., {et~al.} 2012,
  \apj, 761, 34

\bibitem[{{Loinard} {et~al.}(2012){Loinard}, {Menten}, {G{\"u}sten}, {Zapata},
  \& {Rodr{\'\i}guez}}]{Lonard}
{Loinard}, L., {Menten}, K.~M., {G{\"u}sten}, R., {Zapata}, L.~A., \&
  {Rodr{\'\i}guez}, L.~F. 2012, \apjl, 749, L4

\bibitem[{{Magnani} \& {Onello}(1995)}]{xfactor}
{Magnani}, L. \& {Onello}, J.~S. 1995, \apj, 443, 169

\bibitem[{{Mayall}(1949)}]{mayall}
{Mayall}, M.~W. 1949, \aj, 54, R191

\bibitem[{{Nagashima} {et~al.}(2014){Nagashima}, {Arai}, {Kajikawa},
  {Kawakita}, {Kitao}, {Arasaki}, {Taguchi}, \& {Ikeda}}]{detectionC2}
{Nagashima}, M., {Arai}, A., {Kajikawa}, T., {et~al.} 2014, \apjl, 780, L26

\bibitem[{{Nielbock} \& {Schmidtobreick}(2003)}]{NS2003}
{Nielbock}, M. \& {Schmidtobreick}, L. 2003, \aap, 400, L5

\bibitem[{{Pastorello} {et~al.}(2019){Pastorello}, {Mason}, {Taubenberger},
  {Fraser}, {Cortini}, {Tomasella}, {Botticella}, {Elias-Rosa}, {Kotak},
  {Smartt}, {Benetti}, {Cappellaro}, {Turatto}, {Tartaglia}, {Djorgovski},
  {Drake}, {Berton}, {Briganti}, {Brimacombe}, {Bufano}, {Cai}, {Chen},
  {Christensen}, {Ciabattari}, {Congiu}, {Dimai}, {Inserra}, {Kankare},
  {Magill}, {Maguire}, {Martinelli}, {Morales-Garoffolo}, {Ochner}, {Pignata},
  {Reguitti}, {Sollerman}, {Spiro}, {Terreran}, \& {Wright}}]{pastorello}
{Pastorello}, A., {Mason}, E., {Taubenberger}, S., {et~al.} 2019, \aap, 630,
  A75

\bibitem[{{Planck Collaboration} {et~al.}(2011){Planck Collaboration},
  {Abergel}, {Ade}, {Aghanim}, {Arnaud}, {Ashdown}, {Aumont}, {Baccigalupi},
  {Balbi}, {Banday}, \& et~al.}]{PlanckHI}
{Planck Collaboration}, {Abergel}, A., {Ade}, P.~A.~R., {et~al.} 2011, \aap,
  536, A24

\bibitem[{{Pontefract} \& {Rawlings}(2004)}]{PR2004}
{Pontefract}, M. \& {Rawlings}, J.~M.~C. 2004, \mnras, 347, 1294

\bibitem[{{Primiani} {et~al.}(2016){Primiani}, {Young}, {Young}, {Patel},
  {Wilson}, {Vertatschitsch}, {Chitwood}, {Srinivasan}, {MacMahon}, \&
  {Weintroub}}]{swarm}
{Primiani}, R.~A., {Young}, K.~H., {Young}, A., {et~al.} 2016, Journal of
  Astronomical Instrumentation, 5, 1641006

\bibitem[{{Reach} {et~al.}(2015){Reach}, {Heiles}, \& {Bernard}}]{ISMphases}
{Reach}, W.~T., {Heiles}, C., \& {Bernard}, J.-P. 2015, \apj, 811, 118

\bibitem[{{Ritchey}(1902)}]{Ritchey}
{Ritchey}, G.~W. 1902, \apj, 15, 129

\bibitem[{Romano \& Matteucci(2003)}]{isotopesISM2}
Romano, D. \& Matteucci, F. 2003, Monthly Notices of the Royal Astronomical
  Society, 342, 185

\bibitem[{{Schaefer}(1988)}]{schaefer}
{Schaefer}, B.~E. 1988, \apj, 327, 347

\bibitem[{{Schaefer}(2018)}]{distances}
{Schaefer}, B.~E. 2018, \mnras, 481, 3033

\bibitem[{{Sch{\"o}ier} {et~al.}(2013){Sch{\"o}ier}, {Ramstedt}, {Olofsson},
  {Lindqvist}, {Bieging}, \& {Marvel}}]{HCN}
{Sch{\"o}ier}, F.~L., {Ramstedt}, S., {Olofsson}, H., {et~al.} 2013, \aap, 550,
  A78

\bibitem[{{Scott} {et~al.}(1994){Scott}, {Rawlings}, \& {Evans}}]{Scott}
{Scott}, A.~D., {Rawlings}, J.~M.~C., \& {Evans}, A. 1994, \mnras, 269, 707

\bibitem[{{Seaquist} {et~al.}(1989){Seaquist}, {Bode}, {Frail}, {Roberts},
  {Evans}, \& {Albinson}}]{Seaquist}
{Seaquist}, E.~R., {Bode}, M.~F., {Frail}, D.~A., {et~al.} 1989, \apj, 344, 805

\bibitem[{{Shara} {et~al.}(2012){Shara}, {Zurek}, {De Marco}, {Mizusawa},
  {Williams}, \& {Livio}}]{Shara}
{Shara}, M.~M., {Zurek}, D., {De Marco}, O., {et~al.} 2012, \aj, 143, 143

\bibitem[{{Shore} \& {Braine}(1992)}]{SB1992}
{Shore}, S.~N. \& {Braine}, J. 1992, \apjl, 392, L59

\bibitem[{{Snow} \& {McCall}(2006)}]{CloudReview}
{Snow}, T.~P. \& {McCall}, B.~J. 2006, \araa, 44, 367

\bibitem[{{Swope}(1940)}]{swope}
{Swope}, H.~H. 1940, Harvard College Observatory Bulletin, 913, 11

\bibitem[{{Tweedy}(1995)}]{Tweedy}
{Tweedy}, R.~W. 1995, \apj, 438, 917

\bibitem[{{van den Bergh}(1977)}]{bergh}
{van den Bergh}, S. 1977, \pasp, 89, 637

\bibitem[{{{\v{S}}imon}(2002)}]{simon}
{{\v{S}}imon}, V. 2002, \aap, 382, 910

\bibitem[{{Wada} {et~al.}(2018){Wada}, {Yuasa}, {Nakazawa}, {Makishima},
  {Hayashi}, \& {Ishida}}]{wada}
{Wada}, Y., {Yuasa}, T., {Nakazawa}, K., {et~al.} 2018, \mnras, 474, 1564

\bibitem[{{Warner}(1995)}]{CVbook}
{Warner}, B. 1995, {Cataclysmic variable stars}, Vol.~28

\bibitem[{{Weight} {et~al.}(1993){Weight}, {Evans}, {Albinson}, \&
  {Krautter}}]{weight1993}
{Weight}, A., {Evans}, A., {Albinson}, J.~S., \& {Krautter}, J. 1993, \aap,
  268, 294

\bibitem[{{Wilson} \& {Merrill}(1935)}]{detectionCN}
{Wilson}, O.~C. \& {Merrill}, P.~W. 1935, \pasp, 47, 53

\bibitem[{{Wright} {et~al.}(2010){Wright}, {Eisenhardt}, {Mainzer}, {Ressler},
  {Cutri}, {Jarrett}, {Kirkpatrick}, {Padgett}, {McMillan}, {Skrutskie},
  {Stanford}, {Cohen}, {Walker}, {Mather}, {Leisawitz}, {Gautier}, {McLean},
  {Benford}, {Lonsdale}, {Blain}, {Mendez}, {Irace}, {Duval}, {Liu}, {Royer},
  {Heinrichsen}, {Howard}, {Shannon}, {Kendall}, {Walsh}, {Larsen}, {Cardon},
  {Schick}, {Schwalm}, {Abid}, {Fabinsky}, {Naes}, \& {Tsai}}]{wiseSurv}
{Wright}, E.~L., {Eisenhardt}, P. R.~M., {Mainzer}, A.~K., {et~al.} 2010, \aj,
  140, 1868

\bibitem[{{Zwitter} \& {Munari}(1996)}]{V365}
{Zwitter}, T. \& {Munari}, U. 1996, \aaps, 117, 449

\end{thebibliography}

\begin{appendix}
\section{Sample APEX and IRAM spectra}


\begin{figure*}
    \centering
    \includegraphics[scale = 0.33, trim={0 0 0 5cm},clip, page=1]{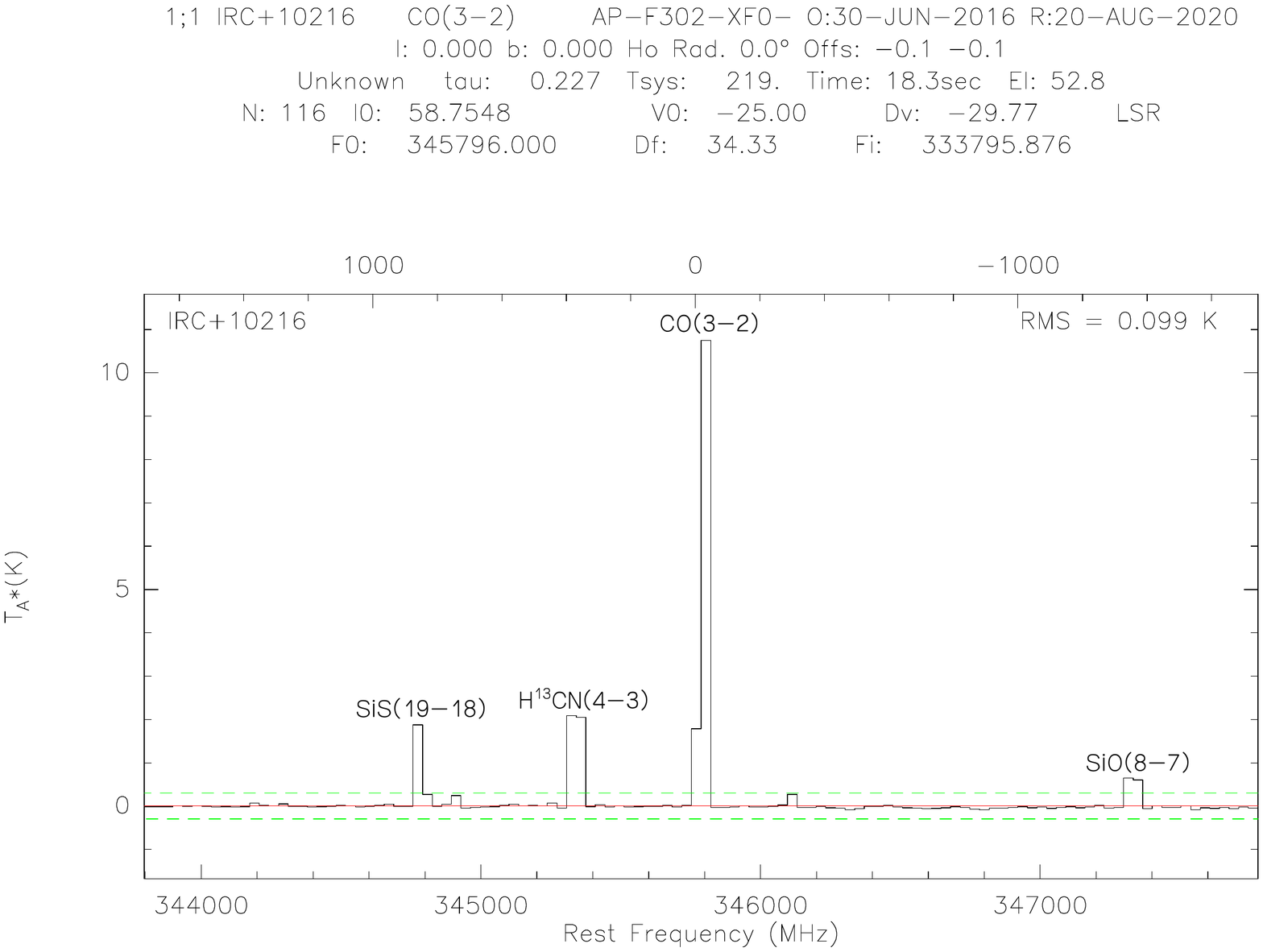}\hspace{-1.5cm}\vspace{-0.5cm}
    \includegraphics[scale = 0.33, trim={0 0 0 5cm},clip, page=11]{all_sm_CO32.pdf}\\
    \includegraphics[scale = 0.33, trim={0 0 0 5cm},clip, page=3]{all_sm_CO32.pdf}\hspace{-1.5cm}\vspace{-0.5cm}
    \includegraphics[scale = 0.33, trim={0 0 0 5cm},clip, page=4]{all_sm_CO32.pdf}\\
    \includegraphics[scale = 0.33, trim={0 0 0 5cm},clip, page=5]{all_sm_CO32.pdf}\hspace{-1.5cm}\vspace{-0.5cm}
    \includegraphics[scale = 0.33, trim={0 0 0 5cm},clip, page=6]{all_sm_CO32.pdf}\\
    \includegraphics[scale = 0.33, trim={0 0 0 5cm},clip, page=7]{all_sm_CO32.pdf}\hspace{-1.5cm}\vspace{-0.5cm}
    \includegraphics[scale = 0.33, trim={0 0 0 5cm},clip, page=8]{all_sm_CO32.pdf}\\
    \includegraphics[scale = 0.33, trim={0 0 0 5cm},clip, page=9]{all_sm_CO32.pdf}\hspace{-1.5cm}\vspace{-0.5cm}
    \includegraphics[scale = 0.33, trim={0 0 0 5cm},clip, page=10]{all_sm_CO32.pdf}\\
    \caption{Sample spectra covering the CO(3--2) transition at a resolution of 30\,\kms. The top left plot shows a reference spectrum of IRC+10216 with its most prominent lines labelled. The velocity scale is given with respect to CO(3--2). The green dashed lines mark 3$\sigma$ noise levels.}
    \label{apex-co32}
\end{figure*}

\begin{figure*}
    \centering
    \includegraphics[scale = 0.33, trim={0 0 0 5cm},clip, page=1]{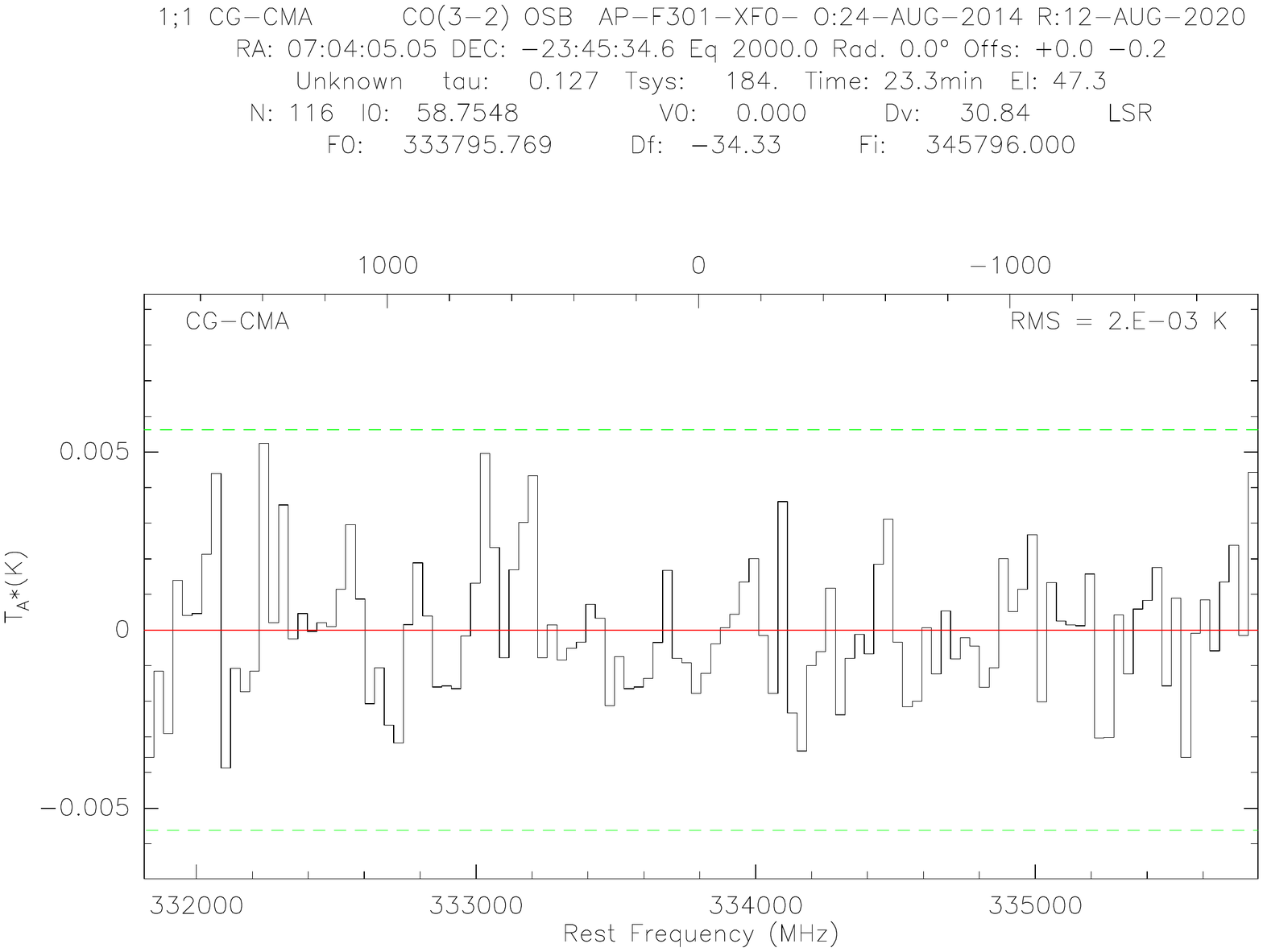}\hspace{-1.5cm}\vspace{-0.5cm}
    \includegraphics[scale = 0.33, trim={0 0 0 5cm},clip, page=11]{all_sm_CO32_OSB.pdf}\\
    \includegraphics[scale = 0.33, trim={0 0 0 5cm},clip, page=3]{all_sm_CO32_OSB.pdf}\hspace{-1.5cm}\vspace{-0.5cm}
    \includegraphics[scale = 0.33, trim={0 0 0 5cm},clip, page=4]{all_sm_CO32_OSB.pdf}\\
    \includegraphics[scale = 0.33, trim={0 0 0 5cm},clip, page=5]{all_sm_CO32_OSB.pdf}\hspace{-1.5cm}\vspace{-0.5cm}
    \includegraphics[scale = 0.33, trim={0 0 0 5cm},clip, page=6]{all_sm_CO32_OSB.pdf}\\
    \includegraphics[scale = 0.33, trim={0 0 0 5cm},clip, page=7]{all_sm_CO32_OSB.pdf}\hspace{-1.5cm}\vspace{-0.5cm}
    \includegraphics[scale = 0.33, trim={0 0 0 5cm},clip, page=8]{all_sm_CO32_OSB.pdf}\\
    \includegraphics[scale = 0.33, trim={0 0 0 5cm},clip, page=9]{all_sm_CO32_OSB.pdf}\hspace{-1.5cm}\vspace{-0.5cm}
    \includegraphics[scale = 0.33, trim={0 0 0 5cm},clip, page=10]{all_sm_CO32_OSB.pdf}\\
    \caption{Same as Fig.\,\ref{apex-co32} but for the CO(3--2)\,OSB setup. The velocity scale is with respect to 333.795769\,GHz.}
\end{figure*}

\begin{figure*}
    \centering
    \includegraphics[scale = 0.33, trim={0 0 0 5cm},clip, page=1]{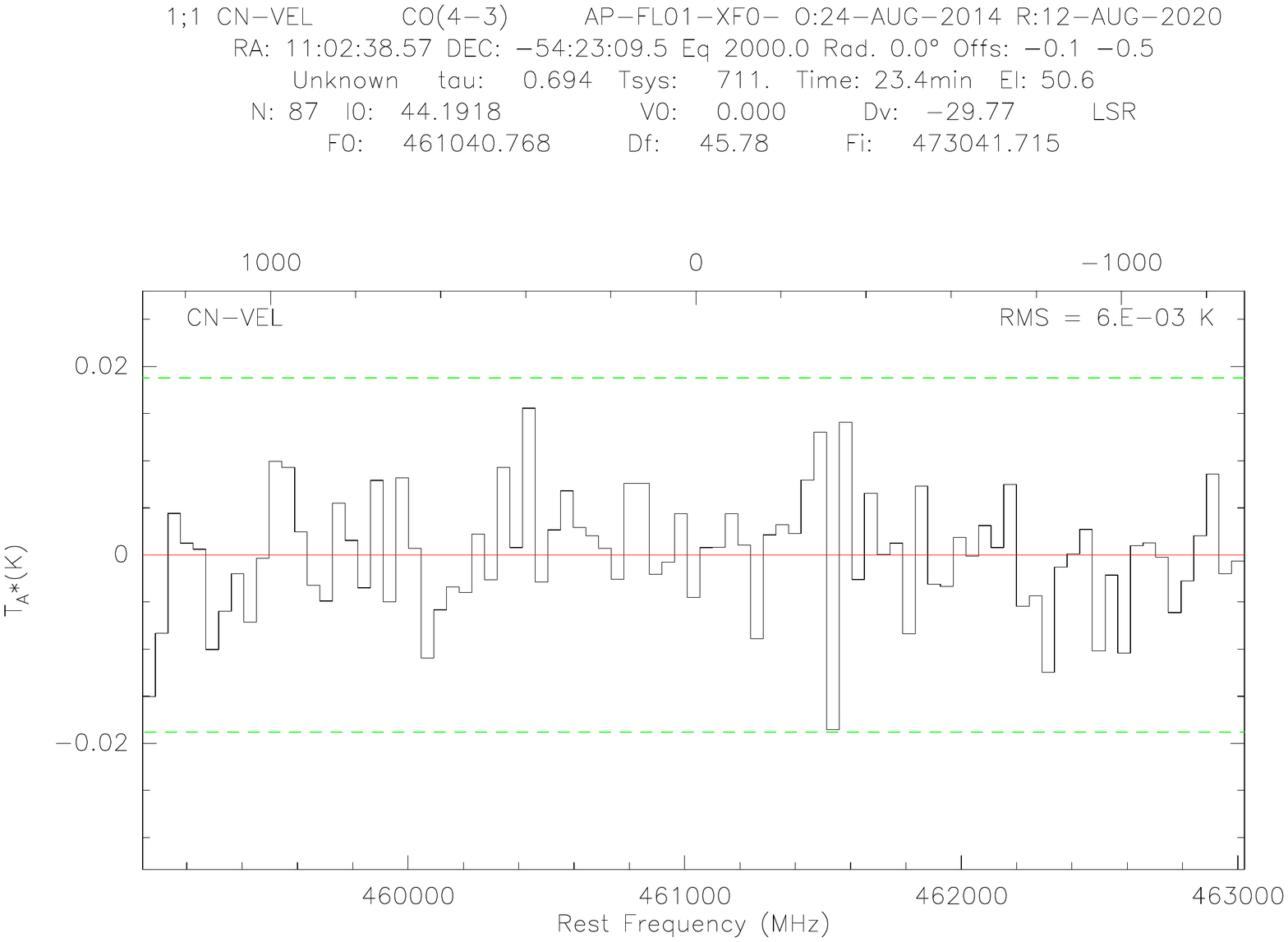}\hspace{-1.5cm}\vspace{-0.5cm}
    \includegraphics[scale = 0.33, trim={0 0 0 5cm},clip, page=11]{all_sm_CO43.pdf}\\
    \includegraphics[scale = 0.33, trim={0 0 0 5cm},clip, page=3]{all_sm_CO43.pdf}\hspace{-1.5cm}\vspace{-0.5cm}
    \includegraphics[scale = 0.33, trim={0 0 0 5cm},clip, page=4]{all_sm_CO43.pdf}\\
    \includegraphics[scale = 0.33, trim={0 0 0 5cm},clip, page=5]{all_sm_CO43.pdf}\hspace{-1.5cm}\vspace{-0.5cm}
    \includegraphics[scale = 0.33, trim={0 0 0 5cm},clip, page=6]{all_sm_CO43.pdf}\\
    \includegraphics[scale = 0.33, trim={0 0 0 5cm},clip, page=7]{all_sm_CO43.pdf}\hspace{-1.5cm}\vspace{-0.5cm}
    \includegraphics[scale = 0.33, trim={0 0 0 5cm},clip, page=8]{all_sm_CO43.pdf}\\
    \includegraphics[scale = 0.33, trim={0 0 0 5cm},clip, page=9]{all_sm_CO43.pdf}\hspace{-1.5cm}\vspace{-0.5cm}
    \includegraphics[scale = 0.33, trim={0 0 0 5cm},clip, page=10]{all_sm_CO43.pdf}\\
    \caption{Same as Fig.\,\ref{apex-co32} but for spectra covering the CO(4--3) transition. Velocities are with respect to the rest frequency of CO(4--3).}
\end{figure*}

\begin{figure*}
    \centering
    \includegraphics[scale = 0.33, trim={0 0 0 5cm},clip, page=1]{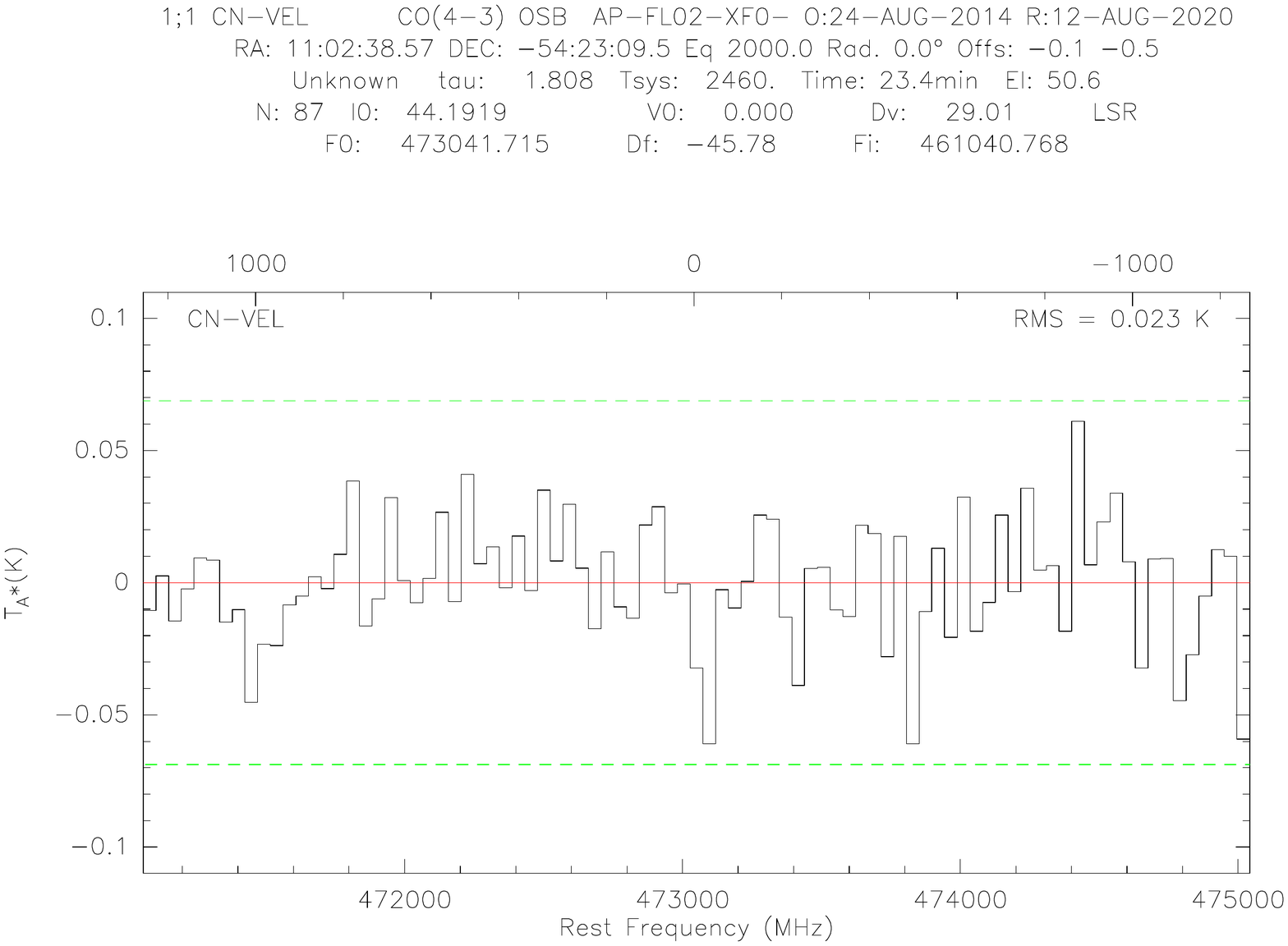}\hspace{-1.5cm}\vspace{-0.5cm}
    \includegraphics[scale = 0.33, trim={0 0 0 5cm},clip, page=11]{all_sm_CO43_OSB.pdf}\\
    \includegraphics[scale = 0.33, trim={0 0 0 5cm},clip, page=3]{all_sm_CO43_OSB.pdf}\hspace{-1.5cm}\vspace{-0.5cm}
    \includegraphics[scale = 0.33, trim={0 0 0 5cm},clip, page=4]{all_sm_CO43_OSB.pdf}\\
    \includegraphics[scale = 0.33, trim={0 0 0 5cm},clip, page=5]{all_sm_CO43_OSB.pdf}\hspace{-1.5cm}\vspace{-0.5cm}
    \includegraphics[scale = 0.33, trim={0 0 0 5cm},clip, page=6]{all_sm_CO43_OSB.pdf}\\
    \includegraphics[scale = 0.33, trim={0 0 0 5cm},clip, page=7]{all_sm_CO43_OSB.pdf}\hspace{-1.5cm}\vspace{-0.5cm}
    \includegraphics[scale = 0.33, trim={0 0 0 5cm},clip, page=8]{all_sm_CO43_OSB.pdf}\\
    \includegraphics[scale = 0.33, trim={0 0 0 5cm},clip, page=9]{all_sm_CO43_OSB.pdf}\hspace{-1.5cm}\vspace{-0.5cm}
    \includegraphics[scale = 0.33, trim={0 0 0 5cm},clip, page=10]{all_sm_CO43_OSB.pdf}\\
    \caption{Same as Fig.\,\ref{apex-co32} but for the CO(4--3)\,OSB setup. The velocity scale is given relative to 473.041715\,GHz.}
    \label{apex-co43-osb}
\end{figure*}


\begin{figure*}
    \centering
    \includegraphics[scale = 0.33, trim={0 0 0 5cm},clip, page=1]{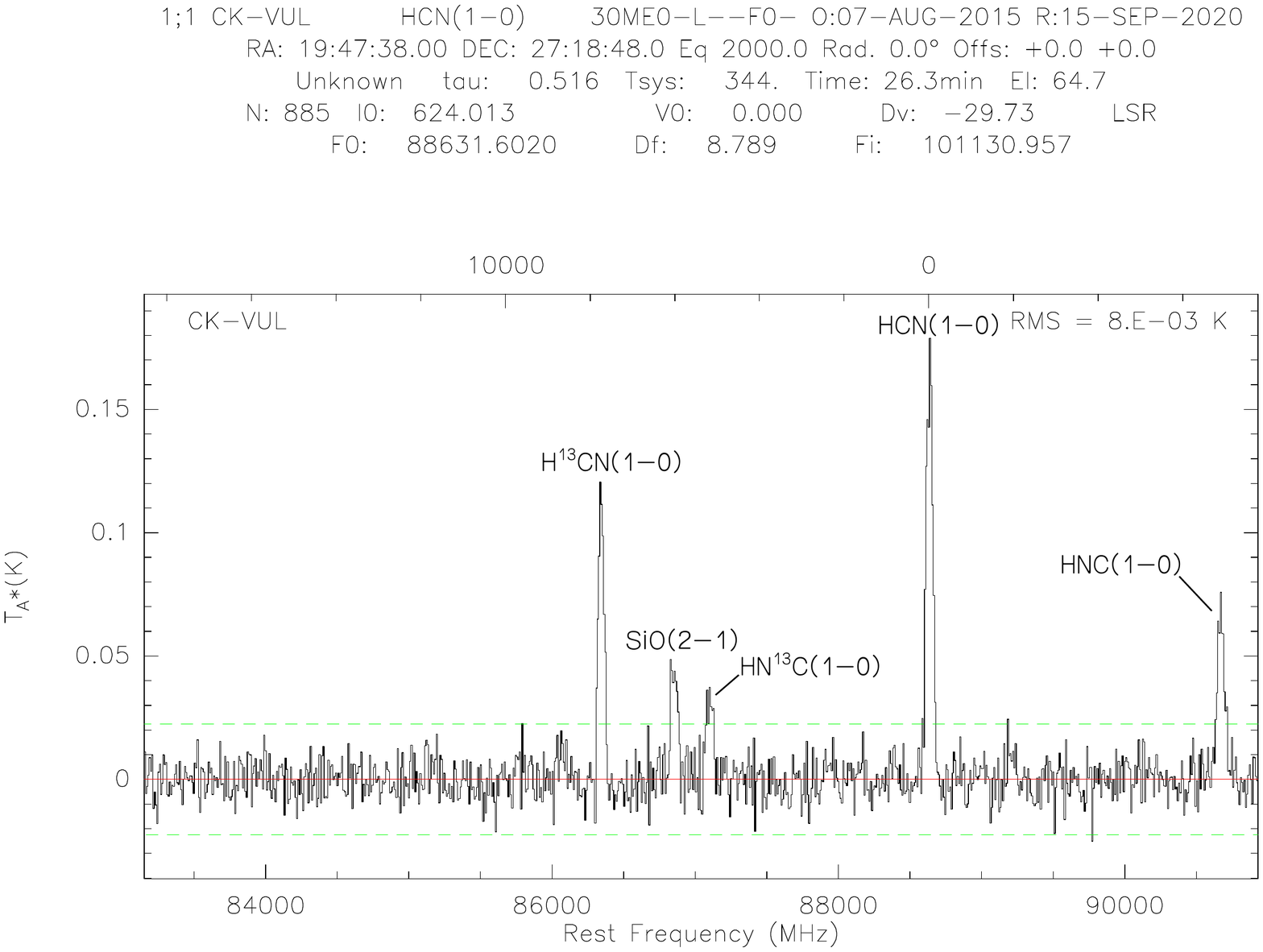}\hspace{-1.5cm}\vspace{-0.5cm}
    \includegraphics[scale = 0.33, trim={0 0 0 5cm},clip, page=2]{FTS_HCN_low.pdf}\\
    \includegraphics[scale = 0.33, trim={0 0 0 5cm},clip, page=3]{FTS_HCN_low.pdf}\hspace{-1.5cm}\vspace{-0.5cm}
    \includegraphics[scale = 0.33, trim={0 0 0 5cm},clip, page=1]{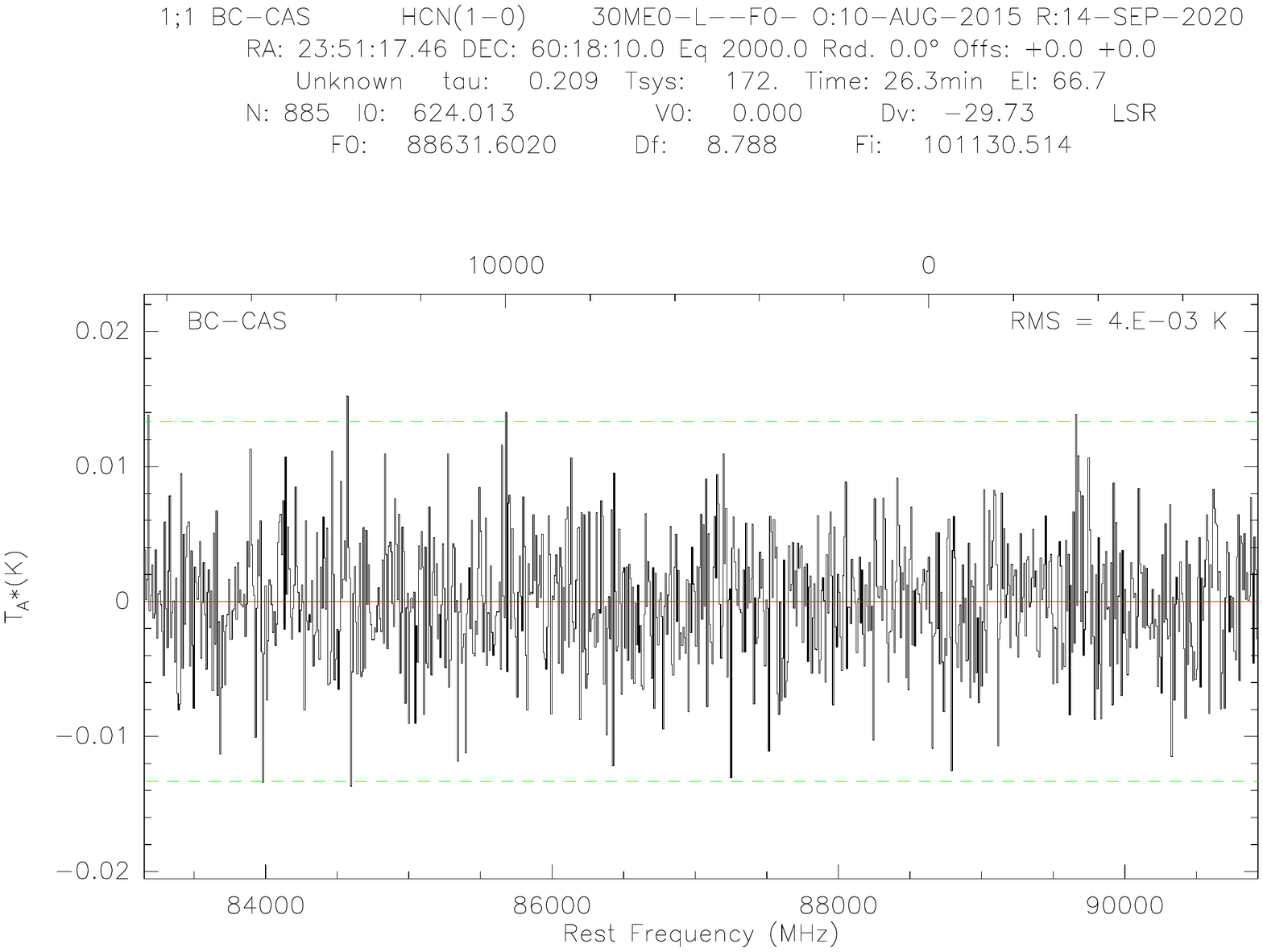}\\
    \includegraphics[scale = 0.33, trim={0 0 0 5cm},clip, page=5]{FTS_HCN_low.pdf}\hspace{-1.5cm}\vspace{-0.5cm}
    \includegraphics[scale = 0.33, trim={0 0 0 5cm},clip, page=6]{FTS_HCN_low.pdf}\\
    \includegraphics[scale = 0.33, trim={0 0 0 5cm},clip, page=7]{FTS_HCN_low.pdf}\hspace{-1.5cm}\vspace{-0.5cm}
    \includegraphics[scale = 0.33, trim={0 0 0 5cm},clip, page=8]{FTS_HCN_low.pdf}\\
    \includegraphics[scale = 0.33, trim={0 0 0 5cm},clip, page=9]{FTS_HCN_low.pdf}\hspace{-1.5cm}\vspace{-0.5cm}
    \includegraphics[scale = 0.33, trim={0 0 0 5cm},clip, page=10]{FTS_HCN_low.pdf}\\
    \caption{Sample IRAM/FTS spectra covering the HCN(1--0) transition and at a resolution of 30\,\kms. The top left plot shows a reference spectrum of CK\,Vul with its most prominent lines labelled. The top axes give LSR velocity with respect to the rest frequency of HCN(1--0).}
    \label{hcn-fts-low}
\end{figure*}

\begin{figure*}
    \centering
    \includegraphics[scale = 0.33, trim={0 0 0 5cm},clip, page=1]{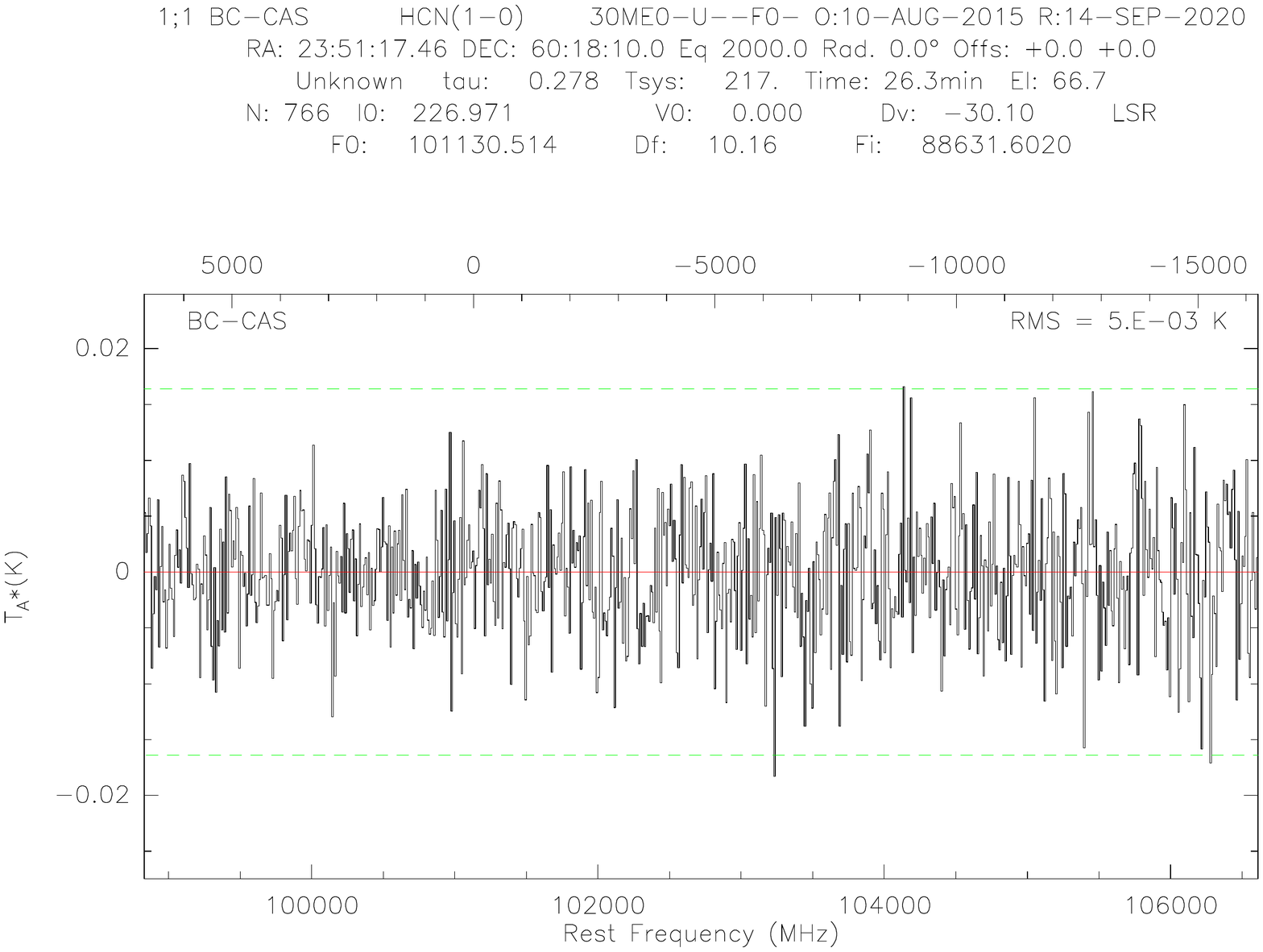}\hspace{-1.5cm}\vspace{-0.5cm}
    \includegraphics[scale = 0.33, trim={0 0 0 5cm},clip, page=2]{FTS_HCN_up.pdf}\\
    \includegraphics[scale = 0.33, trim={0 0 0 5cm},clip, page=3]{FTS_HCN_up.pdf}\hspace{-1.5cm}\vspace{-0.5cm}
    \includegraphics[scale = 0.33, trim={0 0 0 5cm},clip, page=4]{FTS_HCN_up.pdf}\\
    \includegraphics[scale = 0.33, trim={0 0 0 5cm},clip, page=5]{FTS_HCN_up.pdf}\hspace{-1.5cm}\vspace{-0.5cm}
    \includegraphics[scale = 0.33, trim={0 0 0 5cm},clip, page=6]{FTS_HCN_up.pdf}\\
    \includegraphics[scale = 0.33, trim={0 0 0 5cm},clip, page=7]{FTS_HCN_up.pdf}\hspace{-1.5cm}\vspace{-0.5cm}
    \includegraphics[scale = 0.33, trim={0 0 0 5cm},clip, page=8]{FTS_HCN_up.pdf}\\
    \includegraphics[scale = 0.33, trim={0 0 0 5cm},clip, page=9]{FTS_HCN_up.pdf}\hspace{-1.5cm}\vspace{-0.5cm}
    \includegraphics[scale = 0.33, trim={0 0 0 5cm},clip, page=10]{FTS_HCN_up.pdf}\\
    \caption{Same as Fig.\,\ref{hcn-fts-low} but for the upper side band. The top axes give LSR velocity with respect to 101.130514\,GHz.}
    \label{hcn-fts-high}
\end{figure*}

\begin{figure*}
    \centering
    \includegraphics[scale = 0.33, trim={0 0 0 5cm},clip, page=1]{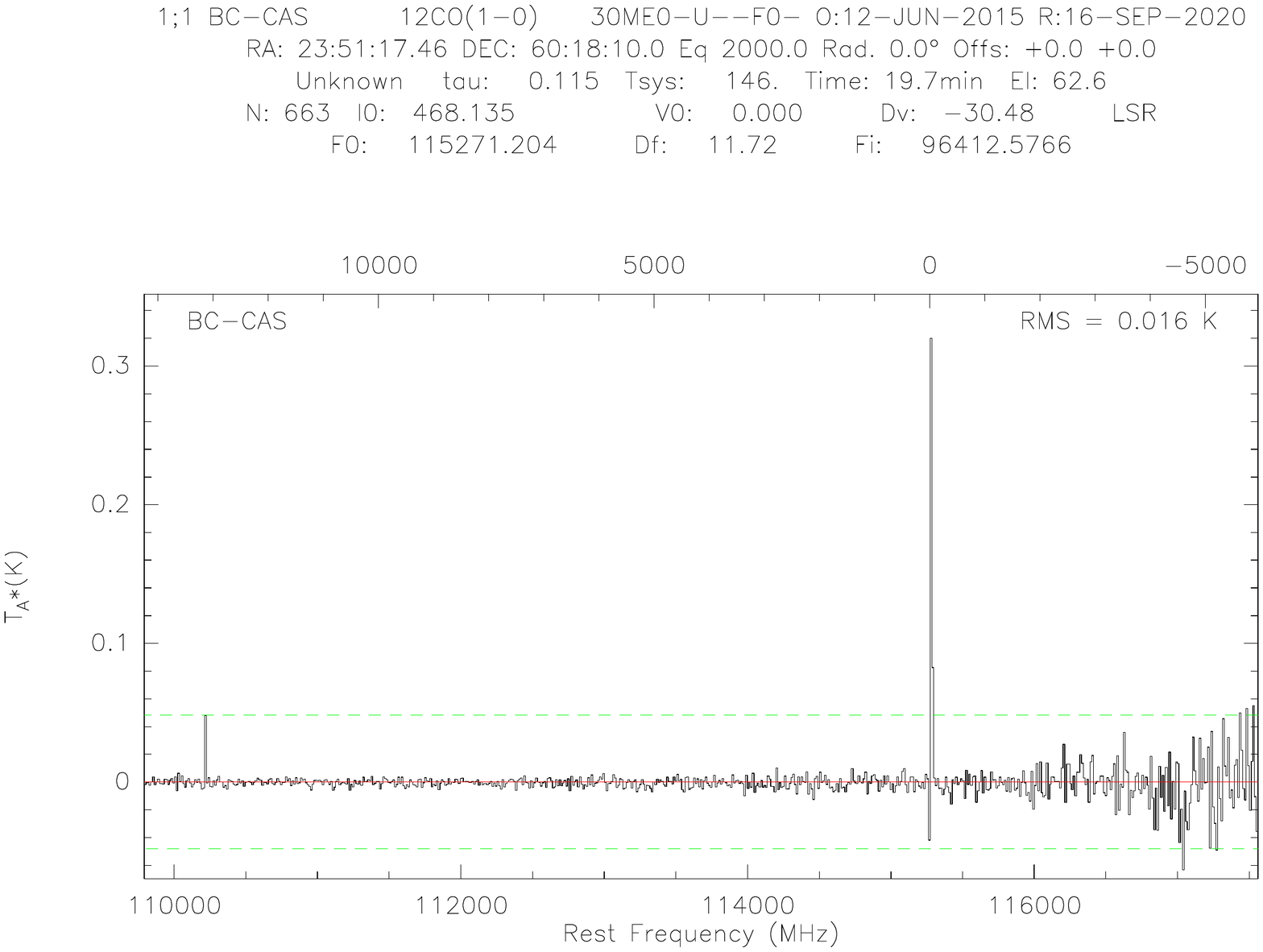}\hspace{-1.5cm}\vspace{-0.5cm}
    \includegraphics[scale = 0.33, trim={0 0 0 5cm},clip, page=2]{FTS_CO_up.pdf}\\
    \includegraphics[scale = 0.33, trim={0 0 0 5cm},clip, page=3]{FTS_CO_up.pdf}\hspace{-1.5cm}\vspace{-0.5cm}
    \includegraphics[scale = 0.33, trim={0 0 0 5cm},clip, page=4]{FTS_CO_up.pdf}\\
    \includegraphics[scale = 0.33, trim={0 0 0 5cm},clip, page=5]{FTS_CO_up.pdf}\hspace{-1.5cm}\vspace{-0.5cm}
    \includegraphics[scale = 0.33, trim={0 0 0 5cm},clip, page=6]{FTS_CO_up.pdf}\\
    \caption{Spectra from IRAM/FTS lower side band  covering the CO(1--0) transition and at a velocity resolution of 30\,\kms. The velocity scale is given relative to the rest frequency of CO(1--0).}
    \label{co-fts-low}
\end{figure*}

\begin{figure*}
    \centering
    \includegraphics[scale = 0.33, trim={0 0 0 5cm},clip, page=1]{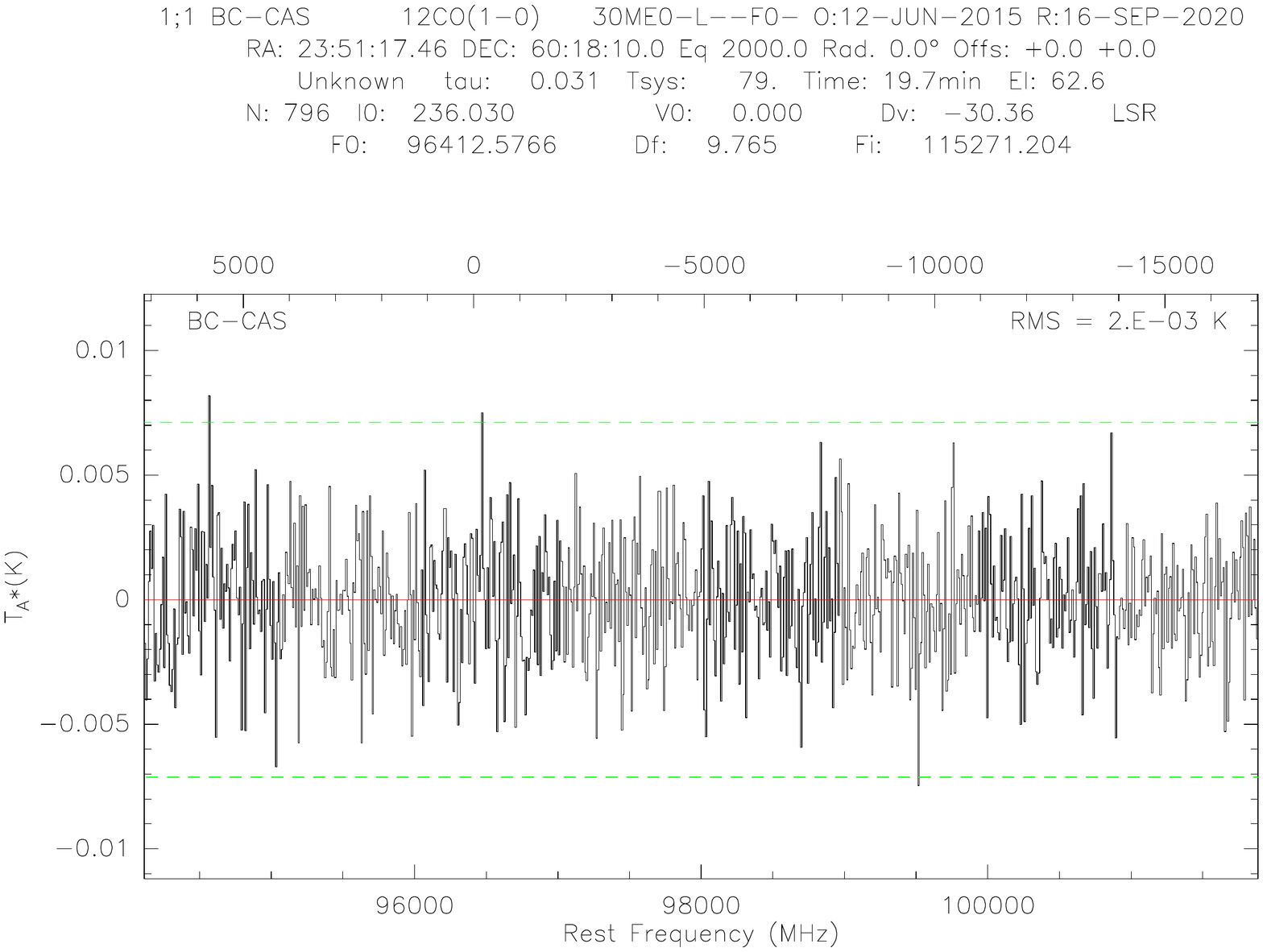}\hspace{-1.5cm}\vspace{-0.5cm}
    \includegraphics[scale = 0.33, trim={0 0 0 5cm},clip, page=2]{FTS_CO_low.pdf}\\
    \includegraphics[scale = 0.33, trim={0 0 0 5cm},clip, page=3]{FTS_CO_low.pdf}\hspace{-1.5cm}\vspace{-0.5cm}
    \includegraphics[scale = 0.33, trim={0 0 0 5cm},clip, page=4]{FTS_CO_low.pdf}\\
    \includegraphics[scale = 0.33, trim={0 0 0 5cm},clip, page=5]{FTS_CO_low.pdf}\hspace{-1.5cm}\vspace{-0.5cm}
    \includegraphics[scale = 0.33, trim={0 0 0 5cm},clip, page=6]{FTS_CO_low.pdf}\\
    \caption{Same as Fig.\,\ref{co-fts-low} but for the lower side band of the same setup. The velocity scale is relative to 96.4125766\,GHz.}
    \label{co-fts-high}
\end{figure*}
\end{appendix}

\end{document}